# A Novel Semantics and Feature Preserving Perspective for Content Aware Image Retargeting

by

## Sukrit Shankar

(CID: 00652983)

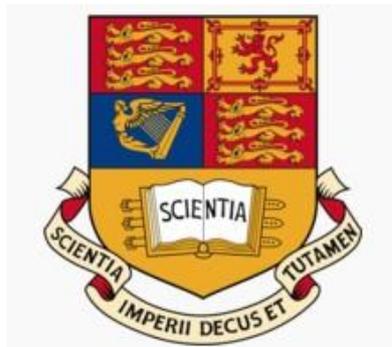

Submitted to the Department of Electrical and Electronic Engineering
in partial fulfilment of the requirements for the Degree of

## Master of Science

(Communications and Signal Processing)

at the

## IMPERIAL COLLEGE of SCIENCE, TECHNOLOGY & MEDICINE

September, 2011

under the supervision of

**Dr. Pier Luigi Dragotti**

# Declaration

I hereby declare that this submission as a part of the requirement for the degree of Master of Science in Communications and Signal Processing is my own work. I have expressed the entire content in my own words, incorporated my own ideas and judgements, and have provided relevant references and citations wherever required.

<div style="text-align: right">Sukrit Shankar</div>

# Abstract

There is an increasing requirement for efficient image retargeting techniques to adapt the content to various forms of digital media. With rapid growth of mobile communications and dynamic web page layouts, one often needs to resize the media content to adapt to the desired display sizes. For various layouts of web pages and typically small sizes of handheld portable devices, the importance in the original image content gets obfuscated after resizing it with the approach of uniform scaling. Thus, there occurs a need for resizing the images in a content aware manner which can automatically discard irrelevant information from the image and present the salient features with more magnitude.

There have been proposed some image retargeting techniques keeping in mind the content awareness of the input image. However, these techniques fail to prove globally effective for various kinds of images and desired sizes. The major problem is the inefficiency of these algorithms to process these images with minimal visual distortion while also retaining the meaning conveyed from the image.

In this dissertation, we present a novel perspective for content aware image retargeting, which is well implementable in real time. We introduce a novel method of analysing semantic information within the input image while also maintaining the important and visually significant features. We present the various nuances of our algorithm mathematically and logically, and show that the results prove better than the state-of-the-art techniques.

# Acknowledgements

I would like to take this opportunity to express my deepest gratitude to my supervisor, **Dr. P.L. Dragotti** for extending his support, valuable suggestions and guidance during the entire course of this dissertation.

The work presented in my dissertation has been made possible due to the benign nature, motivating and involving supervision and impeccable technical know-how of Dr. Dragotti. I have always considered myself very fortunate for getting to work under him during this dissertation, and it shall always remain one of the most memorable experiences of my technical journey. I was highly mesmerized by attending his course on *Wavelets and Applications* and it was very appreciative of him to accept my request for working under his supervision.

Dr. Dragotti in his esteemed stature has helped me develop an insight for analysing things more logically and fundamentally, which undoubtedly are very beneficial assets for any aspiring researcher. Apart from being a supervisor, he has also looked forward in helping and supporting me in my personal concerns, and providing me with immense freedom for carrying out the work. He is undoubtedly the best teacher, researcher, supervisor I have ever come across during the entire course of my living. With a hopeful glance into the future, I earnestly wish of working again with him.

I am grateful to my friends here at Imperial and otherwise, for constantly motivating me to pursue research and sharing unforgettable lighter moments with me. I would like to thank my former colleague Ayush Bhandari (now at EPFL) for providing me with his awesome photographs from Flickr for use in my image processing work.

I would like to express my heartiest gratitude to my parents for keeping utmost faith in me, and motivating in times of need, and specifically my mother in whose renounces resides my identity. Needless to mention, I am writing this narrative because of the Creator.

# Contents

# List of Figures

# 1 Introduction

This dissertation investigates the challenges in the state-of-the-art content aware image resizing techniques, and presents a **novel algorithm** for a globally improved and a real time efficient content aware image retargeting solution.

## 1.1 Motivation

Contemporary world demands the distribution of images to a wide variety of media platforms. With a wide variety of media devices available and demanding content distribution; typically, images need to be targeted to different media sources, and thus various sizes of the same image are often required.

Image retargeting finds one of its major applications in the design of modern web sites. Often, the chosen images for websites need to be adapted to different sizes to fit the layout structure, thereby resulting in image resizing, even sometimes to arbitrary aspect ratios. This can include either the reduction or the enlargement of the original image in one or both directions, or a combination of two in either direction. Also, dynamically changing the layout of a web site in browsers should consider apposite composition of the web sites' contents. This demands efficient image retargeting procedures for rendering visually pleasant images on web pages in different sizes.

Image retargeting is required while creating thumbnails for user browsing. This typically requires the reduction in the original size of the image, unless the image to be thumb nailed in substantially small, in which case, an appropriate algorithm for image enlargement is suitable.

Modern handheld devices and mobile phones often have small display screens, limited resolution, memory and processing power. Although, memory, processing power of mobile/portable devices has advanced to a reasonable extent with the availability of network bandwidths also improving, small screen size (display area) remains to be a persistent feature of such devices. As a result, all the details of images on these devices are cumbersome to view, and one often ends up zooming and panning for viewing different regions of an image. This requires the use of efficient image retargeting techniques.

Image retargeting has also shown its use in photographic applications. Liu et al. [1] depicts the use of cropping and retargeting methods to change the relative positions of the salient regions in the image and thus to modify the compositional aesthetics of the image.

*Image scaling* (also referred to as *homogenous image resizing*) has been in practice since long as an approach for image resizing, and a plethora of algorithms have been developed for resizing images to arbitrary aspect ratios in this manner. However, this very concept of image resizing does not in any way take into account the content information of the image. In elaborative words, each region of the image is homogenously resized by the same factor without considering the visual saliency model of the human visual system, and thus visually important regions in the original image may seem less significant after reduction in the image size. This very drawback with image scaling operation led to the development of the methods that could (automatically or through reasonable user interaction) track in some way, the visually important regions of an image, and then produce a resized version which despite preserving the salient regions more than non-salient regions in the resized version, also presents a visually coherent targeted image. Such methods are referred to as *content aware image resizing* techniques.

As mentioned above, content aware image resizing intends to resize the images taking into account the visual pleasantness of the image from a human viewpoint. *Fig. 1-1* depicts the differences between content aware resizing, homogenous scaling and image cropping (removing outer parts of a chosen rectangular region of an image). It can be clearly seen from Fig. 1-1 that scaling operation renders the boat in the image too thin thereby obfuscating the contents of the boat; cropping results in clipping of the right part of the boat; content aware image resizing preserves the whole structure of the boat and provides a better visually pleasant retargeted image.

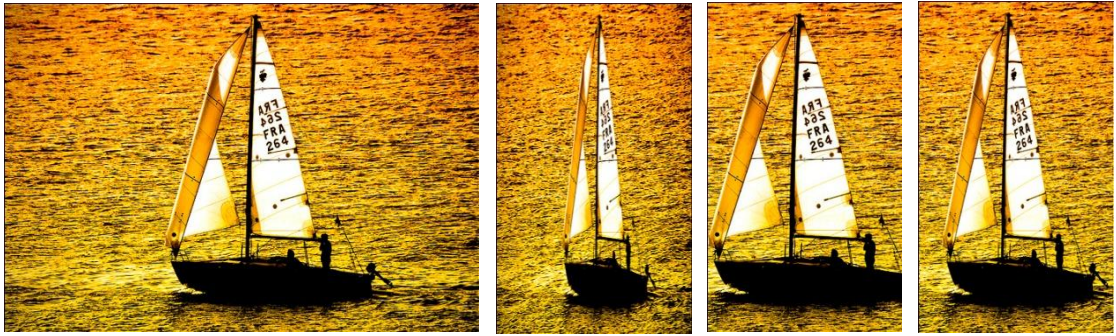

**Fig. 1-1** *(From left to right)* Original Image, Image Retargeted using scaling, Image retargeted using cropping, and Image retargeted using content aware image resizing. *The retargeting is attempted to make the width 40% of the original width with no change in the original image height. Here, we have used our novel algorithm (presented in Chapter 4) for content aware resizing. [Image Courtesy -* http://www.flickr.com/photos/ayushbhandari/1524259093/ *(Sailors at St. Malo)]*

It now becomes clearer that content aware image resizing aims to retarget the images in the way where the importantly visible regions are retained with least possible distortion, while the lesser significant regions are distorted more. It might appear from Fig. 1-1 that one might also reach the content aware retargeting result with manually adjustable cropping and scaling; however, such an operation is not always possible given infinite number of image compositions and in such cases, a generic algorithm for content aware resizing is always useful. This notion shall become more conspicuous in subsequent chapters where we delve into the topic.

With the idea of targeting images on small display devices in mind, some initial attempts were made for image and video retargeting on small display devices. Researchers in [2] - [9] have presented various ways of devising adaptive image and video resizing methods for mobile/handheld devices. These methods have focussed more on the image and video browsing on handheld devices with automatic zooming and panning based on the recognised/ user-specified region of interest. However, as we shall discuss in Chapter 2, more robust globally relevant techniques have evolved, which now drive the forum of content aware media retargeting.

It is noteworthy here, that the use of the terms *visual saliency, important regions (regions of interest),* being used in the context of content aware image retargeting are sort of abstract terms and do not form the definition of some well defined set, since for different images, their definition from the observers' point of view might change. For instance, one might argue that the text written on the boat is more significant than the boat itself subject to varied applications. This very notion has now given the concept of content aware image resizing a novel application. Content aware resizing methods now also find their use when an image needs to be resized for an embedded object's/region's protection. With professional image processing software such as Adobe Photoshop CS5 [10], one often attempts to protect or remove some user defined area in an image. This requires an adaptation of the image retargeting algorithm at hand since the entire problem now gets added with a user defined constraint, and thus can be roughly thought of as resizing different regions of the image constrained by non-rectangular or arbitrary object boundaries, while also preserving image's visual coherence with a plausibly different but meaningful semantics.

## 1.2 Challenges

With various algorithms developed for content aware image resizing in the recent past, the major problems faced can be summarized as follows.

- With various methods (in detail are discussed in Chapter 2) proposed for content aware resizing and the resizing quality depending upon the image semantics, it can be seen that while one method may prove perfect for a certain image and aspect ratio, the other method might totally fail; and vice versa. The amount of distortion and the visual unpleasantness depends on the type of the image and the targeted size desired apart from the retargeting algorithm used. For some images, the retargeting results may look very awkward and in such cases one might argue that the simple scaling operation proves better. *Fig. 1-2 and Fig. 1-3* represent two such cases where most users might usually prefer pick scaling operation for image retargeting. Looking at Fig. 1-2, one might argue that the image already contains too many objects placed near to each other, and content aware resizing method used for retargeting in the figure is more applicable for images having a good amount of non-salient regions [12]. To counter this fact, we have considered Fig. 1-3, which clearly has lesser salient regions than the original image in Fig. 1-2. However, looking at the results in Fig. 1-3, one can say that scaling operation presents more visual pleasantness and lesser awkward semantics at places. Thus, image retargeting algorithms should also make appropriate adjustments according to the ***semantics*** of the input *image* apart from apposite ***feature preservation*** based content aware retargeting. (As we shall discuss in Chapter 2, some researchers have tried to combine different approaches in many ways; however, a computationally effective and a globally reliable method is a challenge to be sought after). Thus, there lies a challenge in devising a globally effective content aware image retargeting algorithm which can cater to most types of images and various aspect ratios.
- Researchers to some extent although have been trying to counter the above challenges; however, while doing so, they typically tend to lose on the computational efficiency of the entire algorithm. Thus, devising a novel algorithm which mitigates the aforementioned problems to a globally reasonable extent while also maintaining a real time computational efficiency of the method is a challenge.

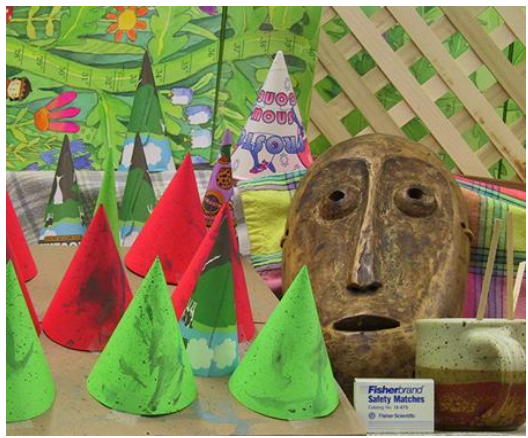

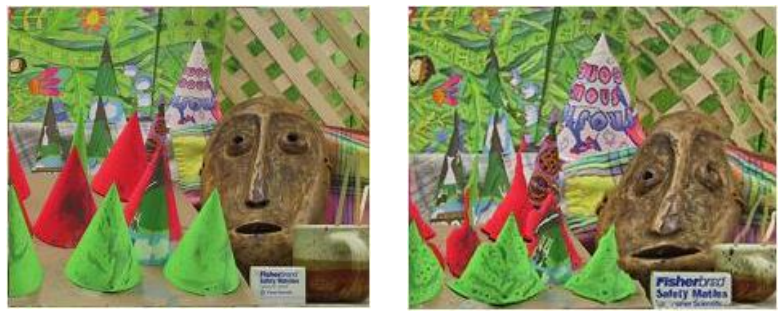

**Fig. 1-2** *(Top)* Original Image *(Bottom – Left to Right)* Image Retargeted with scaling, and Image retargeted with content aware image resizing using [12]. *The retargeting is attempted to make the both the width and the height of the original image reduced by 50%. [Image Courtesy – Middlebury Vision Stereo Dataset (Cones [11])]*

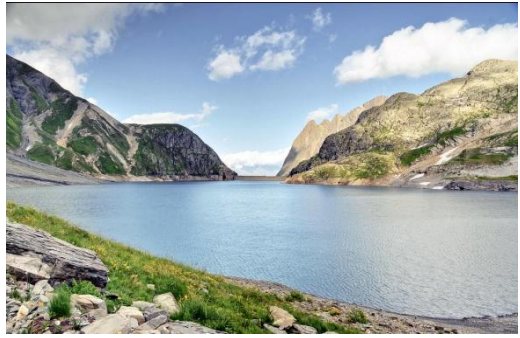

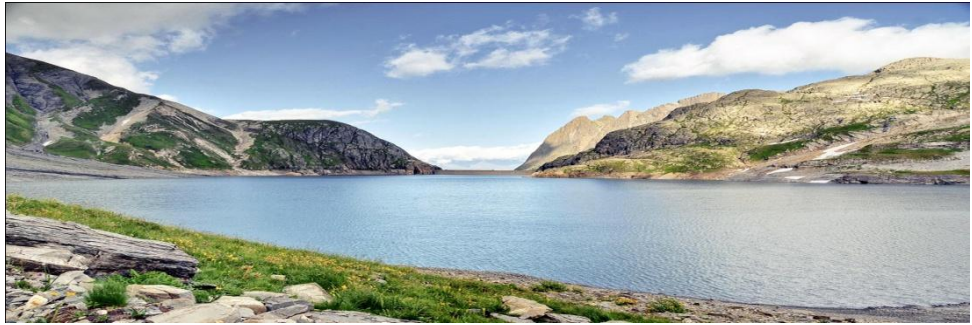

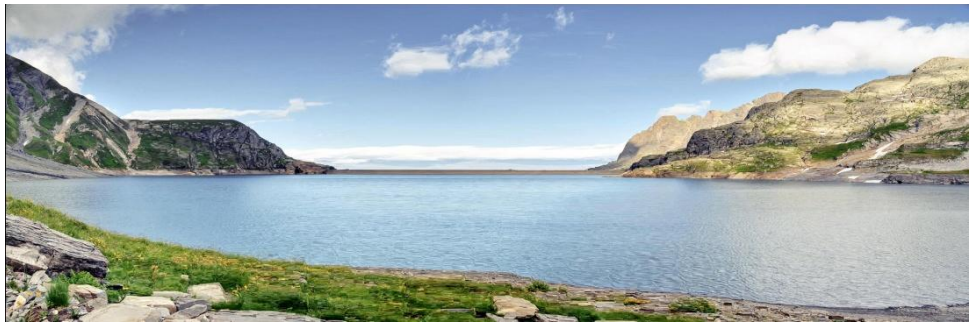

**Fig. 1-3** *(From top to bottom)* Original Image, Image Retargeted with scaling, and Image retargeted with content aware image resizing using [12]. *The retargeting is attempted to make the width 1.95 times the original width with no change in the original height. [Image Courtesy - http://www.flickr.com/photos/ayushbhandari/4846948022/ (Lake Emosson)]*

## 1.3 Contribution

We have followed a holistic approach (delineated in Chapter 3 ahead) for achieving content aware image resizing to counter the aforementioned challenges in a much better way, which caters to images of several types. It is needless to say, that any such algorithm cannot be claimed to be perfect for all types of images, since we are always dealing, although somewhat virtually rather than directly, with the concepts of image semantics, image saliency, regions of interest, etc from a computer vision point of view. It can only be argued that the proposed algorithm fundamentally is adequate to process a more varied variety of images for more varied targeted aspect ratios, with conformal evidence in the relevant results.

In this dissertation, we have developed a novel and a real time efficient algorithm for content aware image retargeting, and have shown with the simulation results that our algorithm proves better than the present state-of-the-art for a variety of images (details given in Chapters 3 and 4). Our major contribution in this dissertation work can be summarized as follows.

➢ A novel technique is devised to confer the semantics of the image with regard to the content aware resizing operation in specificity. This is altogether different from typical classification of image

> pixels in pre-defined semantic sets. Our approach considers the saliency and the non-saliency of the original image content keeping in mind the targeted size desired. This provides better results (as we shall show in Chapters 3 and 4).

> Our algorithm depends on the non-homogenous mesh based image warping method, and we have devised novel and more robust cost functions for the purpose of content aware retargeting. Also, we have tried to keep the computational complexity of the algorithm within a feasible limit so that a C/C++ version of the algorithm is well implementable in real time.

> As mentioned previously that no algorithm can be deemed as perfect given various types of images, various desired sizes, and varied user based choices of defining regions of interest, (This is typically the case when a person wants to morph some image to highlight some objects of his/her interest more than others, and typically does not do what most users would do) we introduce some adjustable parameters which can be used to make our system interactive for justifiable user interaction. The user parameters in no way make the algorithm less generic or expect something to be only specified by the user; rather, it adds more flexibility to users' demands.

## 1.4 Outline of the dissertation

*Fig. 1-4* outlines the major chapters in this dissertation with their titles, and brief summary of their content. This is essentially the way this dissertation stands organized and know-how of the way the dissertation should be approached by a reader.

| Number | Chapter Title | Summary of Chapter's Content |
|:---:|:---:|:---:|
| 1 | Introduction | Current Chapter |
| 2 | Overview of Content Aware Image Retargeting Techniques | The chapter provides a comprehensive overview of the content aware resizing techniques, advantages and limitations of various methods, and relevant algorithmic connections during the entire research. The chapter is much more than a mere adaptation from the previous literature in the sense that it also contains our own analysis of the various methods with visible evidence. |
| 3 | Novel Perspective for Content Aware Image Retargeting | The chapter describes our novel algorithm delineating our contribution of Section 1.3. It provides complete mathematical and logical analysis of our algorithm, and image snippets that help to better visualize various notions presented in the algorithm. |
| 4 | Results and Discussions | The chapter contains the final simulation results of our algorithm for vast variety of images taken from various sources, and compares to the state-of-the-art techniques. |
| 5 | Future Work and Conclusions | The chapter comprises of the concluding remarks and suggestions for future research work in this area. |

**Fig 1-4** Outline of the dissertation. It gives the titles of various Chapters included in the dissertation along with a brief description of their contents.

# 2 Overview of Content Aware Image Retargeting Techniques

This chapter presents a comprehensive overview of the content aware image retargeting techniques. Although, a review of the image retargeting techniques appears in [13], it lacks some of the recent improvements in this domain and does not unfold the nuances and the various pitfalls of the discussed algorithms so as to help the reader get a research oriented and a thoughtful insight. We start with defining the image retargeting problem in a generic manner and continue with discussing the evolution of this domain since years and the advantages/disadvantages of each of the algorithms designed. We delve into the open issues and problems that various state-of-the-art algorithms have presented, and thereby provide a platform for the reader to appreciate our novel algorithm and the associated results and discussions in Chapters 3 and 4.

## 2.1 Formulation of the Image Retargeting Problem

The image retargeting problem starts with converting a digital image *I* of dimensions *m x n* (*m* rows and *n* columns) to a target image *I'* of dimensions *m' x n'* (*m'* rows and *n'* columns). The dimensions given here specify the number of rows and columns in the image only, and we are not very concerned with the RGB or the gray scale nature of the image while retargeting. From a more technical and a programmatic point of view, while a gray scale image will have dimensions of *m x n,* an RGB image shall have the corresponding dimensions of *m x n x 3;* the factor 3 being there for the *red (R), green (G), blue (B)* components of the image. However, for a content aware image retargeting problem, we are required to scale a gray scale image into a gray scale one and an RGB image into a RGB one, with change in the number of rows and columns of the image only.

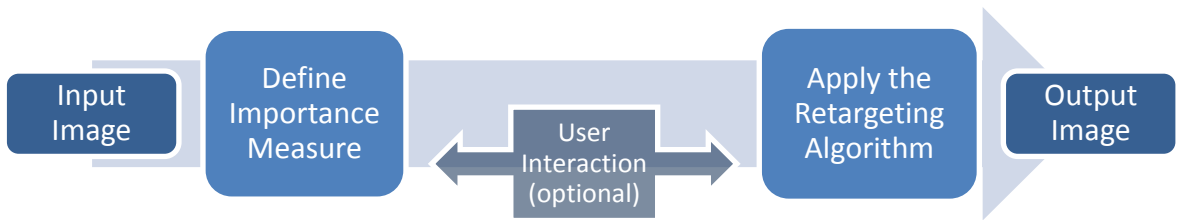

**Fig. 2-1** Formulation of the Image Retargeting Problem. A desired targeted size is generally (unless the approach is implemented as an interactive one) defined before input image is processed.

Fig. 2-1 shows diagrammatically the formulation of the image retargeting problem in a generic manner. Here, *user interaction* is optional since the image retargeting problem is often designed to be automatic; however, given the infinite number of image compositions combined with the various applications, an optional user interaction adds more flexibility. The term *user interaction* typically has the following meanings.

> ➢ A user interaction towards the definition of the importance map is essentially the specification by the user of the regions he/she perceives as important for their application or otherwise. Such an interaction can either be through specifically mentioning the sampling points for the desired important regions or through the specification of some adjustable parameters provided by the interface. The adjustable parameters often are provided in a way so that their values draw a trade off between the mostly perceivable salient regions and the mostly perceivable non-salient regions of the image. Thus, through the parameter adjustment, the user might be able to protect the salient regions more than the others and vice versa. However, the adjustable parameter may not correspond to the exact need of the user and in such a case; the former case of an exact specification of the feature region(s) suffices.

- ➢ A user interaction towards the retargeting algorithm can be in two ways. First, the user can specify certain constraints for the retargeting problem, such as outlining an object of interest to be protected. In such a case, the retargeting problem (which is typically constrained to maintain the rectangular shape of the entire image) is additionally constrained not to disturb some arbitrary boundary of an embedded object. *Fig. 2-2* shows such a scenario. A somewhat looser constraint is sometimes specified as having the aspect ratio of the selected object preserved if not, along with the original size. Second, the user can specify the various adjustable parameters which form a part of the image retargeting algorithm. Although not all retargeting algorithms consist of adjustable parameters, some of the algorithms present the user with such an option. The parameters help the user get slightly different results for different values.

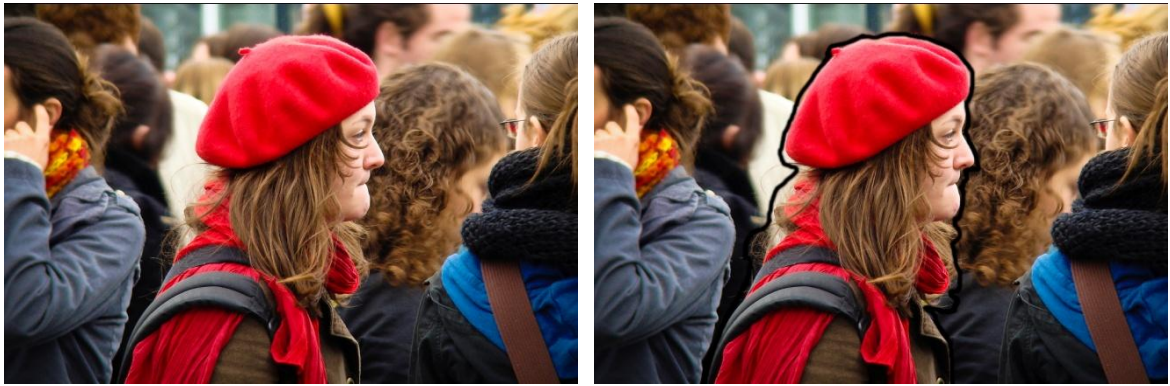

**Fig. 2-2** *(Left to Right)* Input image with the typical constraints of maintaining the rectangular shape of the image, Input image with an additional user specified constraint of protecting the girl with the red hat while retargeting (marked with black outline). *Sometimes the constraint can be loosely specified to have the aspect ratio of the selected object preserved instead of the original size of the object. [Image Courtesy - http://www.flickr.com/photos/ayushbhandari/2042304502/ (Red Hat is Not Usually Linux)]*

As is evident from *Fig. 2-1,* the recognition of an importance measure is the first step in any image retargeting problem. Although, we shall discuss the various importance measures used in the image retargeting domain along with the corresponding algorithms in which they were first used / proposed, it becomes worthwhile here to mention a top level abstraction of the class of importance measures.

There are essentially two categories of approaches to automatically estimate the importance map of an image (we reiterate that the term *importance measure* is not a well defined set, and the definition is always subjective apropos to human context and attention), viz. bottom-up methods, and top-down methods. While bottom-up methods are based on low-level features, top-down methods make use of classifiable semantic information [15]. Low level features typically include edges at various orientations, color, and intensities. Semantic information generally comprises of faces, human/animal bodies, text, objects of interest such as sky, trees, road, grass, water, mountains, etc [16]. Top-down approaches are sometimes combined with bottom-up saliency estimation approaches to generate the importance measure.

Once the retargeting is done, it becomes essential to have some sort of criteria to compare the input image *I* with the output image *I'*. For an image retargeting problem, a subjective or qualitative criteria of comparison suffices within the constraints of the requirement that the content in *I'* should reflect the important regions in *I* with the geometry and the structure in *I'* free from visual artefacts as far as possible [14]. There are many quantitative criteria as well for such comparisons, which have been used by some researchers (discussed later in this Chapter) for forming cost functions in the process of image retargeting. However, such comparisons are often deemed as not very reliable and it would not be wrong to say here, that a reliable objective comparison from an image retargeting problem is still sought after.

## 2.2 Scaling for Image Retargeting

Content aware image retargeting using scaling operation is in general not a very useful approach. We had briefly pointed out in *Chapter 1* that the scaling operation results in not preserving the important regions of the image. Although, there exist no algorithms which in a real sense, talk about the content preservation of an image during the homogenous image resizing operation; some methods talk about the minimization of the loss of information and the preservation of features like noise and blur during image reduction.

Researchers in [17] have proposed an optimal spline based algorithm for image resizing with arbitrary (non integer) scaling factors. The method minimizes the loss of information in least squares sense and is shown to outperform the interpolation based approaches achieving a reduction of aliasing and blocking artefacts and improvement in the overall signal to noise ratio.

Researchers in [18] have proposed an algorithm for preserving blur and noise during image reduction, and show the application of their method particularly for creating efficient image thumbnails. The method outperforms the other methods for image thumbnail creation which employ low pass filtering and sub sampling. The authors claim that their method helps to distinguish between the high quality and the low quality originals. Fig. 2-3 shows an adaptation of the image thumbnail generation result from their paper, along with the result of image retargeting using [12]. It can be seen that while scaling approaches in their entirety have tried to preserve some of the features of the original image, but they are rather quite far for quantifying the real essence of content aware image retargeting. Thus, such methods have not been very applicable for the purpose.

It is noteworthy here that we are not trying to draw a comparison between the content aware image retargeting techniques such as given in [12] and the scaling variants given in [18]. We are rather trying to posit the fact that the scaling variants for image retargeting have been designed to preserve different types of features during image resizing, which are quite different from what one might typically require for content awareness oriented image resizing.

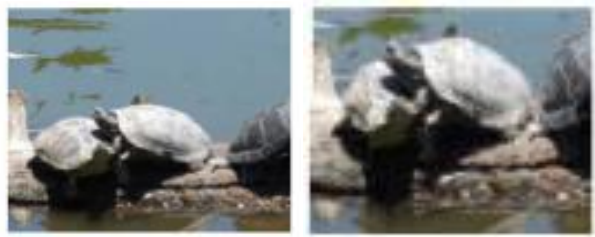

**Fig. 2-3** *(Left to Right)* Image Thumbnail that better represents blur and noise [18] *[Image adapted from [18] (© IEEE)]* , Image retargeted using content awareness oriented technique of [12]

## 2.3 Rapid Serial Visual Presentation (RSVP)

Rapid Serial Visual Presentation (RSVP) is a technique of displaying pictures on a temporal basis. It is useful particularly in situations where there is too much content information to be displayed and it is difficult to display everything in full at single instant of time. This notion has been used for image retargeting especially for mobile devices. The underlying concept is to automatically display the important regions of an image sequentially (one at an instant of time) so that the user is able to see all the regions zoomed in, although in over a period of time.

Researchers in [19] have estimated regions of interest by combining basic saliency maps along with face and text detectors to determine a path for browsing through the image contents. It is given by a sequence of panning and zooming operations, inspired by the RSVP technique. Researchers in [9] follow a similar approach with the determination of an optimal path to display the maximum information in minimum amount of time.

Researchers in [8] have followed an approach that avoids panning of the images displayed. They only tend to display the images sequentially by detecting important regions using, and then either crop or rescale to fit the size of the device. The salient regions are determined by segmentation while considering heuristics such as size, position in the image, and relationships between neighbouring regions. While avoiding panning speeds up the process of display, it deteriorates the smoothness of the display.

Researchers in [2], [4] – [7], [20], [21] have also implemented various methodologies for content aware image and video browsing. All of them employ some optimized method of image browsing after detecting salient regions of an image using either bottom-up saliency methods or top-down methods or a combination of both. However, such methods do not offer the advantage of seeing the entire image in a single display, and thus, such methods are often not classified under the banner of content aware image retargeting. By image retargeting (as we shall later see through the course of this dissertation), we aim to target the entire image so as to view all the salient regions of the original image in the targeted image as significantly as possible.

*Fig. 2-4* presents a scenario of RSVP browsing of images after detecting regions of interest in the original image.

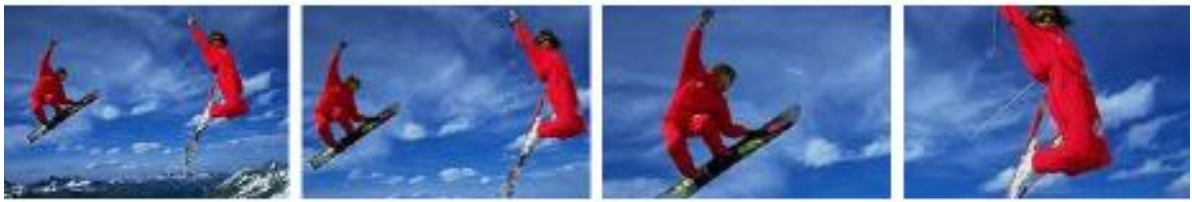

**Fig. 2-4** *(Left to Right)* Original Image, Browsing Sequence 1, Browsing Sequence 2, Browsing Sequence 3 (The technique used is that of sequential display without panning inspired by RSVP technique as proposed in [8]). *[Images Adapted from [8] (© ACM)]*

## 2.4 Content Aware Cropping

A variety of approaches have been proposed for content aware image retargeting using cropping. Usually, cropping is done to select a window region in an image, out of which everything is removed. When the window region is selected keeping in mind the region of interest (according to human perception), the approach is termed as *content aware cropping*. While, some of these approaches are totally automatic, some of them are semi-automatic in the sense that they demand user interaction for hinting the regions of interest. The user interaction should only be hinted in content aware cropping based algorithms since with software like Adobe Photoshop available since years and supporting cropping operations; it barely makes sense to crop the images after obtaining the region of interest completely from the user.

Apart from the criterion of user based interaction, the approaches for content aware cropping can also be segregated on the basis of whether the approach performs segmentation of the input image before cropping or not. Quite often, the cropping window is adaptively decided (keeping in mind the targeted size) based on some sort of saliency map, which may be combined with semantic information. However, there are approaches that perform a region based segmentation of the image as a first step, and then try to target the image to the desired size using cropping and/or scaling.

Below, we briefly discuss both the fully automatic and the semi-automatic approaches of content aware cropping, while covering segmentation and non-segmentation based approaches as well. As we shall see, semi-automatic approaches for content aware cropping have not been designed to involve image segmentation as one of the steps.

Researchers in [22] have proposed a method of automatically selecting the most appropriate cropping window based on the basic bottom-up saliency map of [30]. They also use face detection for images with human faces, thereby considering semantic information along with the saliency measure. Once the importance map is detected, they optimize their algorithm for searching the cropping window in a way that maximizes the percentage of salient points inside the window, using a greedy approach. Fig. 2-5 shows an adaptation of their results. It can clearly be seen that the method tries to capture the most relevant portion of the image before cropping.

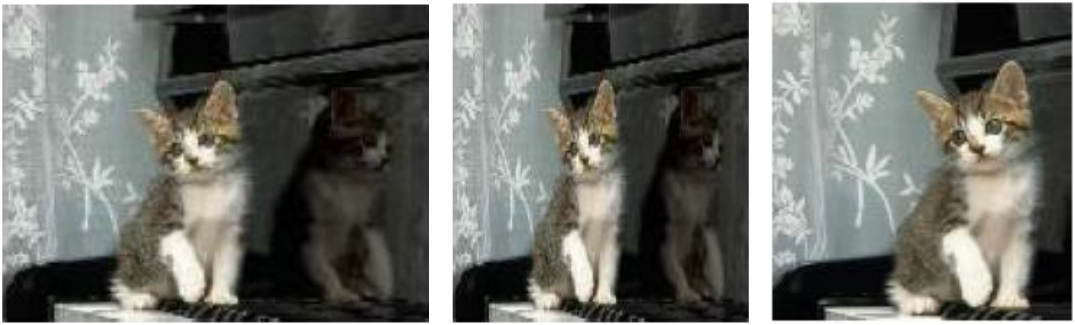

**Fig. 2-5** *(Left to Right)* Original Image, Image after scaling to the targeted size, Image after content aware cropping using [22] *[Images Adapted from [22] (© ACM)]*

Researchers in [24] have proposed a method of content aware cropping by running a particle swarm optimization of an objective function composed of three sub models, viz. composition sub model based on heuristics motivated by rules of thumb in photography, conservative sub model to avoid the picture from being cropped too aggressively and thereby destroying photo composition and a penalty sub model to prevent faces from being cropped off in the finally cropped image. Face detection and the region of interest detection (saliency detection) act as inputs to the optimization framework, where an energy function is defined for each of the aforementioned sub models.

Researchers in [3] propose content aware cropping based on the semantic information. The image is first classified into one of the semantic categories, viz. landscape, close-up or other, and then different algorithms are applied for cropping the image depending on the semantic type. An exhaustive semantic labelling and application of algorithms based on the same is not done in the algorithm. The images classified under the *landscape* category are not cropped, but are rather simply adapted through uniform scaling to fit the targeted size. The images classified under the category of *close-up* are processed with [30] to obtain a saliency map. The saliency map is then converted to binary for identifying the saliency regions, and a single relevant region is obtained, considering the bounding box that comprises of all the saliency regions thus identified. The image is then cropped and adapted with respect to this region. The images classified under the category of *other* are further segregated into the images containing faces and not containing faces based on a face detection mechanism. Such images are cropped based on saliency if no faces are detected, else skin color and face regions are also considered along with the most salient region while cropping and adapting the input image to the targeted size.

Researchers in [25] have proposed to crop images based on a saliency map computed by analyzing similarities between neighbourhoods in the input image. Given a zoom factor and the targeted aspect ratio, the cropping window that maximizes the average saliency is chosen as the result. The algorithm may however not always lead to an acceptable cropped window.

Researchers in [26] have attempted to find the best crop based on saliency maps obtained from a combination of models presented in [30] and [25]. Researchers in [27] have proposed a method automatically cropping an image based on a quality classifier that assesses whether the cropped region is agreeable to users or not. The quality classifier is statistically built using large photo collections available on websites with users' quality scores. The quality classifier is applied to each of the candidate regions for cropping, and the candidate with the highest quality score is selected as the final cropped region.

Researchers in [23] present a technique based on the image segmentation. The input image is segmented based on the texture and color consistency by employing fuzzy k-means clustering method. Entropy and the relative position and area of the candidates with respect to the original image are then used to determine if a segmented region is more or less interested to the user. Another method based on image segmentation has been proposed in [31]. Researchers in [31] propose a non-photorealistic method for image retargeting. Their algorithm first applies the mean-shift algorithm to segment the source image into different regions. The adjacent regions are then combined based on their spatial distribution of color/intensity. In order to identify important regions, a saliency map using [30] is generated along with the detection of facial regions. If the specified size contains all the salient regions, the image is simply cropped from there; else the important regions are removed from the image, and the resulting holes are filled using inpainting. *Fig. 2-6* presents an adaptation of the results from [31]. It is noteworthy to mention here that the segmentation based techniques depend heavily on the accurate segmentation of the input image, without which the output image might result in unacceptable distortions.

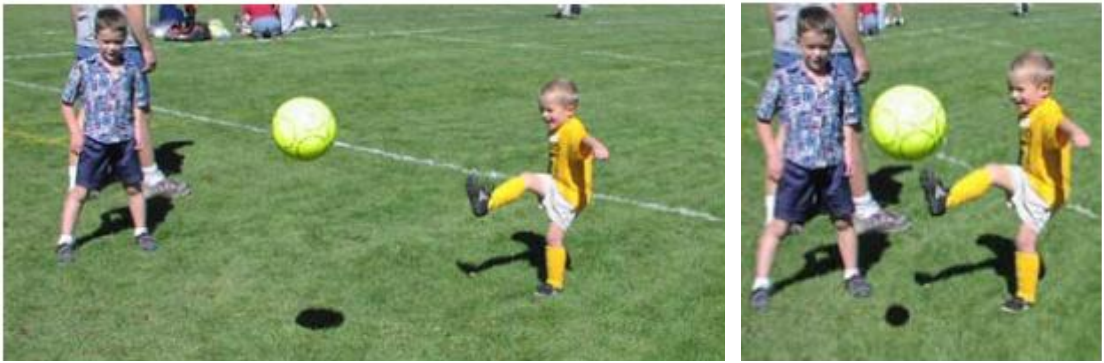

**Fig. 2-6** *(Left to Right)* Original image, Retargeted Image using algorithm of [31] *[Images adapted from [31] (© ACM)]*

There are other techniques that demand some user interaction for content aware cropping. The method proposed in [28] require users to look at each image for a few seconds, and during this period, their system records eye movements of the users. Using these recordings, the system identifies the important image content and then automatically generates crops of the targeted size. Researchers in [29] describe another system for semi-automatic image cropping. The user in their method is first required to select a point of interest in the input image and then a few cropping candidates placing the point of interest according to photography rules-of-thumb are suggested. The user then is required to pick the desired cropping region with the option of adjusting the zoom level, while retaining the point of interest at the selected location.

Having discussed many applied techniques of content aware cropping for image retargeting, it is worthwhile to mention here that all methods discussed barring that of [31] do not try to accurately identify relevant image content beyond cropping. In elaborative words, the image content is not analysed in a way where one can retain the visually attentive or salient regions of the image without cropping (totally removing) some other regions of the image. We shall dimension this notion from a slightly different perspective when we discuss multi operator retargeting methods towards the end of this chapter. The approach of [31] has rather applied a more sophisticated technique before deciding on cropping and also uses inpainting in case cropping does not seem desirable. Thus, the technique somewhat caters to the need of image retargeting while analysing the entire content of the image and avoiding cropping as far as possible. However, as previously mentioned, the segmentation based approaches rely heavily on the accurate segmentation of the image and the combination of the adjacent regions. Thus, these methods cannot be termed as very reliable for a variety of images.

As we shall see from the next section onwards, there have evolved better techniques for content aware image retargeting that analyse the entire content of the image, and generally don't result in removal of any of the parts of the image. Such methods have formed the basis of future research in the area of content aware image resizing as they fundamentally and practically prove feasibly accurate for a more varied set of images.

## 2.5 Seam Carving and Variants

Researchers in [32] proposed the technique of seam carving for content aware image resizing. The technique has inspired many variants upon the original approach and produced some decent range of results with low computational complexity. We start with a brief review of the basic seam carving approach proposed in [32].

The approach of seam carving is a pixel based approach for content-aware resizing and aims to remove pixels from the input image in a judicious manner, such that the image content after removal of the pixels blends amongst each other to provide minimal visual distortion. With the energy of the image *I* defined as in *equation (2-1)*, the removal of pixels cannot be done in a straight forward way such as removing the columns/rows with minimum energy, removing the pixel with least amount of energy in each row/column or globally removing of the pixels with the minimal energy regardless of their positions. Thus, a more holistic approach was adopted for such a pixel based content aware image retargeting. *Fig. 2-7* shows the visual distortions caused by removal of pixels in various straightforward ways. While global removal of pixels with lowest energy and removal of pixels with least energy in each row totally distort the image composition, the removal of column with minimal energy also gives slight visual distortions. As we shall see now, the approach of the removal of a column/row with minimal energy was extended to a *seam* removal with minimal energy after running an energy cost function minimization.

$$e_1(\mathbf{I}) = \left|\frac{\partial}{\partial x}\mathbf{I}\right| + \left|\frac{\partial}{\partial y}\mathbf{I}\right| \tag{2-1}$$

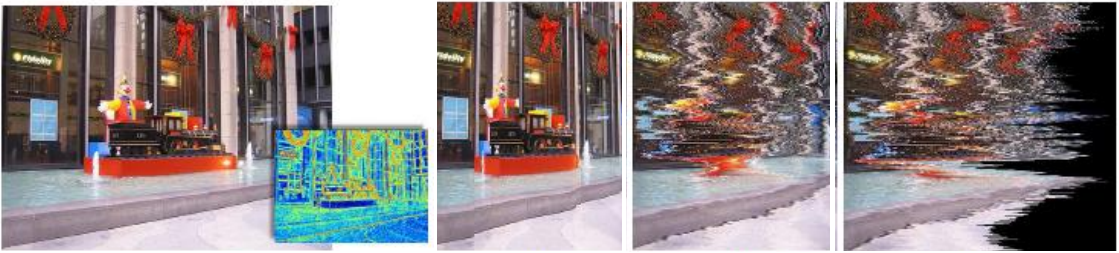

**Fig. 2-7** *(Left to Right)* Original Image with the associated energy map, Image after removing columns with minimal energy, Image after removing the pixel with the least amount of energy in each row, I1mage after globally removing the pixels with the lowest energy irrespective of their positions within the image *[Images adapted from [32] (© ACM)]*

Thus, the following approach was proposed in [32] termed as *seam carving*.

> - For an input image *I* of size *m x n,* the vertical seams and the horizontal seams are defined as 8-connected path of pixels from top to bottom or from left to right respectively. Mathematically, the vertical and the horizontal seams can be represented by *equations (2-2) and (2-3)* respectively given below.

$$\boldsymbol{s^x} = \{\boldsymbol{s^x_i}\}_{i=1}^{m} = \{(x(i), i)\}_{i=1}^{m}, \text{ s.t. } \forall i, |x(i) - x(i-1)| \leq 1 \tag{2-2}$$

$$\boldsymbol{s^y} = \{\boldsymbol{s^y_j}\}_{j=1}^{n} = \{(j, y(j))\}_{j=1}^{n}, \text{ s.t. } \forall j, |y(j) - y(j-1)| \leq 1 \tag{2-3}$$

> - Similar to the removal row or the column with minimal energy, a seam removal with minimal energy also has a local effect on the resultant image. Through seams, it is just ensured that after removing pixels, the contents blend well with the surroundings so as to give the minimal visual distortion. The approach works well since due to the local effect of seam removal, the visual impact is only noticeable along the seam path keeping rest of the image intact. The task is now to find the optimal seam for removal in each of the iterations. The optimal vertical seam $s_{opt}$ is found out by minimizing the seam cost of *equation (2-4)*. The horizontal seam is found out in a similar manner.

$$s_{opt} = \min E(s) = \min \sum_{i=1}^{m} e(\mathbf{I}(s_i)) \qquad (2\text{-}4)$$

The optimal seam is found out using the dynamic programming approach. The image is first traversed from the second row to the last row (assuming that we are considering the vertical seam removal and thus reduction of the width of the input image only) and a cumulative energy term given by *equation (2-5)* is calculated for all possible connected seams for each entry *(i,j)*.

$$M(i,j) = e(i,j) + \min(M(i-1,j-1), M(i-1,j), M(i-1,j+1)) \qquad (2\text{-}5)$$

Note that that in *equation (2-5)*, appropriate adjustments are made for dealing with the boundary cases. Such adjustments are trivial and often the non-existing term for boundary cases is neglected. Once, the cumulative energy function is calculated for each of the entry in the input image, a back tracing step follows starting from the last row, and finding the minimum cumulative energy in each row upwards. The pixels thus found out form the components of the optimal seam which is removed for reduction of the input image. The back tracking scenario considering the cumulative energy values and thereby finding the pixels of the optimal seam is depicted in *Fig. 2-8*. Please note that *Fig. 2-8* is for demonstration purposes and it is assumed that the values in the energy map are not normalized to a maximum of 1. Also, the energy map is not a binary image (the image containing only two values), but rather is a gray scale image.

> Authors in [32] proposed many energy measures for the image instead of the simple measure specified in *equation (2-1)*. They tested the proposed approach with the saliency measure of [30], entropy energy measure, histogram of gradients and Harris corner measures of [33]. Their analysis indicated that the energy function defined in *equation (2-1)* and the histogram of gradients gave the best results. However, a single energy function may not in many cases prove good for a variety of images.

> Seam carving procedure can be applied for both the vertical and horizontal seams; therefore authors in [32] also introduced a criterion for an optimal selection of the order of the removal for horizontal and vertical seams, i.e. removing all vertical seams first and then horizontal seams might not prove optimal for every image, considering the energy function of the input image and thus, an additional cost function specifying the order of the removal was proposed in [32].

> The seam carving methodology was also defined for the image enlargement and in such cases, also, the criteria for selection of the adequate seams both in horizontal and vertical directions remains same as above. However, instead of seam removal, seam insertion operation is applied. The seam insertion operation essentially inserts relevant number of artificial seams into the input image. The artificial seams are inserted by duplicating the pixels of the optimal vertical/horizontal seam found by averaging them with their left - right / top - bottom neighbours respectively.

| 30 | 20 | 18 | 16 | 20 | 15 |
|---|---|---|---|---|---|
| 35 | 25 | 22 | 25 | 22 | 27 |
| 40 | 30 | 28 | 25 | 26 | 32 |
| 46 | 35 | 30 | 33 | 32 | 36 |
| 50 | 38 | 32 | 35 | 36 | 42 |
| 54 | 42 | 48 | 38 | 40 | 45 |

**Fig. 2-8** Depiction of the back tracking in seam carving algorithm proposed in [32]. Each of the boxes represents pixels of the input image with the values indicating the cumulative energy at the respective pixel. The bottom row represents the bottom row of the image (back tracking starts from the bottom most row). The yellow coloured pixels represent the pixels for consideration in each row for choosing the minimal value (a seam is defined as an 8-connected path), and the green pixels represent the pixels of the optimal seam thus found. The dashed boundaries indicate that there is more image content along those directions, and that we have only presented a plausible portion of the image.

*Fig. 2-9* and *Fig. 2-10* represent the above concepts of seam carving in a more intuitive considering the entire image, showing the optimal seams found by using the aforementioned methodology, and the energy as well as the cumulative energy maps.

Authors in [32] also applied their basic approach for seam carving for the purposes of object removal from an image. The process of object removal can be thought of as content aware fill (upon the area that is selected to be removed by the user) and finds its efficiency with the seam carving methodology since it happens to be pixel based. We shall see later while discussing the warping based approaches for content aware image retargeting that the applications like object removal or object protection are most easily achieved in pixel based (also sometimes termed as *discontinuous* approaches) methods.

Since the advent of the seam carving methodology, various variants of the basic approach mentioned above have been proposed. The major adaptation to the work of [32] was proposed in [12], where a new criterion was described for calculating the seam cost. The seam carving algorithm introduced in [12] was termed as *improved seam carving* and has formed the basis of almost all variants introduced thereafter, with a real time implementation also surfacing in Adobe Photoshop CS4 and CS5[10]. The approach proposed in [12] can be briefly described as follows.

> ➢ Instead of using the backward tracking procedure, researchers in [12] use a *forward energy criterion*. The new criterion does not choose to remove the seam with the least energy in the image; rather it chooses the seam which after removal would insert the minimal energy in the image. After the removal of a seam, the inserted energy is generally and most dominantly due to the new edges formed. *Fig. 2-11* depicts the scenario of the formation of new edges when a seam (or generally speaking some sequence of pixels within the image) is removed. Based on this concept, a novel seam cost was proposed as given in *equation 2-6,* where $P(i,j)$ is a suitable energy measure. Minimizing this criterion gives the optimal vertical seam. The case for the horizontal seam is devised in a similar manner.

$$M(i,j) = P(i,j) + \min \begin{cases} M(i-1,j-1) + |\mathbf{I}(i,j+1) - \mathbf{I}(i,j-1)| + |\mathbf{I}(i-1,j) - \mathbf{I}(i,j-1)| \\ M(i-1,j) + |\mathbf{I}(i,j+1) - \mathbf{I}(i,j-1)| \\ M(i-1,j+1) + |\mathbf{I}(i,j+1) - \mathbf{I}(i,j-1)| + |\mathbf{I}(i-1,j) - \mathbf{I}(i,j+1)| \end{cases} \quad \text{(2-6)}$$

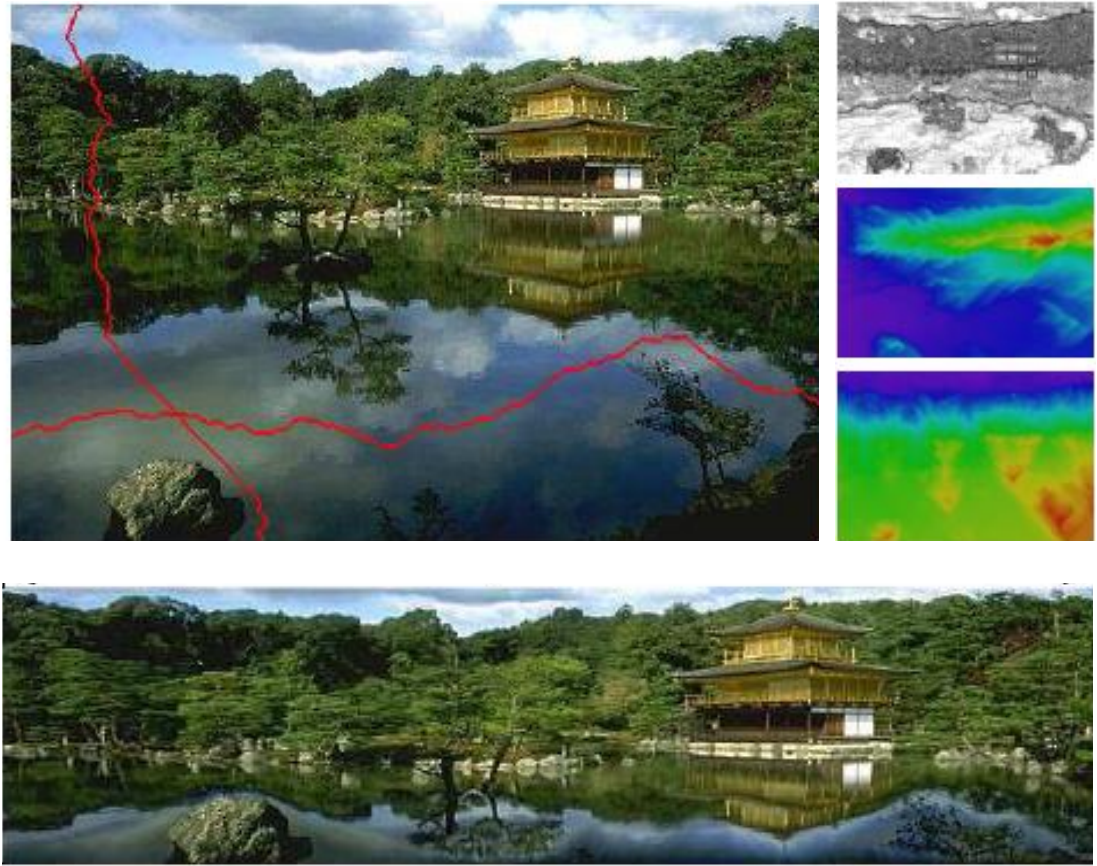

**Fig. 2-9** *(Top row) (Left)* Original image showing an optimal vertical and a horizontal seam (Many optimal seams are generally found for achieving the targeted size), *(Right – Top to bottom)* Energy function used (Magnitude of gradient), vertical cumulative energy map to calculate optimal vertical seams, horizontal cumulative energy map to find optimal horizontal seams. *(Bottom row)* Retargeted result using method of [32] for extending the width and reducing the height of the original image *[Images adapted from [32] (© ACM)]*

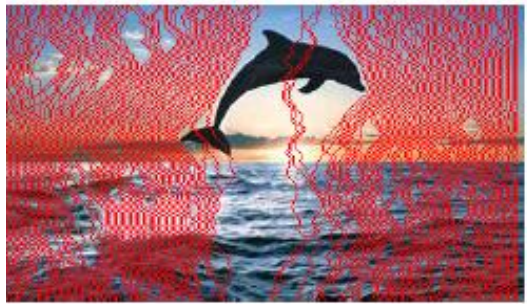

**Fig. 2-10** Overlay of the optimal seams found by [32] on the input image. The seams thus found can be used for either removal or duplication for insertion *[Images adapted from [32] (© ACM)]*

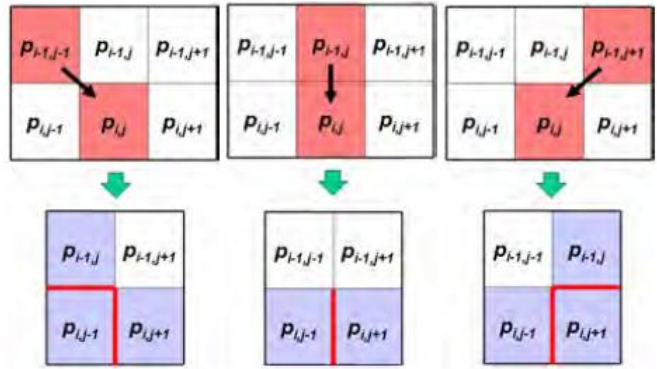

**Fig. 2-11** Depiction of the new edges formed after removal of the pixels (belonging to a seam or otherwise) from two rows of an image *[Images adapted from [12] ((© ACM)]*

The suitable energy measure $P(i,j)$ given in *equation 2-6* has been used by many researchers for enhancing results upon the improved seam carving algorithm of [12]. Before discussing the intrinsic dynamics of such enhancement approaches used, we show by *Fig. 2-12* and *Fig. 2-13* that improved seam carving has for most type of images proven to be better than the basic seam carving approach of [32].

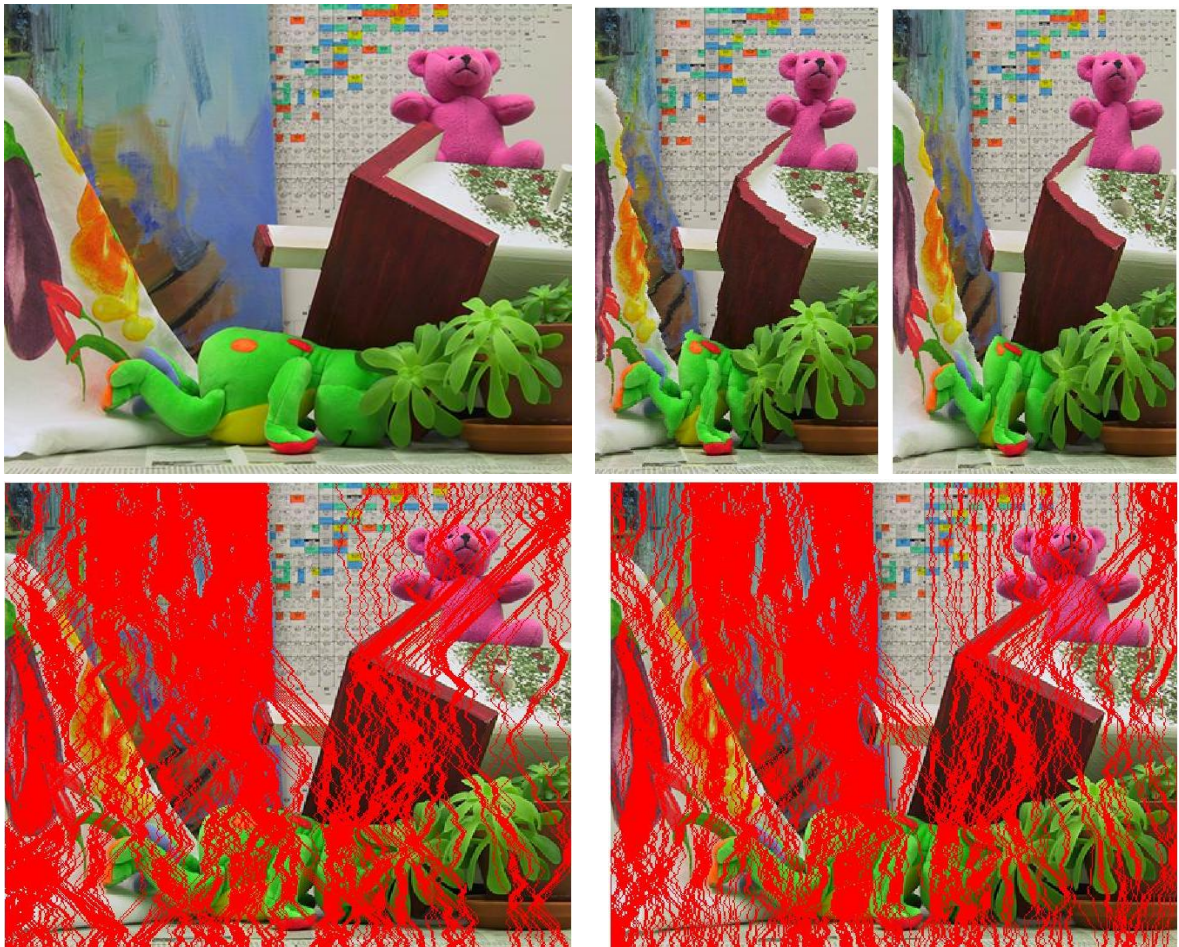

**Fig. 2-12** *(Top Row – Left to Right)* Original image *[Image Courtesy – Middlebury vision Dataset [11] (Teddy)]*, Image retargeted using [32], Image retargeted using Improved seam carving of [12]. *(Bottom Row – Left to Right)* Original Image overlaid with vertical seams detected by [32], Original Image overlaid with vertical seams detected by [12]. The original image is retargeted for a desired size of same height but 50% of the original width.

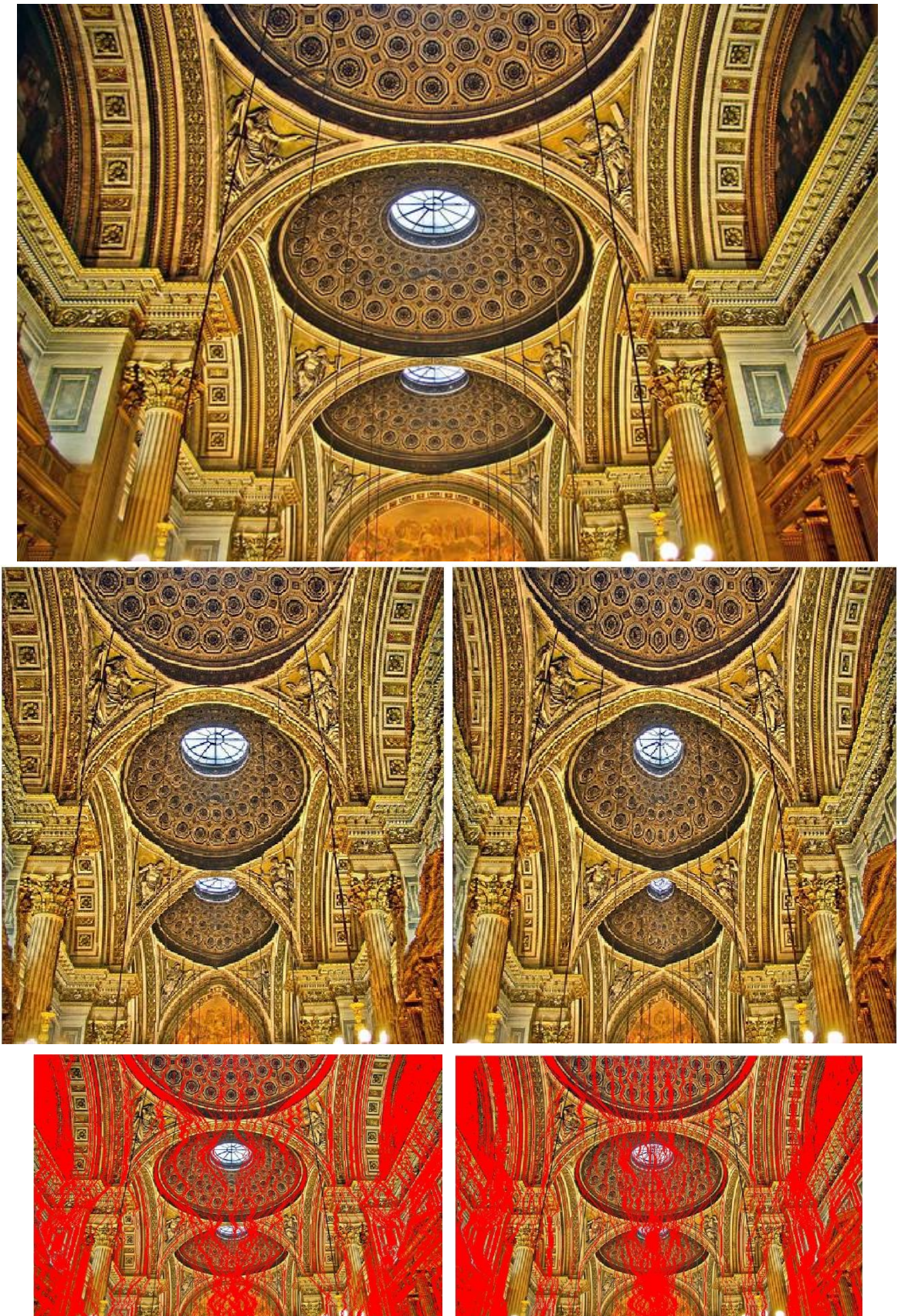

**Fig. 2-13** *(Top Row)* Original image *[Image Courtesy – http://www.flickr.com/photos/ayushbhandari/2164111513/ (Ceiling of the L'église Sainte-Marie-Madeleine)] (Middle row- Left to Right)* Image retargeted using [32], Image retargeted using Improved seam carving of [12]. *(Bottom Row – Left to Right)* Original Image overlaid with vertical seams detected by [32], Original Image overlaid with vertical seams detected by [12]. The original image is retargeted for a desired size of same height but 60% of the original width.

The authors in [12] also propose to formulate the problem of finding the optimal seam as a graph cut optimization. Although, dynamic programming is also very useful for the method, graph cut optimization is very optimally extendible to video.

The images depicted in the above two figures give a clear indication of the fact that the seam carving procedures cause good amount of distortion to the images. Please note that from now on, the term *seam carving* in this narrative shall refer to the improved seam carving method of [12]. The distortion due to seam carving in a very intuitive way can be attributed to the pixel based discrete kind of approach, which is bound to cut through the important objects and structures within the image. For this reason, initially seam carving was supposed to be used for the images containing sufficient non-salient regions (given the desired targeted size), so that most of the optimal seams' pixels are contained within the non-salient regions only. However, there occur many images such as in *Fig. 2-12* and *Fig. 2-13* which contain a lot of important content throughout the image and thus it becomes difficult to find the non salient regions for seam carving. However, with varied saliency maps, researchers have demonstrated many enhancements in the seam carving results.

One of the basic approaches that we depict here is to add a term in $P(i,j)$ which can protect the important edges of the  objects better than otherwise. One of the very easy implementations for this approach is when we have also the depth maps available for the images. With the advent of Kinect depth cameras which capture the information about the depth of each pixel captured from the camera and displays as an additional intensity depth map, we can use such depth maps to value the important edges more than the fine textures in the image. Although there are methods as in the papers [34] – [38] which can estimate the depth information given a single image. However, such methods have not yet proved to be very accurate when compared to what is given by depth cameras, and so cannot be termed as very reliable to use. Also, since we are using the depth information to segregate important edges from the textures and are not particularly interested in using the depth information, applying depth estimation techniques for content aware resizing relating to such purposes does not make much sense. It is that when the depth images are available, one can straight away use the depth maps to get better results with seam carving. We shall discuss in more detail the use of depth maps when we talk about uniform scaling and texture mapping in warping based approaches in the next section.

For $P(i,j)$, we use the *L1* gradient of the depth image associated with the input image. The results in comparison to [12] are shown in Fig. 2-14, Fig. 2-15 and Fig. 2-16. These figures also include the associated depth maps, whose information have been used for running simulations. Using the *L1* gradient of the input image in place of the gradient of the depth map has proven not to yield any better results.

It can be seen from these figures that even after using some additional image information like that of using depth maps with seam carving, there are quite perceptible distortions in the resultant images. As mentioned before, the distortions in the seam carving methodology are quite intuitive given the fact that it is a discontinuous pixel based approach. Before we go on to discuss the other variants of seam carving, it is worthwhile to mention here that most of those methods have been applied to the images for which there can be detected significant homogenous regions for the seams to divert through. More generically, applying any sort of saliency maps, visual attention models, etc, although improves results for content aware image retargeting using seam carving; such methods cannot take away the inherent discontinuous nature of the seams. Also, it is noteworthy that a seam has been defined as an 8-connected path, i.e., each pixel of the seam can get connected with one of the pixels out of the adjacent pixels in the bottom or the top row. Thus, one might argue that seams in certain circumstances cannot take a totally deviant path to avoid an object/structure cut, since the next pixel of the seam is constrained. We would like to mention here that relaxation of this constraint although might look to make the seam carving approach more widely reliable, this is not the case in practice. Also, the relaxation of the constraint of the seam being an 8-connected path poses stricter and more complex constraints on the forward energy criterion, and thus the entire minimization problem becomes all the more computationally inefficient.

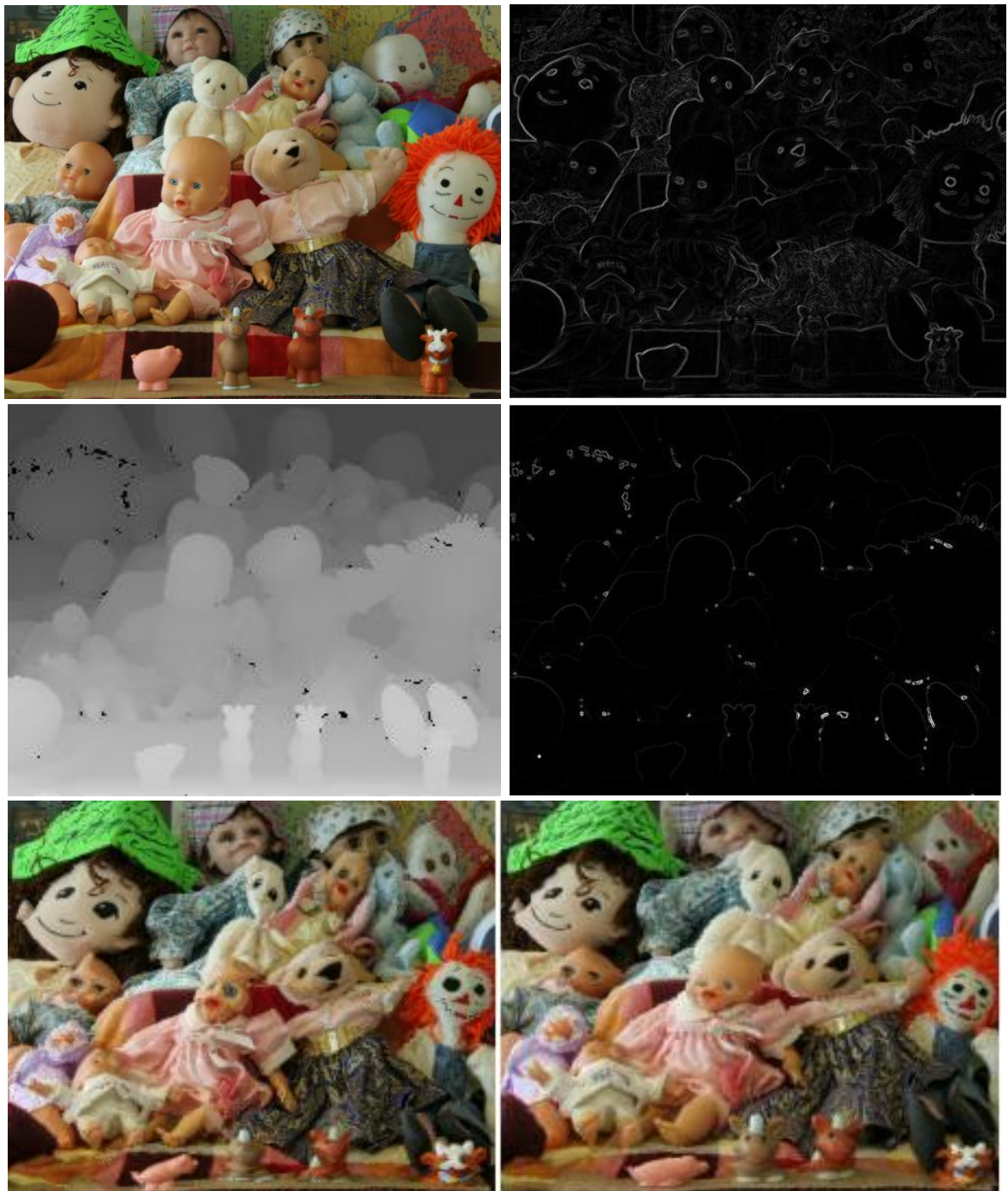

**Fig. 2-14** *(Top Row- Left to Right)* Input Image *[Image Courtesy – Middlebury Vision Dataset [39, 40] (Dolls)], L1* gradient of the input image *(Middle Row- Left to Right)* Depth Map of the input image *[Image Courtesy – Middlebury Vision Dataset [39, 40] (Dolls)], L1* gradient of the depth map *(Bottom Row- Left to Right)* Seam carving result using [12], Seam carving result using modified seam cost function with gradient of the depth map. The retargeting is done to reduce both the width and height of the original image by 50%. (*Here, the images have been uniformly scaled to fit to appropriate sizes for display purposes*)

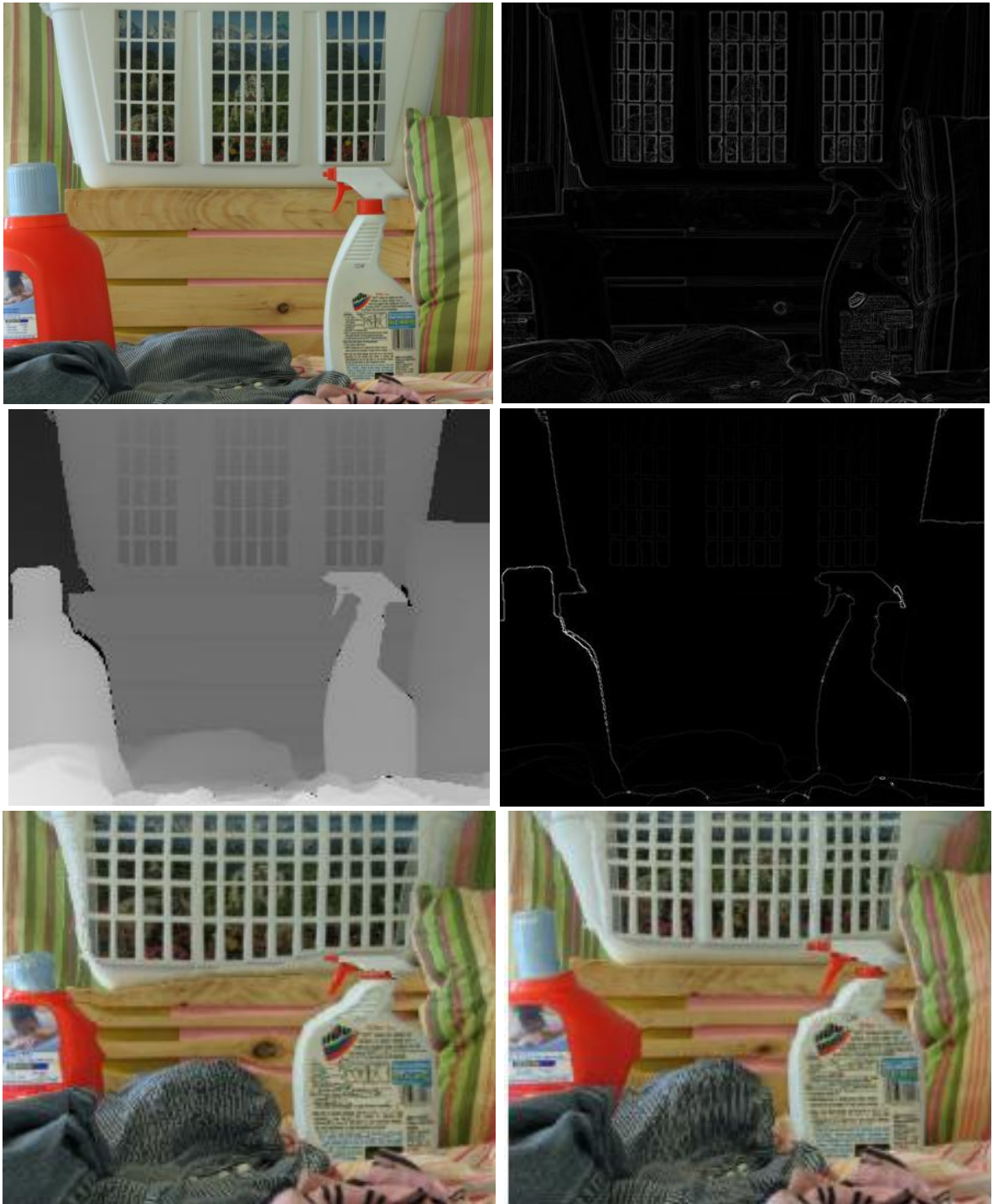

**Fig. 2-15** *(Top Row- Left to Right)* Input Image *[Image Courtesy – Middlebury Vision Dataset [39, 40] (Laundry)]*, L1 gradient of the input image *(Middle Row- Left to Right)* Depth Map of the input image *[Image Courtesy – Middlebury Vision Dataset [39, 40] (Laundry)]*, L1 gradient of the depth map *(Bottom Row- Left to Right)* Seam carving result using [12], Seam carving result using modified seam cost function with gradient of the depth map. The retargeting is done to reduce both the width and height of the original image by 50%. (*Here, the images have been uniformly scaled to fit to appropriate sizes for display purposes*)

Researchers in [41] have proposed an extension to the work of seam carving of [12] by the methodology of importance diffusion. Their method regards the neighbours of the seams being removed as more important that the other surrounding regions, and thus they increment the importance of the pixels adjacent to the optimal seams found. Their underlying idea relies on the fact that the seams to be removed provide important contextual information that must be preserved, and so their inherent significance should be replicated somewhere in some way even after removal. Importance diffusion helps to preserve the context information in the removed seams by adding it to its neighbours. The authors of [41] have shown that their method produces better results than that of [12] and sometimes even simple column or row removal using importance diffusion can yield acceptable results. *Fig. 2-16* shows an adaptation of the results of [41].

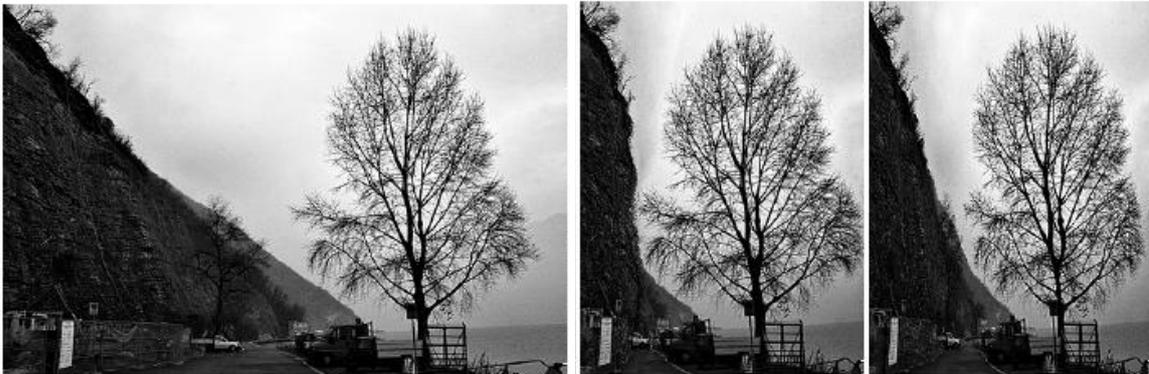

**Fig. 2-16** *(Left to Right)* Original image, Image retargeted using [12], Image retargeted using importance diffusion method of [41] *[Images adapted from [41] (© ACM)]*

While the method of [41] relies on preserving the information content in the removed seams, some other researchers as in [42] – [45] have targeted on to the saliency map modifications for improving the seam carving method.

Researchers in [42] propose a method for perceptual seam carving considering both the face map and the saliency map within the human attention model, and combine it with the forward energy criterion of [12]. They also propose a switching scheme between seam carving and resampling, inspired by the fact that it might not always be possible for an image to get content-aware resized by removal of seams in absence of sufficient non-salient regions. Although, this method may also fall into the category of multi operator methods, we mentioned this method here, since the proposed approach essentially contributes to the modified attention model and their criteria for switching between seam carving and uniform scaling is very loose. We shall later see while discussing multi operator based image retargeting methods that more reliable and typically computationally complex criteria are required for deciding on the switching between seam carving and scaling to yield globally acceptable results.

Researchers in [43] proposed to combine a segmentation based approach for an enhanced saliency model with the forward energy criterion of [12] to achieve better results. Specifically, a pre processing stage confirming protection for seam carving using fuzzy segmentation coupled with neural network skin detection is introduced in their approach. The approach is deemed to have a low computational cost and is implementable in real time.

Researchers in [44] modified the forward energy criterion of [12] to also include color as opposed to only gray scale gradient energies. The authors show that the new saliency based modification tends to result in more contiguous salient regions and is more robust to noise components within the images. However, their method works well for the images where the regions can be easily segmented based on the difference in their color components.

Researchers in [15] have proposed a content aware saliency map computation and combining it with the forward energy criterion; they show the improvement of the results over that of [12]. Their method of computing saliency tends to also include the important contextual regions near the salient objects. A saliency map is determined considering local low level factors such as contrast and color, along with the global distinctiveness that suppress frequently occurring features. Their saliency map estimation also incorporates visual organization rules, which hints at several possible centres of gravity about which a visual form is organized, and high level factors such as human faces. They present image retargeting through seam carving as one of the major application of their saliency map. *Fig. 2-17* presents an adaptation of the results of [15] and comparison to that of [12].

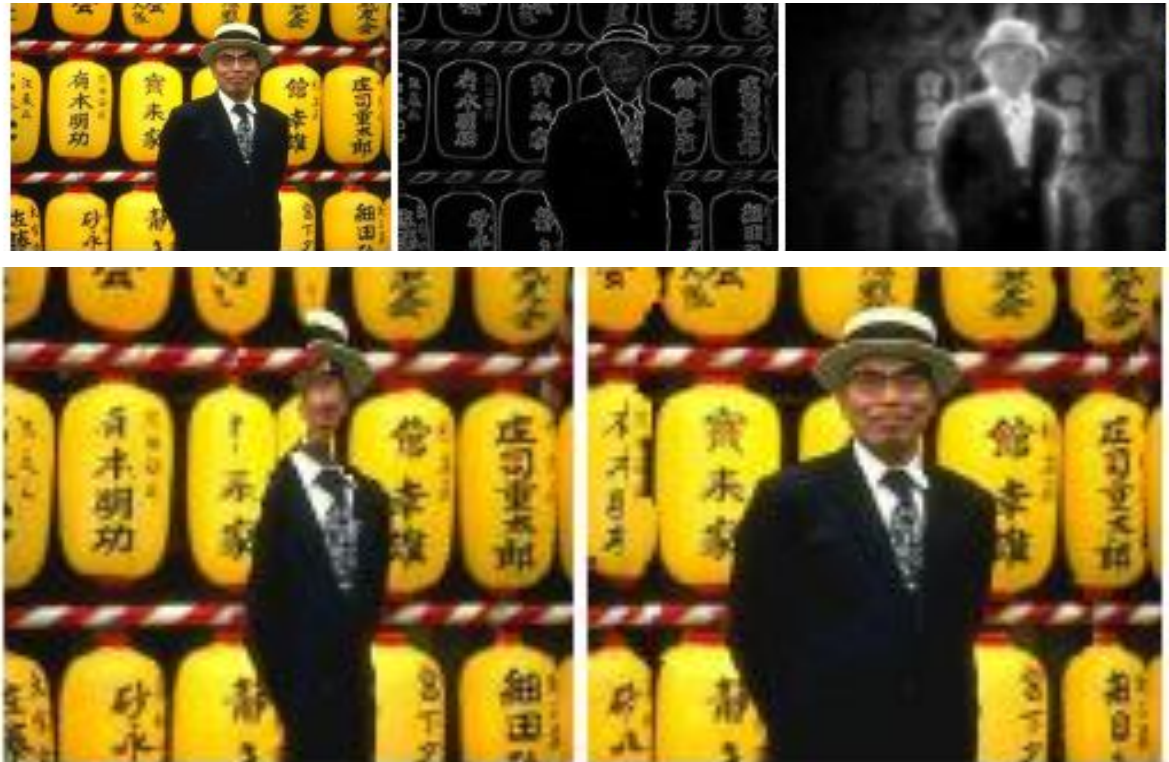

**Fig. 2-17** *(Top Row – Left to Right)* Original Image, Gradient Energy based saliency map used for seam carving in [12], Proposed saliency map of [15] *(Bottom Row – Left to Right)* Retargeting result of [12], Retargeting result with the proposed saliency map of [15] using seam carving forward energy criterion of [12]. The original image is retargeted so that there is no reduction in the height of the original image while the width gets reduced to 60% of the original width. The results shown here are uniformly scaled for display purposes and for the readers to be able to view the retargeted results more clearly *[Images adapted from [15] (© IEEE)]*

One of the other major classes in the enhancement of the seam carving algorithm has been the introduction of the wavelets for characterizing the local information of the image. The works presented in [45] – [49] target the input image by modifying the energy function based on the wavelet decomposition of the image. Although, all the works are slightly different from one another, the type of the images they all can cater to for plausibly distortion less resizing is same.

An approach that improves upon the wavelet based approaches and thus more or less the improved seam carving method of [12] has been recently proposed in [50]. The authors present one of the nearest adaptations to the classical seam carving works of [32] and [12] which involved backward and forward seam carving criteria. It has been already mentioned that a simple addition of the gradient energy function to the forward energy criterion does not yield significantly better results, and using the gradient of depth images has been more useful. Extending on similar notions, researchers in [50] propose an absolute energy cost function by taking into account the energy gradient along the seams being removed. They accomplish this by incorporating the energy gradient alongside the backward and forward energy cost functions of [32] and [12] into the dynamic programming process for finding the optimal seam. They show in their paper that the results produced were better than that of [12] and wavelet based methods, and their method attempted for the seams not to cut through significant objects in a more reliable manner. *Fig. 2-18* shows the results of [50] in comparison to some of the wavelet based methods (which were shown to produce better results than that of [12]).

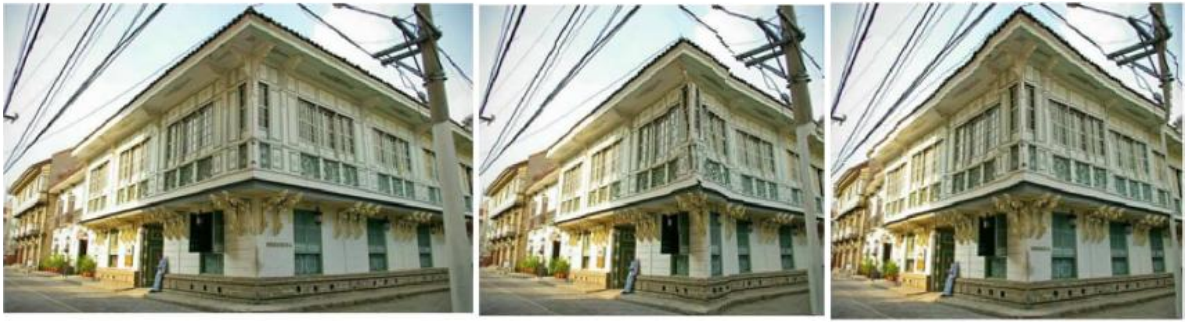

**Fig. 2-18** *(Left to Right)* Original image, Image retargeting result of [46], Image retargeting result of [50] The retargeting is done to reduce the original width to 70% with no reduction in the original height. *[Images adapted from [50] (© IEEE)]*

While the methods introduced above reduce the visual distortions caused by seam carving to a good extent, it can be argued that these procedures cannot be termed as very reliable for preserving the geometry in the images where the homogenous regions are scarce. As mentioned before, researchers in [42] suggested the combination of seam carving and scaling for the same. Extending this, researchers in [51] have proposed an adaptive seam carving method combining concepts of bi-directional search, importance diffusion, face and Hough line detection. switching between scaling and seam carving, and extending seams to *streams*(seams wider than a pixel width). Their work can be termed as one of the few approaches that combine various measures for edge, object boundary protection using seam carving and where the adaptive importance map distributes the distortion in both directions, even when the targeting is done only in one direction.

While seam carving has evolved as a powerful approach for content aware resizing, it fails for many types of images and for different aspect ratios. We shall slowly uncover the advantages / disadvantages of the other state-of-the-art methods before summarizing the globally unaddressed potential problems needing to be researched or currently being researched at the end of this chapter.

## 2.6 Warping based Approaches

Warping-based approaches for content aware image retargeting can be termed as a separate class of approaches complementally somewhat opposite to seam carving. While seam carving is a discontinuous pixel based approach, warping is often referred to as a continuous approach for image retargeting [14]. The underlying concept is almost the same as in the seam carving techniques, in the sense that the non-salient (unimportant) areas are modified so as to contain more distortion in comparison to the distortion of the more important or the salient regions. Since the approach is a continuous one, one can easily make out that both the salient and the non-salient objects can be retained with their original structure preserved in most cases, assuming that the target size is not very absurd given the nature of the image. This is very much unlike the seam carving strategies where the seams have to cut through different structures in the image irrespective of the nature of the input image and the desired aspect ratio, thereby distorting the geometry of the images in most cases. As we shall see now, warping although tends to preserve both the salient and the non-salient regions of the input image, excessive retargeting can sometimes make some of the regions negligibly small resulting in content removal. Several warping based methods have been proposed, which make use of different constraints for optimization. In general, warping methods tend to produce smoother results than the methods previously discussed.

Researchers in [52] have proposed a technique that warps the image in a similar way as a fisheye lens, by employing a piecewise linear warping, with the initial importance map being derived from a contrast difference based approach. However, the method has a limitation in assuming that there lies only one region of interest (important or the salient region) in the entire image.

Researchers in [53] have proposed a method for mapping textures into different surfaces so as to avoid the distortion of important features. However, the method requires user interaction for the specification of the important areas, and the deformation of different regions is constrained to be a similarity transformation, within a Laplacian image editing optimization framework.

In [54], authors have proposed a method for retargeting images (in particular to reduce both or either of width or height) and extend the work to retarget videos. They solve a sparse linear system of linear equations to find out the new pixel locations, which essentially specify the warped points. The authors employ the constraints that specify the relative position of the output pixels with respect to their neighbours. The relative position is measured with the help of an initial importance map (saliency map) that combines the $L2$ gradient energy and face detection.

Many authors have tried to build upon the work of [54] in directions where one is able to preserve the edges, important feature lines and structures in a better way. Researchers in [55] have extended the work of [54] by adding constraints so as to prevent the distortion of most visible lines. In [56], researchers have proposed a method for multi rate sampling of the input image based on an importance map, composed of an edge map (obtained from the boundaries of the result of the basic mean shift segmentation), the Canny edge detector [67], and artificial edges determined using the relative positions of the salient regions. The method then finds the sampling rates of different regions so as to minimize the edge distortion. In [57], researchers employ an approach of partitioning the image into vertical/horizontal strips based on an initial importance map, which they choose to be the sum of $L1$ gradient energies. Each strip is then resampled based on the frequency content of each region.

Some other researchers have formulated the problem of content aware image retargeting using warping as a label assignment to pixels. Researchers in [58] use a bottom-up saliency and face detection to form their importance map. They then label the pixels in the range of 0 to 1 for final retargeting, after solving a integer programming formulation by linear programming relaxation. Researchers in [59] use trellises of possible image edits and impose the image retargeting problem as that of integer dynamic programming. Researchers in [60] have proposed to retarget the images by using a weighted average of gradient energy, and saliency.

Perhaps the most important formulation of the image retargeting problem using warping based techniques has been the mesh grid optimization approach. The techniques in general represent the input image as a mesh (mostly quad mesh or a Delaunay triangular mesh) and target the image by finding nonlinear warping functions that resize the mesh to the desired size. Constraints are sometimes set on the control points of the mesh so as to minimize the distortion of the important content. The targeted image is finally rendered after obtaining the set of new control points for the mesh, and the output is represented as a warped mesh.

Researchers in [61] published the first major adaptation to the aforementioned approach using a quad mesh grid. They propose the importance map as being the product of the energy gradient map and the saliency map of [30] for weighting the quad grids. The initial quad mesh can be regular or non uniform depending upon the importance map, with the underlying motivation being that more control points should be placed in the regions of interest. They define the two dimensional shape deformation function, the grid line bending function and the smoothing function for their optimization and achieve quite useful results. Here, we do not delve deep into the function definitions for the mesh based warping methods since we shall discuss these in better intuitive detail in Chapter 3 while proposing our own novel energy functions and our own novel framework. The method has the major disadvantages of not being able to preserve the diagonal edges in the image and not being robust to unnecessary content removal (fold over problems where a quad grid may get scaled to the extent of getting invisible). The warping is done using homogenous scaling from the initial set of mesh points to the new set by solving the optimization problem. The method is well implementable in real time as well. The mesh based methods offer an advantage of being able to distribute the distortion in both the directions, even when the targeting is done in one direction only. This way, the methods prevent distortion for images which may contain more of salient content in the direction of retargeting than the other. Also, the method adapts automatically to uniform scaling of the salient and the non salient regions in case the image cannot be content aware retargeted. Such adaptation merely depends on the importance map and the energy functions formulation. While the approach has major advantages over the discontinuous approach, it has its own limitations and might prove inferior in some cases to seam carving. One of the typical examples of failure of the approach of [61] is that it tends to uniformly scale some of the important portions of the image rather than properly content-aware resizing, in order to minimize distortion. Thus, salient features may sometimes look really small as compared to their surroundings in the final result, which is not the case in seam carving. Thus, there is a trade off between the preservation of important features and resizing so as to make the salient regions really well noticeable. *Fig. 2-19* presents such an adaptation of the results from [61]. One can clearly see in *Fig. 2-19* that the approach of [61] although minimizes visual distortions which were there in seam carving approach, it makes the overall size of the ship smaller as compared to what is in the results of seam carving. *Fig. 2-20* shows some cases where the method of [61] fails horribly along with the corresponding seam carving results which fail too.

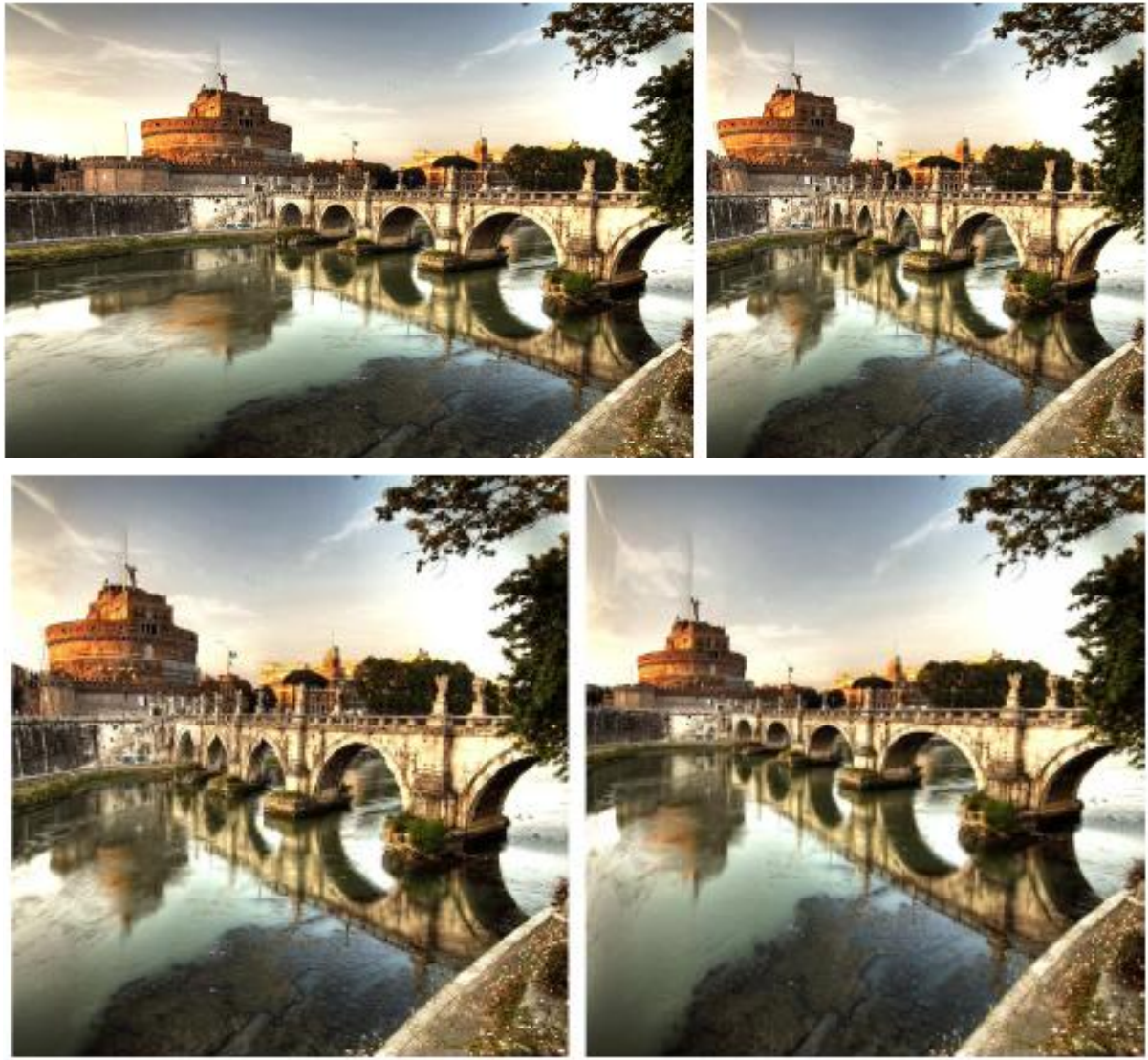

**Fig. 2-19** *(Top Row – Left to Right)* Input Image, Image retargeted using improved seam carving of [12] with the saliency map proposed in [61] instead of only *L1* gradient energy as used in [12]. *(Bottom Row – Left to Right)* Image retargeted using seam carving of [12] with no change in the saliency map; Image retargeted using the approach of [61]. The retargeting is done to make the width 60% of the original with no change in the original height. *The images in the bottom row are scaled uniformly for display purposes in order to show the distortions more clearly. This does not obfuscate the algorithmic results in any way. [Images adapted from [61] (© ACM)]*

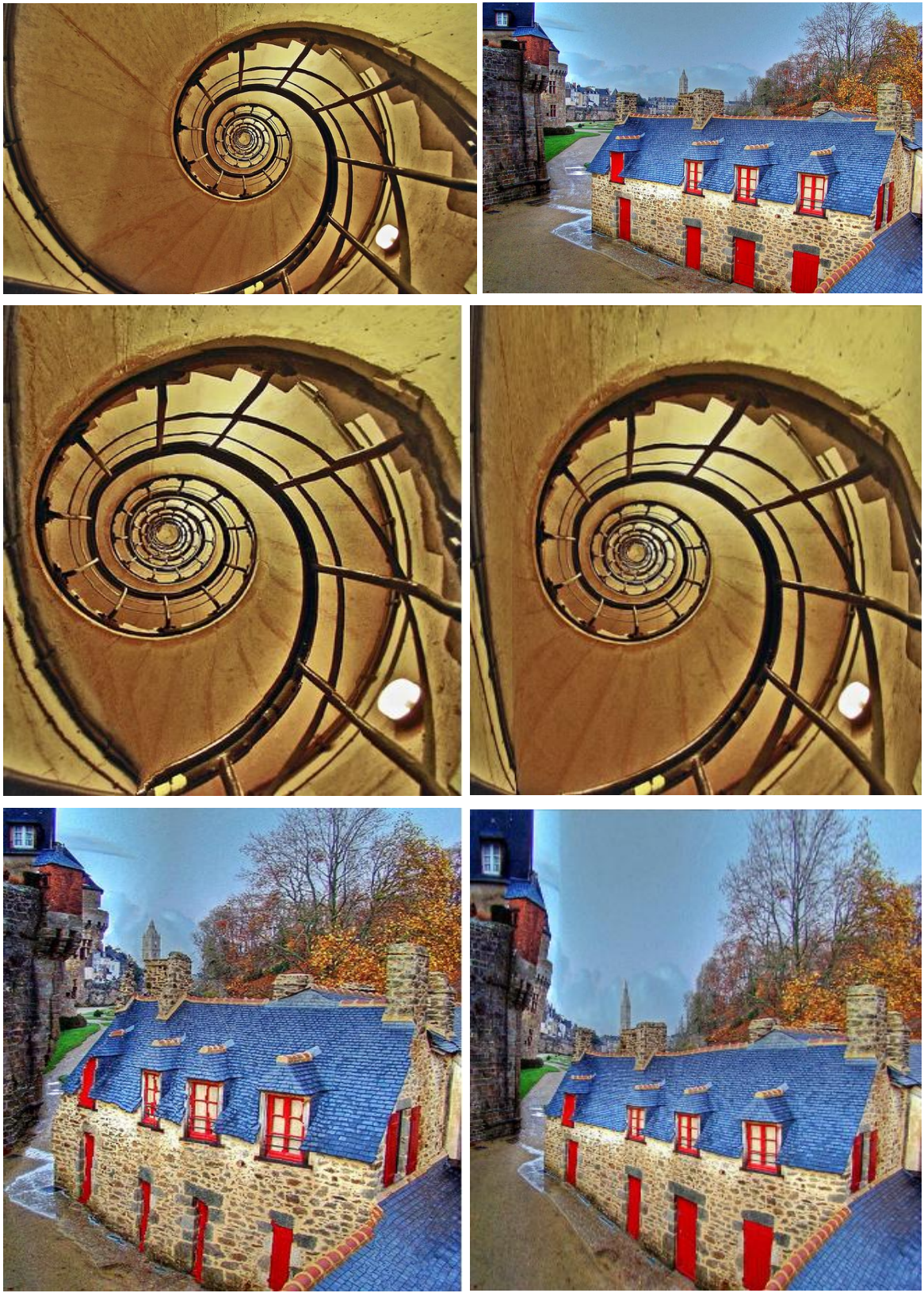

**Fig. 2-20** *(Top Row – Left to Right)* Input Image *1 [Image Courtesy – http://www.flickr.com/photos/ ayushbhandari/2164951646/ (Spiral Stairway to the top of Arc de Triomphe)]*, Input Image 2 *[Image Courtesy – http://www.flickr.com/photos/ayushbhandari/2397969443/ (House with Red Doors, Vannes)]. (Middle Row – Left to Right)* Retargeted result of input image 1 using [12], Retargeted result of input image 1 using [61]. *(Bottom Row – Left to Right)* Retargeted result of input image 2 using [12], Retargeted result of input image 2 using [61]. *The images are retargeted for around 40% reduction in width with no change in the original heights. The retargeted results are scaled for display purposes.*

Researchers in [62] introduced the concept of two coloured pixels (TCP) and use a quad mesh with TCP split edges for image warping. The image is tessellated into quads and each quad is split by an edge into two regions of constant color thereby forming a TCP split quad mesh. The split is done so that the TCP best approximates the colours of the original pixels within the quad. This TCP method is shown to preserve the diagonal edges better than the method of [61].

Researchers in [63] use a quadratic distortion energy function for mesh optimization by defining the control points of a regular mesh grid over the important edges. Constraints mainly deal with the importance map and preservation of the edges. The quadratic energy distortion function is shown to have a closed form minimization solution thereby avoiding the need for an iterative method.

In [64], researchers proposed the use of curve edge trapezoidal meshes instead of quad meshes and show that it helps to achieve better results than that of [61]. The results mainly improve upon preserving very small regions of the image.

Researchers in [65] proposed a mesh based parameterization method using a triangular mesh grid for optimization purposes. They keep the mesh edge lengths around salient objects (computed from the importance map combining saliency maps, face detection algorithms, Hough line detection and optional use input) are constrained to be rigid, and the remaining edge lengths are computed mesh parameterization. Their method produces impressive results as compared to those of [61] mainly because of their new functions for the optimization procedure. *Fig. 2-21* shows an adaptation of their results in comparison to those of [61]. One can observe that their method still suffers from visual distortions, especially regarding the important feature lines.

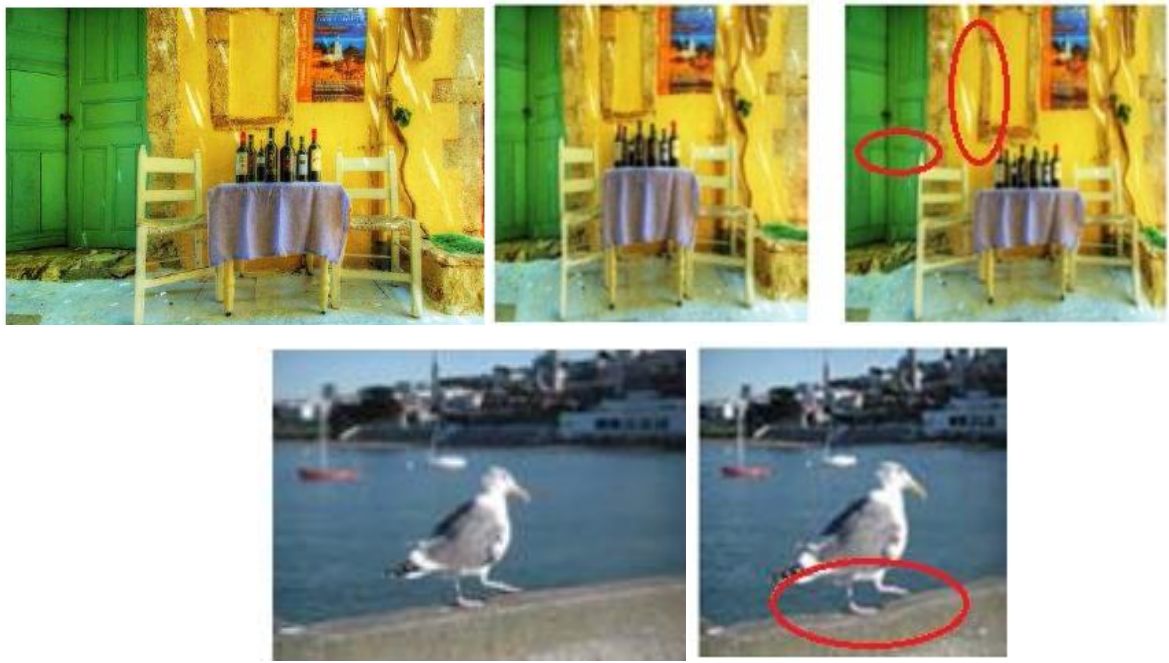

**Fig. 2-21** *(Top Row – Left to Right)* Input Image, Retargeted Image using [61], Retargeted image using [65]. *(Bottom Row – Left to Right)* Input Image, Retargeted image using [65]. Feature line distortions for the method of [65] are marked in red ellipses. *[Image adapted from [65] (© IEEE)]*

Researchers in [66] proposed to preserve both the linear and the curved features of the image using a regular Delaunay triangular mesh. They use the Canny edge detector [67], graph based visual saliency map [68] (which provides the saliency map considering the vision from a human context), max margin Hough Transform for detecting the object boundaries and allow optional user interaction for sampling areas of interest. Their results prove better than that of [61]; however, visual distortions are still not full confronted. With the triangular mesh based methods (methods of [65] and [69] use triangular meshes), one of the important considerations is to avoid the triangular flip over during optimization of mesh control points to maintain the topology of the image. Thus, this forms a mandatory constraint to be satisfied when dealing with the Delaunay triangular mesh for content aware warping.

One of the advantages of content aware warping is when the retargeting is to be done for enlargement in one or both directions. *Fig. 2-22* shows the result of content aware image enlarging using seam carving of [12] and using mesh based warping of [61]. It can be seen that while seam carving attempts to enlarge the homogenous regions (since optimal seams are found in the homogenous regions and they are replicated for enlarging), warping enlarges the regions of interest. An alternative to this effect of seam carving has been proposed [32], where one needs to first uniformly upscale the image and then apply seam carving for reduction to the targeted size. However, this approach for enlarging is not very efficient and fully automatic as compared to enlargement using mesh based warping. Also, one can notice here that during enlarging, since the seam carving approach tends to interpolate in between the seams, the salient objects (such as the corner boy in *Fig. 2-22*) may become blurred. This is not the case in the warping based methods; however, the method of [61] suffers from an advantage of plausible content removal which can be seen in the following figure towards the top end of the retargeted image. The content removal in case of [61] takes place in the other direction (as that of retargeting) since warping based methods are able to distribute the distortions in both directions.

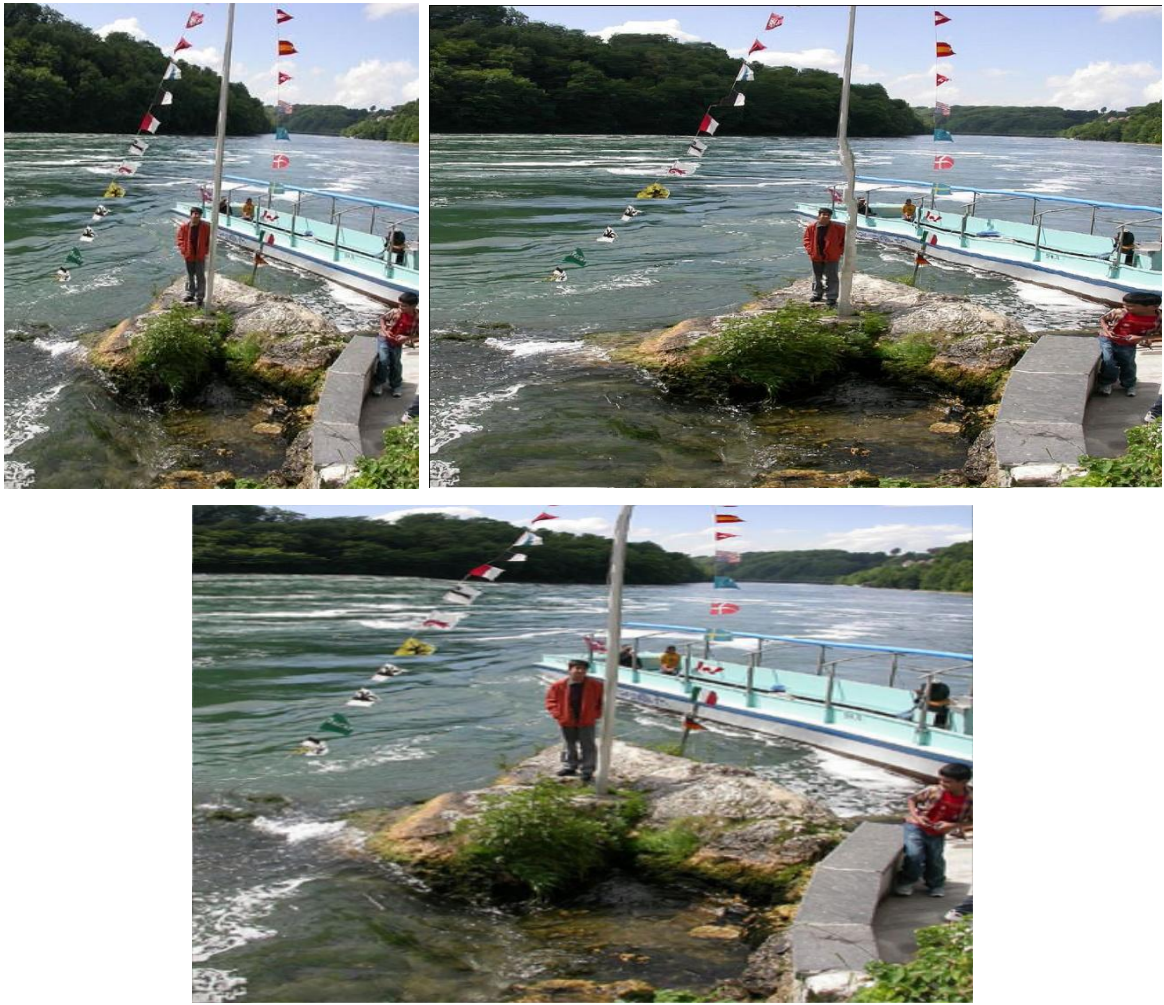

**Fig. 2-22** *(Top Row – Left to Right)* Input Image [Image Courtesy – http://www.flickr.com/photos/ayushbhandari/184367571/ (Schaffhausen, Switzerland)], retargeted image using [12]. *(Bottom row)* Image retargeted using [61]. The retargeting is done to make the original width double with no change in height.

Using the above inherent advantage in warping based methods, researchers in [69] introduced a content-aware zooming operator for high resolution image visualization on small displays. Their method allows user interaction to control the trade-off between distortion and content aware zooming of the important regions. The method uses an adaptive view-dependent mesh representation for warping.

## 2.7 Patch based Approaches

Image retargeting achieved through the manipulation of image patches forms the core of the patch based methods. The algorithms typically use the measure of distances between image patches, thereby aiming to minimize the same between the input image and the targeted image. However, such measures that can be deemed as relevant are still being explored by researchers [70]. With the currently available methods, the approach is being pursued for the purpose of content aware image resizing.

Researchers in [71] proposed a bidirectional similarity measure between images, comprising of both completeness and coherence measures. The measures are computed from the image patches. The completeness aims to measure the presence of all visual features in the target patch that were present in the source patch, while coherence aims to adhere to the fact that the used transformation has not created new visual artefacts. The method scales the image in small steps with each step minimizing the error. Thus, the method is iterative. It can also incorporate importance functions to retain high level features such as faces in the target image.

Researchers in [72] introduced the patch transform for image editing tasks. Their approach suggests breaking the image into patches and then reconstructing the image after retargeting by rearranging these patches. The constraints are specified by the user regarding locations of some patches. Their process is modelled in a framework involving the Markov Random Fields and the model is solved by belief propagation. Due to this, the approach is computationally complex. One of the other disadvantages is the inability to preserve local salient structures for preserving the overall global context.

Researchers in [73] proposed a randomized approach for finding approximate nearest neighbour matches between image patches using a fast algorithm. Their algorithm employs a random search among a sequence of randomly selected candidates, and the best matches found are propagated among the the neighbourhood patches.

One of the most widely used patch based methods for content aware image retargeting has been proposed in [74]. The researchers in [74] propose a shift map method where the relative shift of each pixel in the output image from its source in the input image is specified. This way, the image segments are removed after performing a global optimization on the discrete graph representing the output image. This method is sometimes considered as a generalization of seam carving since it adds the flexibility to remove wider seams in a single step. The method can be compared to the stream carving method of [51] which attempts to give a similar effect. However, the patch based approach of [74] is more generic.

The major disadvantage with the patch based approaches is that they might lead in the content removal in the images. While, they might preserve the global effectiveness of the scene at times; in doing so, they might affect the local saliency of the image, and it looks very awkward when some of the content is removed from the image. *Fig. 2-23* depicts one such scenario of patch based methods.

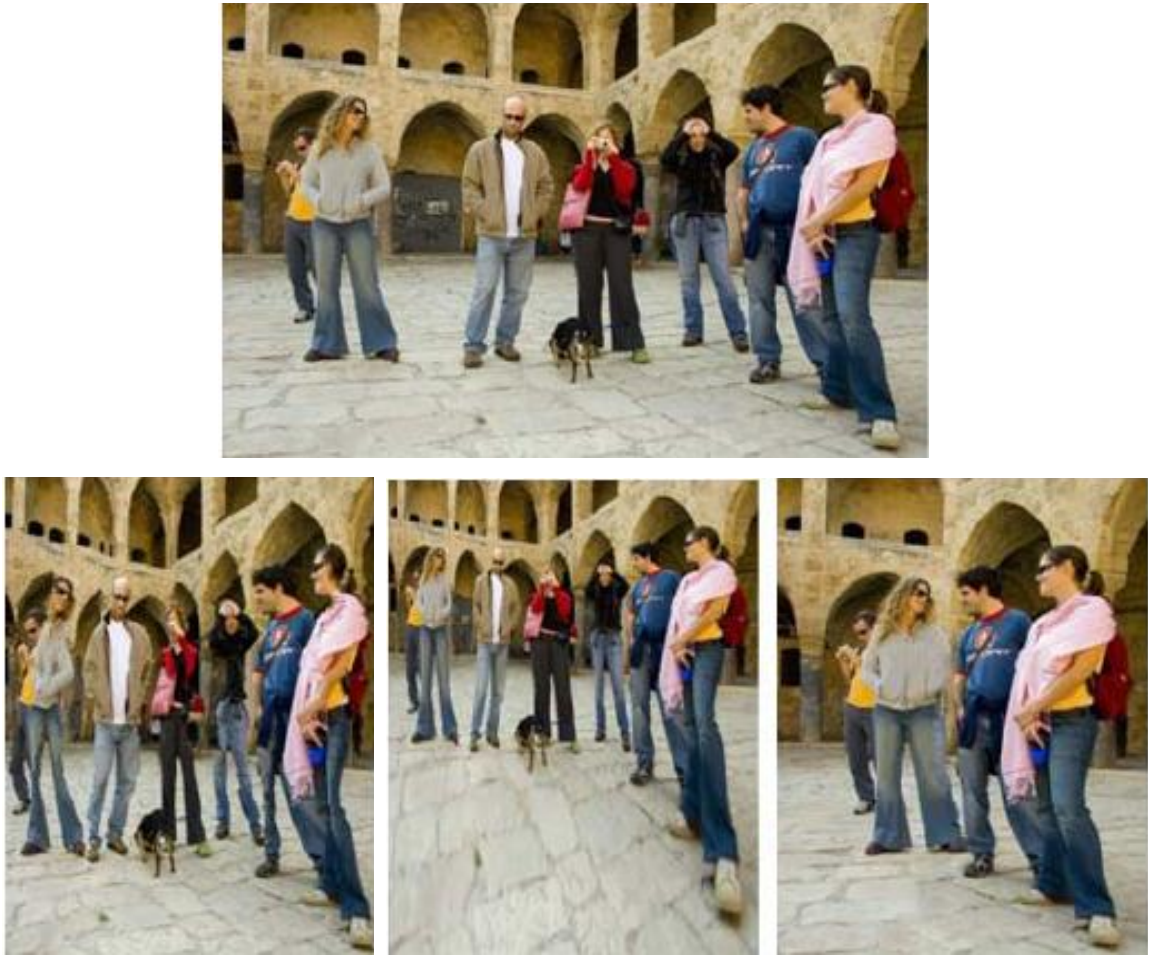

**Fig. 2-23** *(Top Row)* Input Image *(Bottom Row – Left to Right)* Image retargeted using seam carving of [12], Image retargeted using warping approach of [61], Image retargeted using shift map editing (This results in the removal of many objects of interest). Retargeting is performed for reducing the original width by 50% with no change in the height.
*[Images adapted from [74] (© IEEE)]*

## 2.8 Multi Operator Methods

As previously mentioned, it is more or less important to consider multiple operators for preserving visual distortion during content aware image retargeting. The aforementioned retargeting methods have their own sets of advantages/disadvantages and one might argue that a single operator may not perform well in every case.

Researchers in [75] presented a user study that hinted the users generally preferring to combine different retargeting operators to obtain visually pleasing results, instead of approving the results of any single operator. We already presented a brief overview of initial work [42] on such notion while discussing seam carving. However, more reliable methods have been proposed for multi operator based retargeting.

In [75], authors also proposed an approach to combine the operators of seam carving, scaling and cropping. Their aim is to find an optimal combination of the three operators for best visual results by maximising the similarity between the input image and the targeted image. The optimization is performed using dynamic programming and is guided by a bidirectional image similarity measure, which is a separable adaptation of the patch based bidirectional similarity measure.

Researchers in [76] also proposed to combine seam carving and uniform scaling to minimize visual distortions. Their image distance function is a modified one as compared to that of [75] and is a combination of a patch-based bidirectional image Euclidean distance (IMED), dominant color descriptor (DCD) similarity, and seam energy variation. The technique is essentially removing a seam at each step and then uniformly scaling this image to the desired targeted size, and measuring the similarity. The process is iterative and the optimization aims to stop seam carving at the point where the image similarity begins to deviate. The proposed approach has been shown to be useful in both reduction and enlargement. They show that they are able to better achieve the content aware image retargeting while preserving image structures than [75]. Researchers in [77] introduced a continuous seam carving operator and combined it with uniform scaling to get the resized image.

There has been an attempt to combine the methods of seam carving and warping through the involvement of the concept of semantic extraction from an image. Such a method has been proposed in [78]. However, the method cannot be deemed as reliable since it does not give a clear account of the semantic extraction procedures which forms the core of the algorithm. Also, the researchers in [78] consider only some images from sports videos for proving the worth of their algorithm, and thus their method is not very generic which can be applied to a vast variety of images.

Having discussed the multi operator based methods; we strongly posit the fact that the methods of [75] and [76] are the two best result producing methods in the category. *Fig. 2-24* and *Fig. 2-25* show an adaptation of the results of [75] and [76] respectively. However, they are computationally complex and cannot be termed as globally reliable since the state-of-the-art image distance measures are still in their novice state. Although, researchers in these papers have formed novel cost functions to add to their optimization framework, the methods suffer from an inherent disadvantage of combining seam carving and uniform scaling (cropping is also combined in [75]). Thus, it is very intuitive that one might end up uniformly scaling the image more than content aware resizing the image in order to preserve the visual content. On the other hand, one might also be not able to avoid visual distortion due to unavailability of very accurate image distance measures. We discuss these pitfalls along with the warping based approaches (where scaling becomes inherent in the optimization framework) with more detail in Chapter 3 while proposing our novel framework.

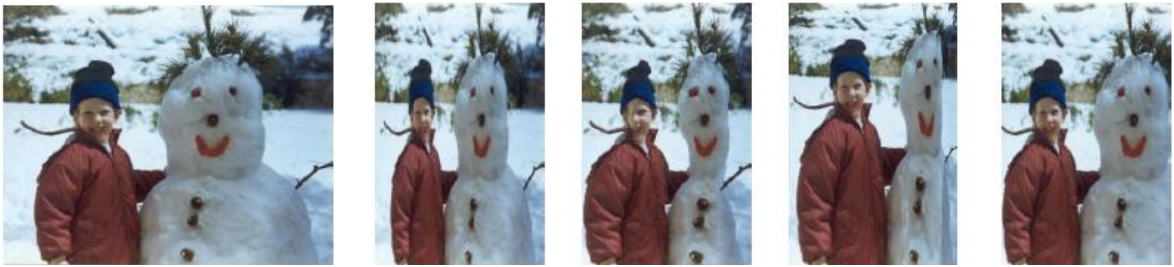

**Fig. 2-24** *(Left to Right)* Input Image, Retargeted image using uniform scaling throughout, Image retargeted using seam carving of [12], Image retargeted using warping of [61], and Image retargeted using multi operator approach of [75]. Retargeting is done to reduce the original width by 50% with no change in the height of the input image. *[Images adapted from [75] (© ACM)]*

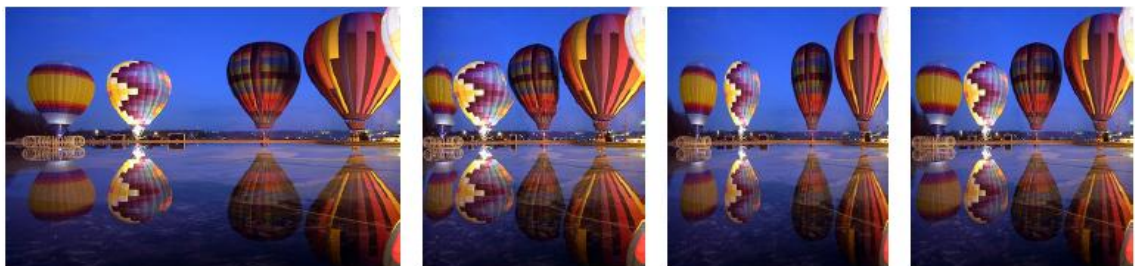

**Fig. 2-25** *(Left to Right)* Input Image, Image retargeted using seam carving of [12], Retargeted image using uniform scaling throughout, and Image retargeted using multi operator approach of [76]. Retargeting is done to reduce the original width by 50% with no change in the height of the input image. *[Images adapted from [76] (© ACM)]*

## 2.9 Symmetry based Approaches

There have been recent trends in dealing with the symmetry structures while doing content aware image retargeting. Researchers in [79] proposed an initial approach to incorporate linear and rotational symmetry constraints within the optimization framework of mesh based warping. *Fig. 2-26* shows an adaptation of their results.

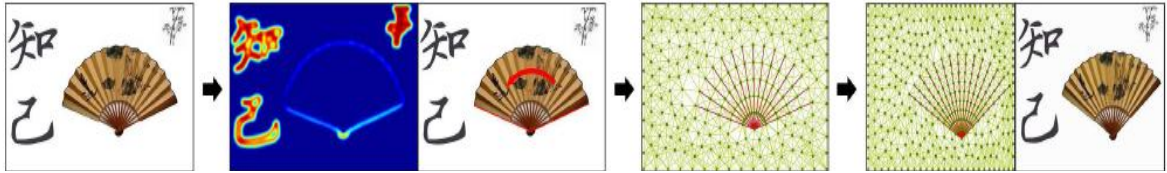

**Fig. 2-26** *(Left to Right)* Input Image, Image analysis using feature maps as suggested in [79], Formulation mesh according to [79], Optimized mesh and the final result after retargeting using approaches of [79]. The retargeting is done for a reduction in the width of the input image by 30% with no change in the height. *[Images adapted from [79] (© ACM)]*

   A more comprehensive approach to symmetry summarization for image retargeting has been proposed in [80]. Their method is able to target the different types of translational symmetries much more reliably than in [79]. They segment the image into symmetrical and non symmetrical regions, and warp the non-symmetrical regions with a constrained mesh grid optimization with the symmetrical structures being resized using their cell summarization method. They also cater for the discontinuities at the boundaries of the symmetric and the non symmetric regions. An adaptation of their results is given in *Fig. 2-27*. Their results prove more or less better when compared to the shift map and the patch based methods, which have proved more useful for images with symmetry.

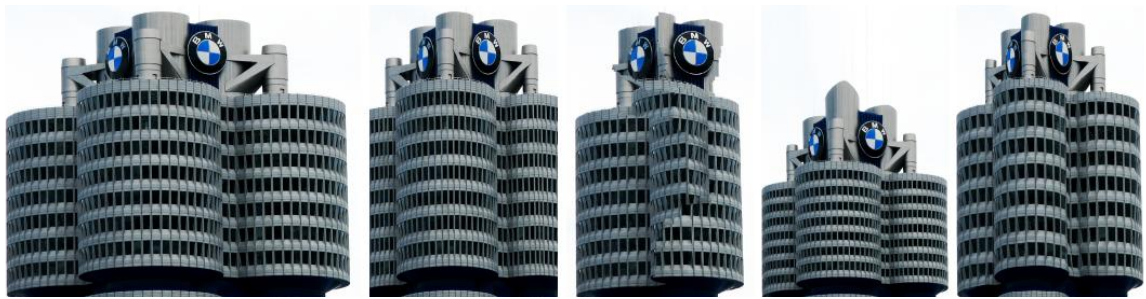

**Fig. 2-27** *(Left to Right)* Input Image, Retargeted Image using multi operator approach of [75], Retargeted Image using shift map method of [74], Retargeted Image using warping method of [61], Retargeted Image using symmetry summarization method of [80]. The retargeting is done so as to reduce the width of the input image by 50% with no change in the original height. *[Images adapted from [80] (© ACM)]*

   The method of [80] is one of the initial and foundation methods of considering translational symmetry cases within an image for content aware image retargeting. However, an inherent problem in the approach considering the symmetry within the input image is the segmentation of the image into symmetrical and non symmetrical regions. Also, it is not always possible for lattice structures proposed in [80] to identify the symmetry regions when the occlusions occur. Due to this, this area remains a vastly open potential research domain for further exploration regarding efficient identification of different types of symmetries and efficient image segmentation while ensuring minimal visual distortions.

## 2.10 Summary

Having discussed the various aforementioned techniques for content aware image resizing, one can easily see that no single technique can be termed as most appropriate given a variety of images. Broadly speaking, one can term the warping based procedures (inspired lately by the works of [61] and [65]) as potential methods for providing distortion less results, in case they are modified or introduced with some novel perspective keeping in mind the possible infinite number of image compositions. It is also important to mention here that researchers have proposed/used various importance measures (which form the first step for any resizing approach) and have depicted an improvement in the results. With the modified/combined saliency maps discussed above and some of the measure like [81], [82] available, some enhancements can be achieved given the type of the image. However, such enhancements may not be very reliable over a global set of images.

We summarize the following major aspects that need to be considered during content aware image retargeting for carrying out potential research in future. Note here that we don't consider as of now discussing the fallacies in the symmetry based methods since they are still in their novice state of research and bear the fact that images with different types of symmetry may not occur very frequently. We mainly concentrate on the non-symmetry based procedures, since such methods are useful for all kinds of images, and the present state-of-the-art methods suffer from various visual drawbacks given different kinds of images and desired aspect ratios.

- ➢ Approaches for understanding the semantics of the image should be developed in ways that are computationally efficient and suitable for content aware image retargeting. A semantic driven approach shall aim to better preserve the inherent meaning in the image content.
- ➢ Approaches serving an efficient trade off between the feature protection, content awareness, uniform scaling and semantic preservation need to be developed for better content aware image resizing. Some researchers have also hinted at introducing the topological definitions with the saliency maps. However, any modification in the saliency maps should be done keeping in mind the tradeoffs the entire framework can accurately and efficiently serve.

We shall see in much more detail the ways to tackle the above major fallacies in the state-of-the-art content aware resizing techniques, while discussing our novel framework for the same in the next Chapter.

# 3 Novel Perspective for Content Aware Image Retargeting

This chapter discusses the novel framework that we have developed for content aware image resizing during the course of this work. The work essentially presents a novel perspective of analysing the content aware image retargeting problem with the major constraints of semantics and feature preservation without visual distortion taken into consideration.

We present our entire approach in a sequential manner constructing the framework step wise step while alongside citing and visually depicting the appropriate reasons for introducing/considering a particular sub-approach.

## 3.1 Importance Measure

As mentioned at various places in Chapter 2, the importance measure is one of the major considerations in the any of the approaches for content aware image retargeting. We use various measures to quantify the importance of different types of images keeping in mind the trade off between the feature preservation and the visual distortion while also doing content aware resizing. With these various measures, we form a novel kind of saliency map for use in our algorithm. The following text sequentially mention the procedure of forming the new saliency map (Note that we take the input image as *I* and the output image (desired after retargeting) as *I'*.

We first compute the **Graph based Visual Saliency (GBVS) Map** of [68] for the input image *I* to get *G*. The visual saliency map in other works is generally used as that of [30]. The warping based method of [61] also uses the saliency map of [30] before combining it with the *L1* gradient energy function. The work of [66] however had proposed the use of the map of [68]. One of the other saliency maps useful for content aware image resizing has been proposed in [15]. However, all of these works depend more or less on the saliency map thus used for their formulation of the problem. Our approach here is quite different in the sense that we don't only rely on this saliency map for our processing, but rather consider many other aspects (apart from mere feature lines detection and *L1* gradient energy) before formulating the final saliency (importance) map for use.

We choose the saliency map of [68] as against the map of [30] since it generally produces results which gives better image visualization from a human context point of view. *Fig. 3-1, Fig. 3-2, Fig. 3-3 and Fig. 3-4* depict this notion for various types of input images. We choose this map against that of [68] since it is also computationally very efficient to calculate the saliency using this approach, while the calculation of the work [68] takes a formidable amount of time for computation. Also, for a variety of images, the results obtained from both are more or less similar. Since, we don't solely depend on this saliency, choosing the saliency computation of [68] achieves our purpose.

The Graph Based Visual Saliency method of [68] uses a Markov chain approach to calculate the activation and saliency values after treating the equilibrium distribution over various image maps. In addition, the algorithm exploits the topographical and parallel nature of the graph based algorithms to compute the desired saliency maps.

It can be seen from the following figures that the GBVS method that we have adopted here always identifies the most salient portion of the image, unlike the measure of [30] which sometimes tends to mark the surrounding regions as similarly salient which are actually not. We emphasize more of this here since we have used this map for multiple purposes (as we shall show in this Chapter).

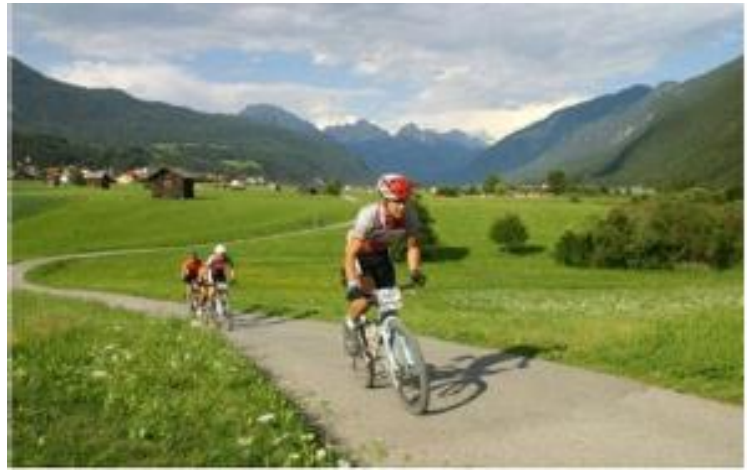

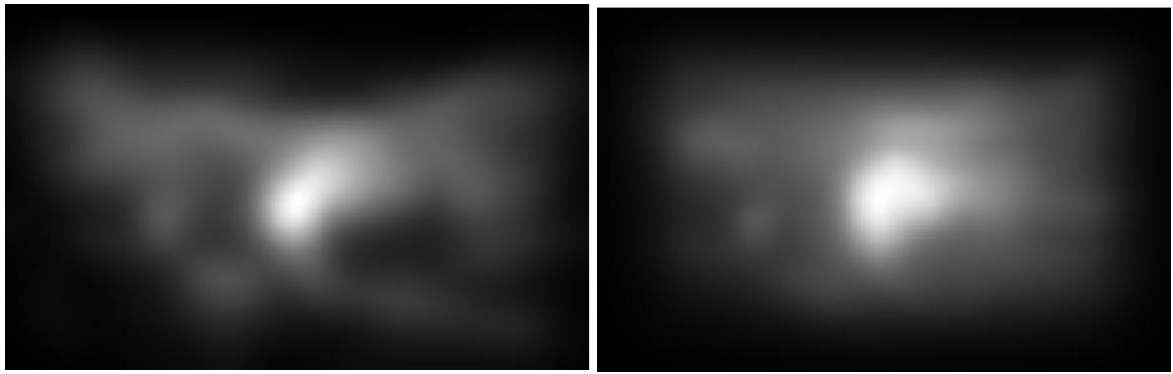

**Fig. 3-1** *(Top Row)* Input Image *[Image adapted from [66] (© Springer)]. (Bottom Row – Left to Right)* Saliency measure using [30], Saliency measure using Graph Based Visual Saliency approach of [68]

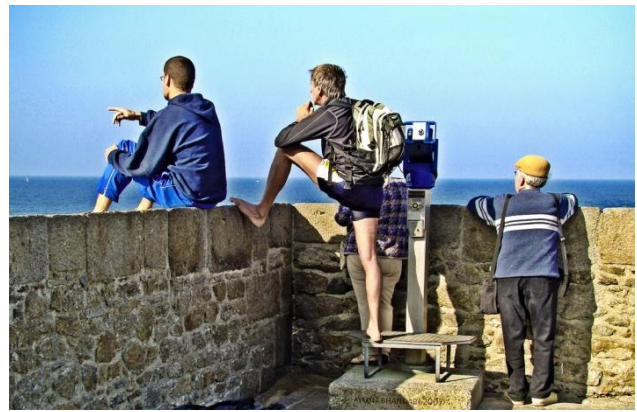

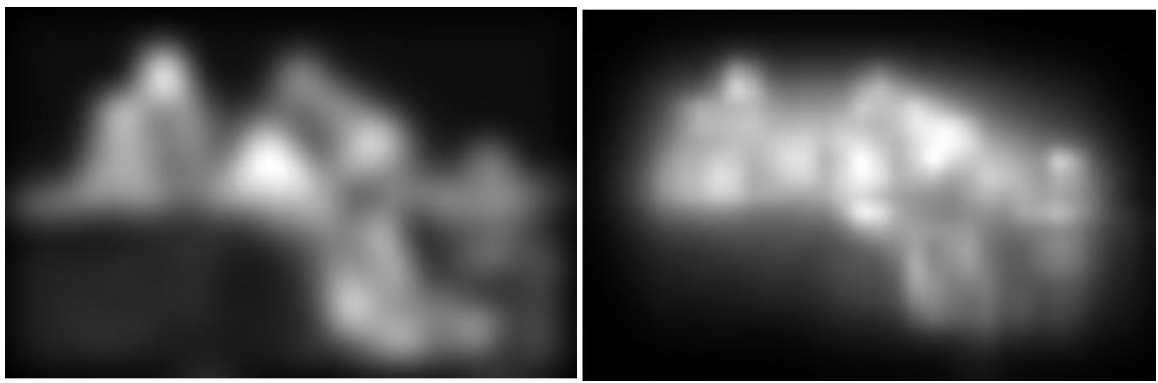

**Fig. 3-2** *(Top Row)* Input Image *[Image courtesy - http://www.flickr.com/photos/ayushbhandari/2054189454/ (The World Tomorrow)]. (Bottom Row – Left to Right)* Saliency measure using [30], Saliency measure using Graph Based Visual Saliency approach of [68]

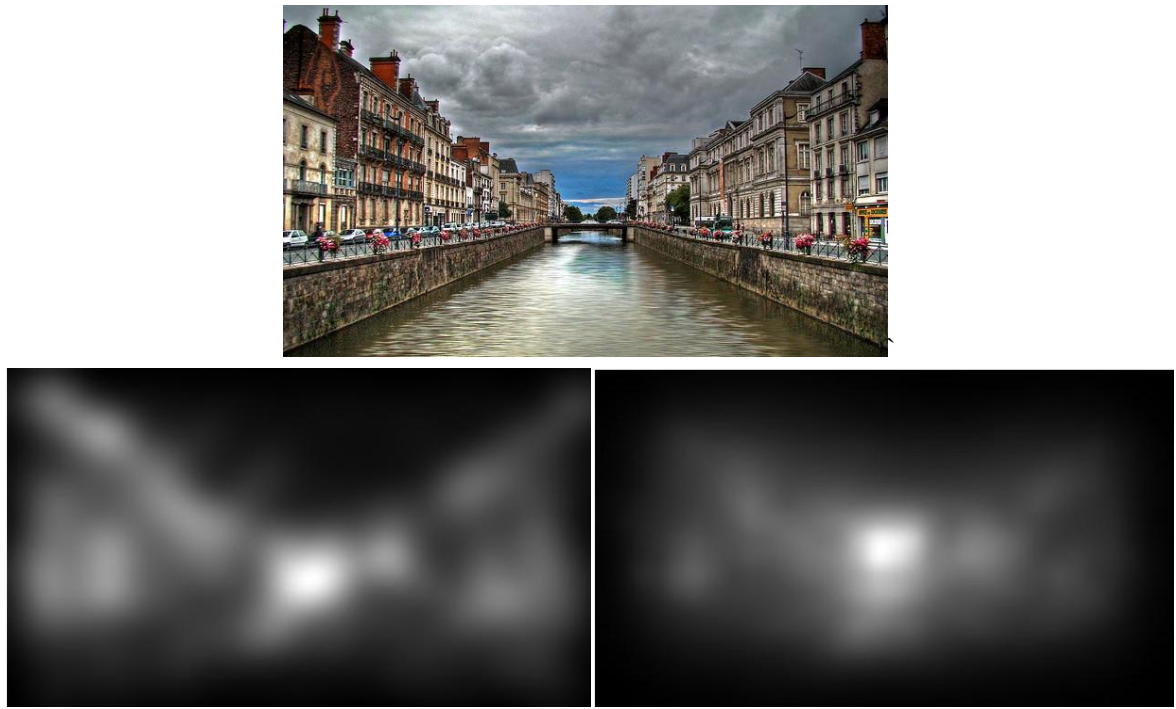

**Fig. 3-3** *(Top Row)* Input Image *[Image courtesy – http://www.flickr.com/photos/ayushbhandari/2330980893/ (Stormy Sky over Rennes!)]. (Bottom Row – Left to Right)* Saliency measure using [30], Saliency measure using Graph Based Visual Saliency approach of [68]

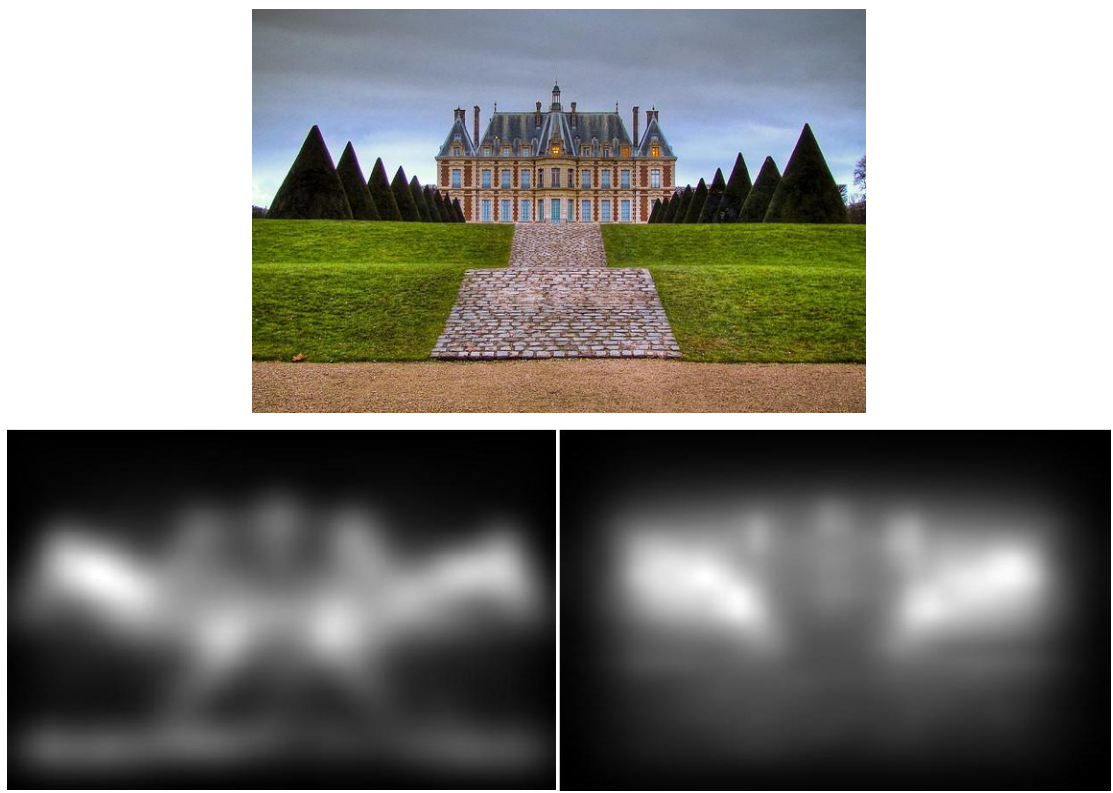

**Fig. 3-4** *(Top Row)* Input Image *[Image courtesy – http://www.flickr.com/photos/ayushbhandari/2172625468 (Château in the Parc de Sceaux)]. (Bottom Row – Left to Right)* Saliency measure using [30], Saliency measure using Graph Based Visual Saliency approach of [68]

Next, we compute the *L1* gradient norm of the input image ***I*** to get ***E***. This is important since we wish to identify the edges in the input image. However, this may identify most of the times some edges that are redundant for our content aware image resizing problem. An example of this is that such an edge detection operator also quantifies the fine textures of the images which are often not that important while content aware resizing an image. In our gradient energy map ***E***, the values are not binary, i.e. the map is a gray scale map rather than being a binary image map.

To solve this problem, we present a thresholding technique to segregate the fine textures and combine it with the above two maps (the exact way of combining is discussed at the end of this sub section).We take the *L2* gradient energy of the input image ***Y***, and for the pixel locations where the values in ***Y*** are less than the mean of ***Y*** multiplied by an adjustable parameter, we make those pixel location in ***E*** as nearly zero. The rest of the values are left untouched. *Fig. 3-5* and *Fig. 3-6* show the result of this technique for various input images. It can be seen that the fine textures in the images are not perceived as significant as compared to major edges.

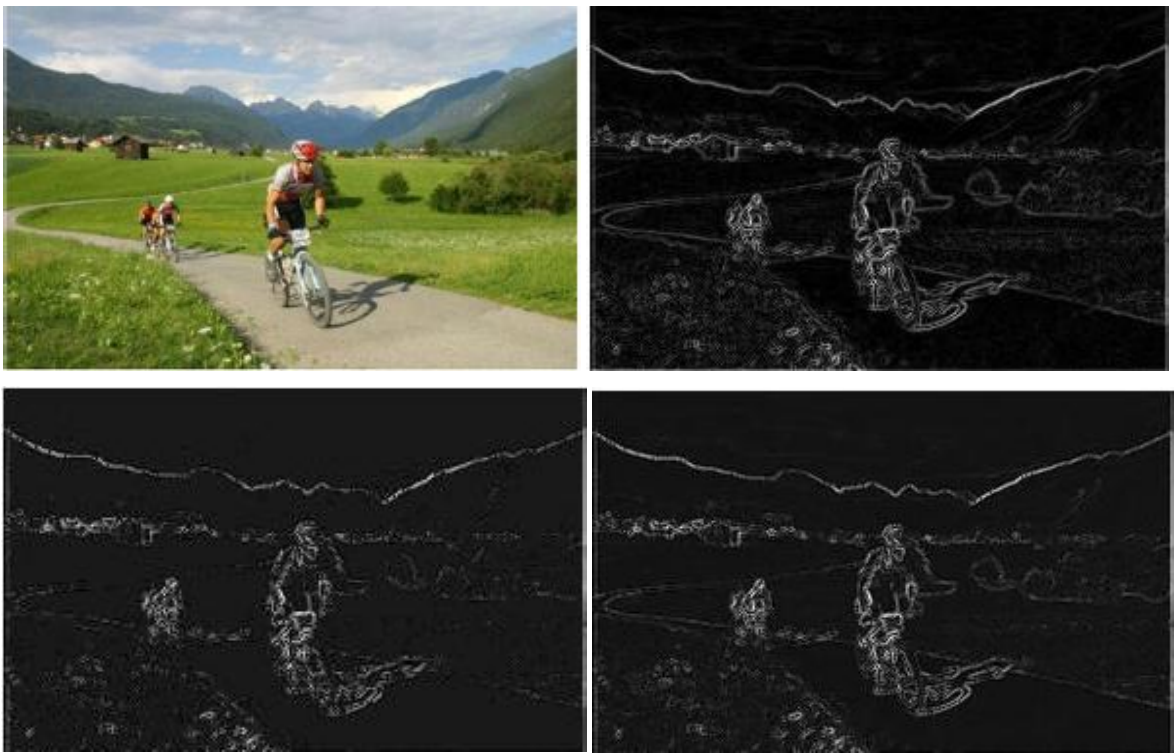

**Fig. 3-5** Depiction of the Thresholding Scheme used. *(Top Row – Left to Right)* Input Image ***I*** *[Image adapted from [66] (© Springer)]*, *L1* gradient of the input Image, ***E*** *(Bottom Row – Left to Right)* L2 gradient of the input image after thresholding ***Y***, Gradient map of input image after combining ***W***

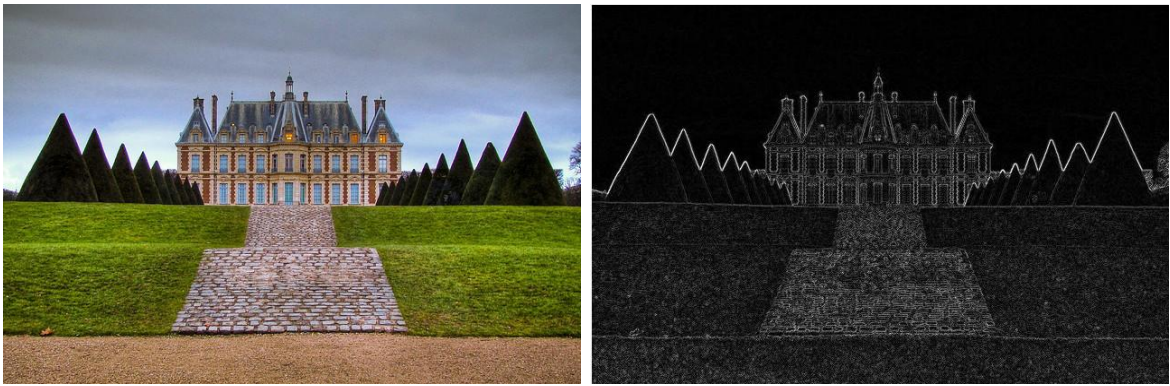

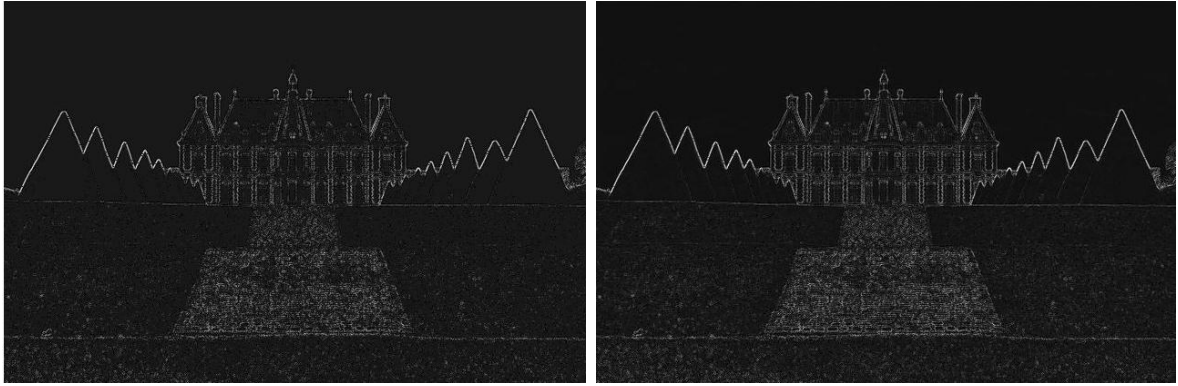

**Fig. 3-6** Depiction of the Thresholding Scheme used. *(Top Row – Left to Right)* Input Image *[Image courtesy – http://www.flickr.com/photos/ayushbhandari/2172625468 (Château in the Parc de Sceaux)]*, *L1* gradient of the input Image **E** *(Bottom Row – Left to Right)* L2 gradient of the input image after thresholding **Y**, Gradient map of input image after combining **W**.

In addition, we make use of the simple Hough transform map for identification of the major feature lines in the input image **I** to get **H**. The Hough transform identifies the major feature lines. It is noteworthy that the simplistic form of the Hough Transform also identifies some redundant lines which are generally very small in their length. Thus, we discard the lines which are generally 1/10 of the diagonal of the image. At this point, we don't really concentrate on the object boundaries since they are automatically taken into account within the region based segmentation approach in our algorithm and thus we have not used a generalized version of the Hough transform for detecting arbitrary shapes [83]. The detection of these feature lines is useful for our algorithm, because we define the energy functions in our framework dealing with such lines specifically (discussed later in this Chapter). Note that there exist some better strategies to make note of the object boundaries; however for the purpose of content aware image retargeting using our framework, this simple technique suffices since we have anyways object boundaries in **E** and that we later quantify the importance of the lines using an adjustable parameter $\mu$ to include in our optimization model.

The above maps thus obtained are combined to form our final importance map. The entire importance map generation procedure is summarized in *Fig. 3-7*. *Fig. 3-8* depicts some of the examples which show our final importance map **M**. Note that we shall delineate in Chapter 4 while discussing the results for various input images the advantages of each of the novel contributions in our framework. Within the importance map generation, the use of the wavelet based technique to make the fine texture regions less salient in the final saliency map is one of the novel contributions which results in better semantics preservation while doing content aware resizing (notion provided with results in Chapter 4).

It is noteworthy to mention here the use of various adjustable parameters as given in *Fig. 3-7*. The typical values of these adjustable parameters are given along with the algorithmic descriptions (at each level and overall). Having said the above, a slight change in these parametric values shall not bring a very noticeable change in the output results. However, a good amount of variation in these values shall bring noticeable changes. For example, making $\alpha$ very small shall not do a significant thresholding resulting is not neglecting the fine texture, which may generally result in visual distortions elsewhere, especially in feature lines. A similar case follows with the parameter $\beta$. Keeping the value of $\gamma$ small generally results in the feature lines and other object boundaries non-homogenously scaled (thus giving slight edge distortions) during our mesh based warping technique, while also giving more importance to the texture.

1) Compute the Graph Based Visual Saliency (GBVS) Map of the input image **I**, to get **G**.
2) Compute the *L1* Gradient Energy of the input image **I**, to get **E**.
3) Compute the Hough transform map from the input image **I**, to get **H**.
4) Compute the *L2* gradient energy of the input image, to get **Y**.
$$E(i,j) = 0.1; \forall Y(i,j) < \alpha * mean(Y)$$

> Here, *i* and *j* represent the indices for representing the image along rows and columns respectively. After thresholding combine as follows:
> $$W = E + \beta Y$$
> *α* and *β* are adjustable parameters. For most cases, we have assigned the following values to these parameters.
> $$\alpha = 1.2 \; ; \; \beta = 1.5$$
> 5) Compute the final importance map for the given input image as *M,* by the following combination rule.
> $$M = \aleph\big(\aleph(G) * \aleph(E) + \aleph(W) + \gamma \aleph(H)\big)$$
> *γ* is an adjustable parameter whose value is typically chosen as $\gamma = 2.$ Here, $\aleph$ is the normalization operator which normalizes all the values of the image between 0 and 1.

**Fig. 3-7** A Method of computing the first Importance Measure for our framework The method considers many different notions while designing the saliency map.

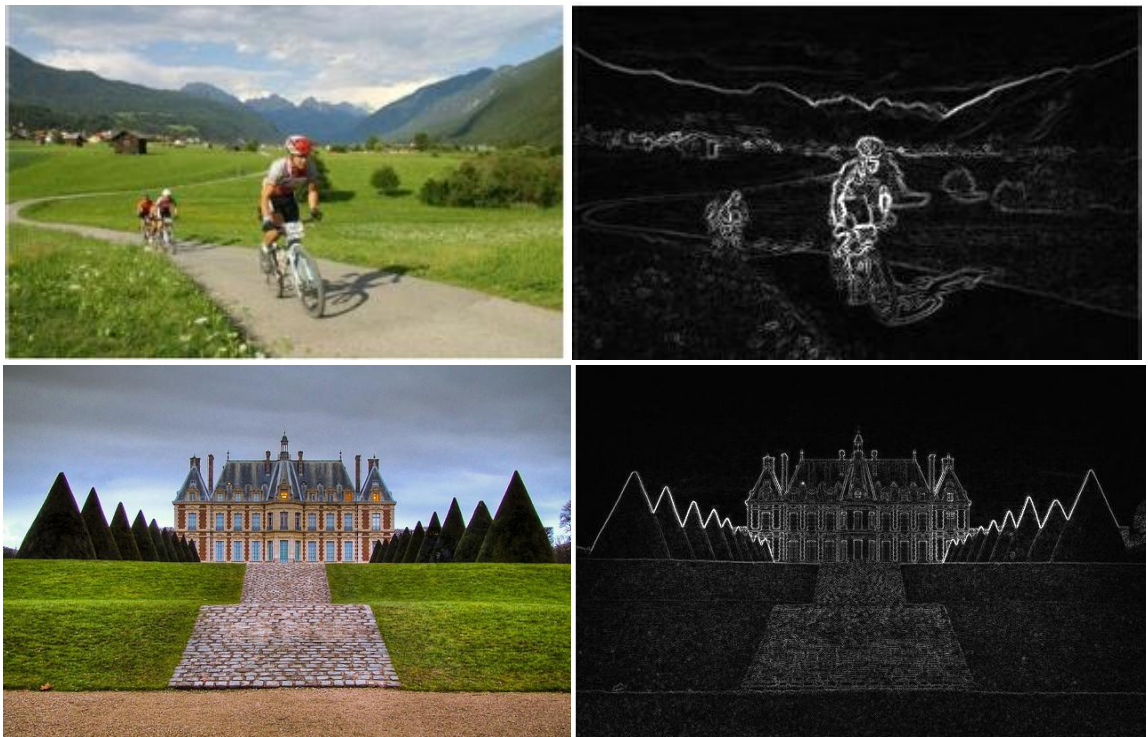

**Fig. 3-8** Depiction of the Final Importance Map generated. *(Top Row – Left to Right)* Input Image *I [Image adapted from [66] (© Springer)],* Importance Map *M* of input image 1 (*Bottom Row – Left to Right*) Input Image 2 *[Image courtesy – http://www.flickr.com/photos/ayushbhandari/2172625468 (Château in the Parc de Sceaux)],* Importance Map *M* of input image 2.

It must be noticed in *Fig. 3-8* that the Hough map has not made much of a difference. This has been a problem for many of the techniques used in this field with the various versions of Hough maps. However, even the state of the art variants of Hough transform for detecting feature lines do not work well. This is where our region based segmentation gains more advantage. Once, the regions are segmented and with the energy functions designed for preservation of aspect to ratios of more salient regions as much as possible, they help to preserve feature lines more efficiently than otherwise. Our results in Chapter 4 shall depict this more conspicuously.

## 3.2 Region Based Color Image Segmentation

We follow an altogether different approach for designing energy functions for warping based content aware image resizing. We apply region based image segmentation first before designing our core formulation of the problem. For region based segmentation, a graph based approach [84] is used. Although, other approaches could also be used for image segmentation, this approach gives us substantial results given our problem at hand, and the way we use the regions. The motivation behind region segmentation is as follows:

➢ In the current warping based techniques, some of the regions are distorted horribly even though the saliency maps identify those regions as salient. This is mainly because the boundaries of such regions are not very much identified while resizing and this results is horrendous results at times. *Fig. 3-9* shows a situation where even though the saliency map identifies the region as important, the warping based method of [61] distorts the region horribly, while preserving the other regions. Thus, a region based segmentation of the image is done here, and the constraints are so formed that the optimization framework always tends to preserve the aspect ratio of the region, if not the original size. This way, we ensure that every semantic region keeps its semantics intact even after resizing. It is noteworthy, that the region based segmentation will not be able to completely segregate the girl's hair from the dark background. However, with region segmentation, the saliency of the other regions decreases and the weighting of the region of girl's hair blended with the background increases relatively. With the criterion of adjacency regions' edged being scaled with almost the same factor, we obtain quite better results (shown in Chapter 4). Also the criterion for the preservation of the original size of the most salient region restricts the hair getting extended since the lower region wishes to preserve the original size.

➢ The other reason for region based segmentation approach is to preserve the aspect ratios of the lesser salient objects, when the target aspect ratio is excessive given the nature of the image. An example of this is shown in *Fig. 3-10.* One can easily observe in the figure that it is really difficult to preserve the semantics of the image given its nature and the targeted size. So, instead of still resizing the image, we follow an approach where we place importance on preserving the aspect ratio of the each of the persons in the image, at the expense that they overall size will not be the same. This preserves better semantics.

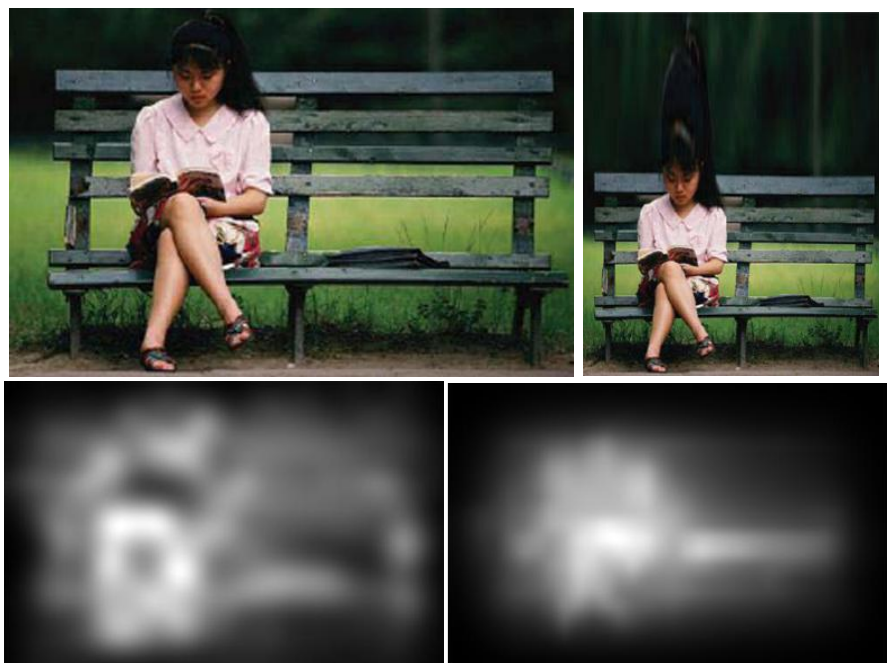

**Fig. 3-9** *(Top Row – Left to Right)* Input Image [Original Image adapted from http://www.asse.com/united_states/thailand.htm], Image Retargeted using [61]. *(Bottom Row – Left to Right)* Saliency map of the input image using [30], Saliency map of the input image using [68] Note that these saliency maps are combined with the *L1* gradient energy for retargeting.

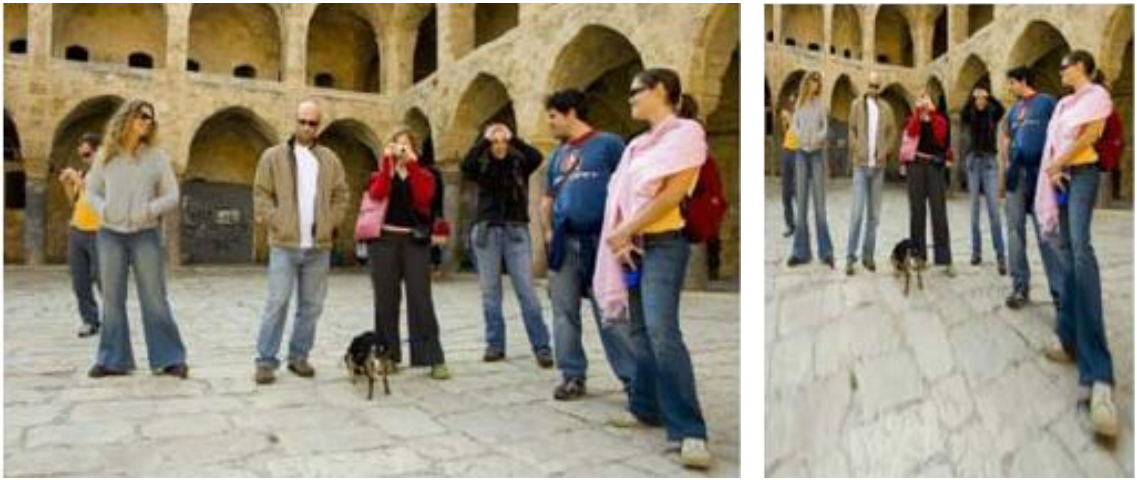

**Fig. 3-10** *(Top Row- Left to Right)* Input Image, Image retargeted using warping approach of [61] Retargeting is performed for reducing the original width by 50% with no change in the height. *[Images adapted from [74] (© IEEE)]*

      We revisit such improvements by citing the results of our algorithm in Chapter 4. Here, we present the motivation behind following novel strategies. While these points drove our motivation for a region based color image segmentation based approach, it is worthwhile to mention here that such an approach cannot be deemed suitable without the formation of suitable energy functions. We discuss all the energy functions used in out optimization framework in Section 3.4. *Fig. 3-11* presents some region based image segmentation results for various input images along with their region based saliency weighted maps $M_R$. The map shows the regions with more saliency as brighter than the others. It is interesting to note that the very small regions detected near the edges are often seen as salient, since these small regions in the saliency map $M$ contain essentially the edges.

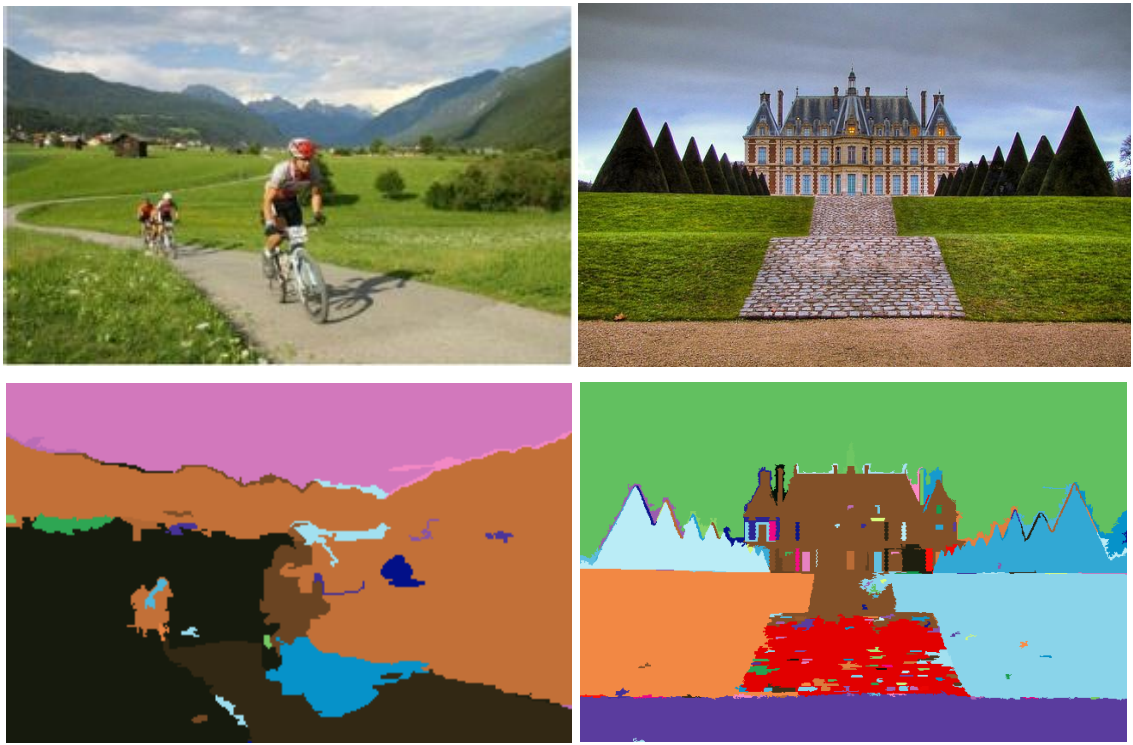

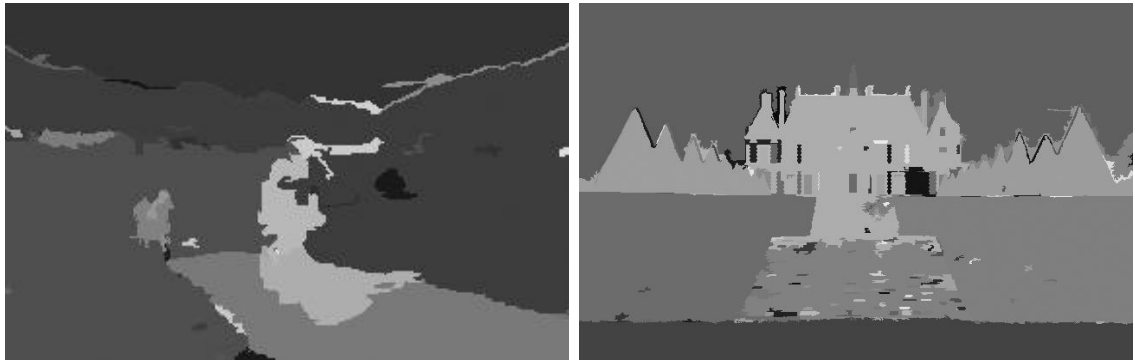

**Fig. 3-11** Depiction of region based segmentation. *(Top Row – Left to Right)* Input Image 1 *[Image adapted from [66] (© Springer)]*, Input Image 2 2 *[Image courtesy – http://www.flickr.com/photos/ayushbhandari/2172625468 (Château in the Parc de Sceaux)] (Middle Row – Left to Right)* Region segmentation of input image 1 (shown in randomly generated colours), Region segmentation of input image 2 (shown in randomly generated colours) *(Bottom Row – Left to Right)* Region weighted map of input image 1 (Brighter indicates more important region), Region weighted map of input image 2 (Brighter indicates more important region).

Having segmented the image into various regions, we multiply it with the above found saliency map *M* so as to weigh the important regions. A mathematical description of this approach can be found in *Fig. 3-12*. Since, we don't tend to detect very small regions, this weighing here as per the regions conveys a very different meaning, and in a way derives the preservation of semantics of the image. This approach helps to weight the different regions of the image (segmented as per the color variations which are very visible to the human eye) rather than weighing the edges and smaller mesh quads and triangles within the image. For example, a region might contain some edges but other non-salient information, while other regions may contain overall salient information. By the proposed approach, we shall weigh the overall region containing edges as less salient (although we try to preserve edges as shown in energy function formulation in Section 3.4) than the other regions, which is in a way desired. Lesser saliency here still preserves the edge orientations and minimizes visual distortions as we shall see in Section 3.4. But, lesser salient regions can be scaled uniformly while maintaining their aspect ratio as much as possible, as compared to the more salient regions whose original size is tried to be preserved as far as possible.

1) Compute Graph Based Color Image Segmentation to obtain the pixels labelled in various regions so segmented. Let the set of segmented regions be represented as follows:

$$\mathbb{R} = \bigcup_{r=1}^{N_R} R_r$$

Here, $N_R$ is the total number of regions segmented with $R_r$ being the individual regions. The algorithm requires a parameter for deciding on the number of segmented regions, which we set typically set as $k = 1000\,;\,\sigma = 0.5$.

2) Form a region based importance map $M_R$ using the following expressions.

$$M_R = \left(\bigcup_{r=1}^{N_R}(R_r * w_{R_r})\right) \forall\, R_r \in \aleph(I)$$

Here, $w_{R_r}$ is the average of the region $R_r$, taken from the importance map *M*.

**Fig. 3-12** Region Based Color Image Segmentation of the input image. After segmentation, the regions are weighted according to the computed importance measure is Section 3.1.

## 3.3 Delaunay tri-mesh based formulation

We use a uniform resolution Delaunay triangular mesh grid for formulating the warping framework in our algorithm. Such a notion has been previously used in the work of [65]. As in other approaches, the initial mesh is formed and the final mesh is obtained after running an optimization framework. Each of the triangles in the original mesh is warped to the destination triangles in the finally computed mesh. We use a uniform scaling for mapping between the triangles of the original mesh and the final mesh rather than texture mapping. Some researchers [61] use uniform scaling while others [65], [66] have used texture mapping. Apart from the implementation issues (discussed in Section 3.5), we also don't use texture mapping since we could not observe any noticeable difference when the warping is done using uniform scaling and when it is done using texture mapping. Also, texture mapping has not proved successful for preserving blur and noise portions in the retargeted image. Thus, for a globally efficient method of mapping between the triangles of the original and the final mesh, a much more sophisticated method needs to be followed, especially for preserving blur in the output images. For normal images which are generally encountered (considered by all researchers involved in the field and by us here), we follow the uniform scaling based warping procedure.

After forming the mesh, the triangles in the mesh are divided into different categories. One major category is the classification of triangles which are contained within different regions. The other category is the triangles which contain most significant pixels of the image according to the weighted importance of each region and the border pixels between two regions. This helps us to smooth the scaling factors for the pixels that are associated with the region boundaries. This categorization is simple since, we have already labelled pixels according to different regions during region segmentation. *Fig. 3-13* shows the categorization of the mesh triangles into different categories thereby forming the associated sets. *Fig. 3-14* shows mesh formation of various input images and classification of the mesh triangles according to the regions and the most important pixels. The mesh used here for the input image (which is warped after solving the optimization framework) is an unstructured mesh generated by slightly randomly perturbing the control points of a structured Delaunay mesh. The mesh is only constrained to sample the points on the image boundaries.

1) Form a uniform initial mesh on the input image *I*, using Delaunay constrained triangular mesh grid. The grid is only constrained within the image boundaries and not constrained anywhere within the image.

2) Divide the entire set of triangles $\mathbb{Q}$ in the initial mesh grid into two categories, thereby forming the following sets.

$$\wp_{R_r} = \bigcup_{\mathbb{Q}} \{q, q_{R_{r_n}}\}; \ M(i,j) \geq \mu \max(M_{R_r}) \ \forall \{i,j\} \in q$$

$$\mathcal{L}_{R_r} = \bigcup_{\mathbb{Q}} q \in R_r \ ; 1 \leq r \leq N_R$$

Here, $q_{R_{r_n}}$ is the set of triangles that have at least two labels of pixels obtained from region segmentation, with one of the them being $R_r$. The parameter $\mu$ is an adjustable parameter chosen as $\boldsymbol{\mu = 0.9.}$

**Fig. 3-13** Division of the triangles of the initial mesh into two categories, one where the triangles are labelled as per the different regions segmented, and the other where the triangles are segregated if they contain the pixels of nearly maximum importance and along the region boundaries.

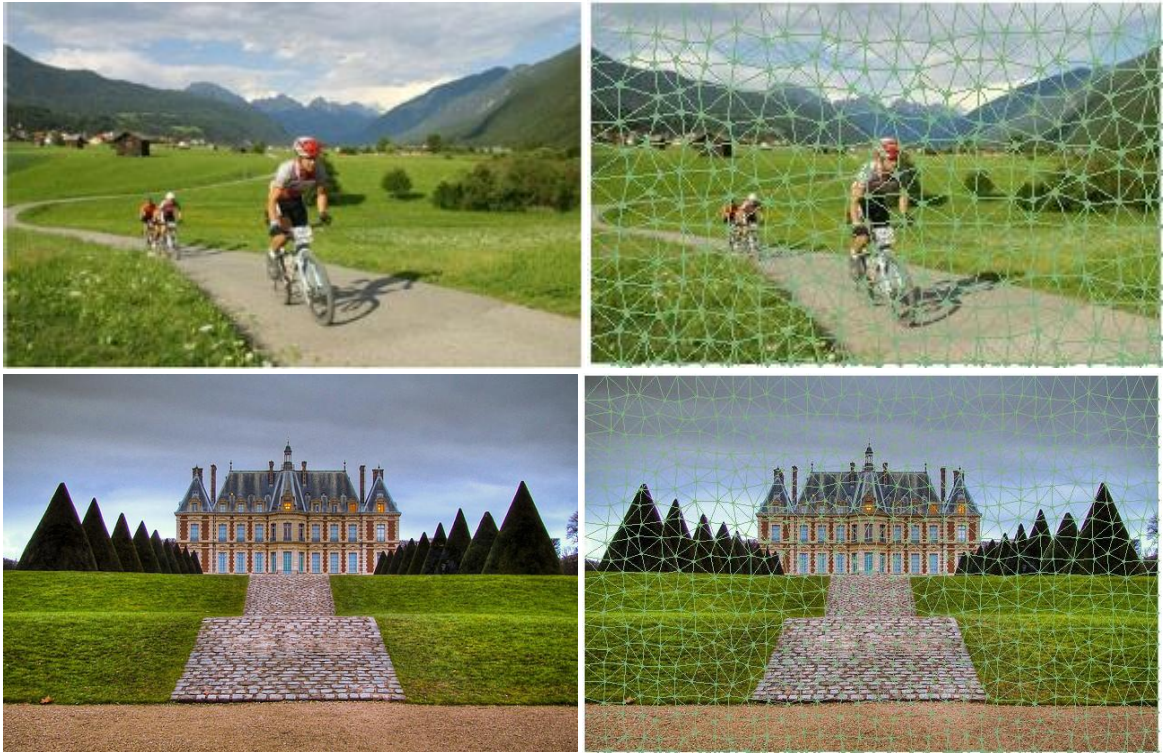

**Fig. 3-14** Triangular mesh overlay *(Top Row – Left to Right)* Input Image 1 *[Image adapted from [66] (© Springer)]*, Delaunay Triangular mesh overlay over input image 1(This is the initial mesh grid used) *(Bottom Row – Left to Right)* Input Image 2 *[Image courtesy – http://www.flickr.com/photos/ayushbhandari/2172625468 (Château in the Parc de Sceaux)]*, Delaunay Triangular mesh overlay over input image 2 (This is the initial mesh grid used). *The division of triangles into the sets formed is mot marked here as it tends to make the overlay quite clustery. However, seeing the region segmented maps given in Fig 3-11, once can clearly get an idea of the triangles belonging to the sets.*

## 3.4 Optimization Framework

Any mesh based (whether a quad mesh such as in [61] or a triangular mesh as in [65]) warping technique for content aware image retargeting involves the solution of a multi grid optimization framework, which is driven by certain energy functions and some constraints. In general, researchers try to keep the functions as simple as possible so that the entire optimization problem is essentially a linear minimization. Regarding the constraints, while some researchers [61] try to keep the constraints simple, others as in [65] make the constraints quite complex thereby resulting in more computational complexity. For all the methods, the novelty lies in the design of the functions created for optimization and obviously the underlying theory behind the design. While, everyone has to satisfy some standard mesh deformation energy functions, the introduction of additional functions (and constraints) helps to achieve different results.

      We design our energy functions driven from the region based image segmentation and the segregation of our triangular meshes into different sets as discussed in previous two sections. However, to start with, we consider the energy functions given by *equations 3-1, 3-2* and *3-3*. The first two equations convey the standard shape deformation energy function used in all methods for image warping while the third equation deals with the a uniform orientation of the edge lengths.

$$E_1 = \sum_{q \in \mathbb{Q}} w_q \sum_{\{i,j\} \in \varepsilon(q)} \left\| (c'_i - c'_j) - \vartheta_q (c_i - c_j) \right\|^2 \qquad \textbf{(3-1)}$$

$$\vartheta_q = \frac{\sum_{\{i,j\} \in \varepsilon(q)} (c_i - c_j)^T (c'_i - c'_j)}{\sum_{\{i,j\} \in \varepsilon(q)} \|(c_i - c_j)\|^2} \qquad \textbf{(3-2)}$$

Here, the set $\mathcal{E}(q)$ is the set of edges of the triangles $q \in \mathbb{Q}$, while $\mathcal{E}$ is the set of all edges contained in the set $\mathbb{Q}$. $w_q$ is the average weight of all pixels within the triangle $q$ according to $M$.

$$E_2 = \sum_{\{i,j\} \in \mathcal{E}} \left\| (c_i' - c_j') - \left( \frac{\|c_i' - c_j'\|}{\|c_i - c_j\|} \right) (c_i - c_j) \right\|^2 \tag{3-3}$$

We now present two novel energy functions for inclusion in the optimization framework, represented by *equations 3-4, 3-5*. *Equation 3-4* presents a function which tries to prevent the different regions of an image from resizing based on the region based saliency map $M_R$. This basically models the concept that we discussed in Section 3.2. Preserving the aspect ratios of some of the salient regions (whose importance is based on the average value of the saliency within the entire region) is one of major constraints of our approach, wherein we try to preserve the original size of the most salient regions, while trying to preserve the aspect ratios of other regions, and the original size as far as possible.

$$E_3 = \sum_{r=1}^{N_R} w_{R_r} \left( \sum_{\{i,j\} \in \mathcal{L}_{R_r}} \left\| (c_i' - c_j') - \left( \frac{p_{ij}'}{p_{ij}} \right) (c_i - c_j) \right\|^2 + \tau \left\| (c_i' - c_j') - (c_i - c_j) \right\|^2 \right) \tag{3-4}$$

Here, $p_{ij}$ is the perpendicular on to the edge *(i,j)* from the opposite vertex in the original mesh, and $p_{ij}'$ the perpendicular from the opposite vertex in the new generated mesh at each iteration. $\tau$ is an adjustable parameter whose value is set for our simulations as 0.4.

As hinted in Section 3.2, while the less salient regions have less importance in preserving the aspect ratios, we don't do this at the expense of the preservation of the edges. Thus, the important edges in all regions should be preserved as far as possible, possible remaining of the same size in the most salient regions, and of lesser size, but the same aspect ratio and orientation in lesser salient regions. This requires that the triangles containing the edges are scales by almost the same factor as their neighbours, constrained with a single region. Where in the edges go across different segmented regions and do not also the boundaries of the regions, the algorithm might result in a bit of distortion of the lines. However, such cases are often not found in the images. Note here that the major noticeable object boundaries are taken care of in the region based segmentation equation trying to prevent the aspect ratio of major objects. *Equation 3-5* represents the function that is solved for optimization at the start of each iteration after the $\vartheta_q$ is found as in *equation 3-2* above and the factors are updated to yield $\vartheta_q^u$. This is for smoothing of only the feature specific regions and thus to avoid distortion. Let $\vartheta_q^u$ be the updated values.

$$A_I = \sum_{r=1}^{N_R} \sum_{q \in (\wp_{R_r})} \sum_{q_n \in (\mathbb{N}(q) \cap \wp_{R_r})} \left( (\vartheta_q^u - \vartheta_q)^2 + \frac{1}{2}(w_q + w_{q_n})(\vartheta_q^u - \vartheta_{q_n})^2 \right) \tag{3-5}$$

Here, $\mathbb{N}(q)$ is the set of the triangles which share their edges with at least one of the edges of $q$. Note that we constrain this smoothing for the triangles containing feature lines only. This ensures that those triangles are uniformly scaled. The initial guess for $\vartheta_q^u$ is all ones (implying uniform scaling criterion), and the stopping criteria is when the difference between all factors between two successive iterations is less than 0.1. Note that this energy minimization is also solved using Newton multi grid solver but without any image or triangular mesh constraints. Only the constraint that we update certain factors is inserted into the method. We reiterate that the factors are updated at the start of every iteration after being found out by *equation 3-2*. The final energy function for optimization is given by the *equation 3-6*.

$$E_o = E_1 + E_2 + E_3 \tag{3-6}$$

Any optimization framework comes with some constraints in general. For our methods, some general constraints have been introduced so as to not allow the resultant vertices (control points of the mesh after optimization) to disturb the rectangular nature of the image, and not to allow flipping and fold over of certain triangles. Flipping is a problem that generally occurs in triangle based mesh warping as against quad mesh based warping. For a triangle, a situation should not occur where the new vertices of a triangle get to the other side with respect to the two other vertices. If that happens, the entire topology of the mesh gets disturbed and one happens to get weird results. Thus, we keep the edge orientations of the triangles same. This also prevents the folding over a certain triangle during resizing. Ideally, we do not want folding over since it results in content removal. We wish to have the area of a triangle to be minimal wherever desired, but not zero. *Equations 3-7, 3-8, 3-9, 3-10* provide the constraints to be included while solving the optimization problem.

$$(c'_{i_x} - c'_{j_x})(c_{i_x} - c_{j_x}) \geq \varepsilon_t \tag{3-7}$$

$$(c'_{i_y} - c'_{j_y})(c_{i_y} - c_{j_y}) \geq \varepsilon_t \tag{3-8}$$

$$(c'_{z_x} - p'_{ij_x})(c_{z_x} - p_{ij_x}) \geq \varepsilon_p \tag{3-9}$$

$$(c'_{z_y} - p'_{ij_y})(c_{z_y} - p_{ij_y}) \geq \varepsilon_p \tag{3-10}$$

Subscripts *x* and *y* indicate the x and y co-ordinates of the corresponding vertex. $c_z$ is the other vertex apart from $c_i$ and $c_j$ in a triangle. $p_{ij}$ is the perpendicular from on to the edge defined by $c_i$ and $c_j$. $\boldsymbol{\varepsilon_t}$ and $\boldsymbol{\varepsilon_p}$ are adjustable parameters whose values are typically set as $\boldsymbol{\varepsilon_t = 0.02}$ ; $\boldsymbol{\varepsilon_p = 0.05}$. It can be clearly seen that both the x-coordinate and the y-coordinate constraints cannot be satisfied if there is a change in the direction of any of the original mesh edges.

One of the major constraints in any optimization model is the initial guess and the stopping criteria. Given the desired size, we take the initial guess of the vertices by homogeneously translating the vertices of the original mesh to the mesh of the desired size. The framework stops when the successive iterations differ only by a margin of 0.5 or less between all vertices.

## 3.5 Implementation

Having explained our entire perspective and the algorithm for content aware image retargeting, we now give an overview of how the code has been implemented. The entire algorithm is given in *Fig. 3-15*. As is conspicuous, we make use of different approaches such as saliency map generation, Hough Transform map generation, region based color segmentation within our algorithm, and build upon them defining our novel energy functions and constraints to get the desired results. So, we use some already available pieces of code and adapt to our algorithmic requirements, while merging with our MATLAB written code. The following points give the major background codes/utilities used for our implementation (Please refer either the previously described sections or *Fig. 3-15* for walking again through our entire algorithm).

- ➢ The Graph Based Visual Saliency (GBVS) Map is calculated using the MATLAB code provided at http://www.klab.caltech.edu/~harel/share/gbvs.php.
- ➢ The region based segmentation based on the graph method has been done using the code provided at http://people.cs.uchicago.edu/~pff/segment/. The code is in C++ and we run the code to get a segmented image, which is then adapted to our MATLAB Code for further processing.
- ➢ The code for the Newton method for multi grid solver has been adapted from the code implemented in [65]. This helped us to run the typical optimization framework with our energy functions. The constraints used in [65] were the same as used by us, since they had also done triangular mesh grid based warping.
- ➢ The Delaunay mesh generation from the set of initial vertices and the final vertices (obtained after optimization) has been done using the built in *DelaunayTri* Class of MATLAB.
- ➢ The lines using Hough Transform has been computed using MATLAB built in functions – *hough, houghlines, houghpeaks.* In a combined way, they use the voting procedure for detection of peaks and thus find the corresponding lines. A discussion on its accuracy and corresponding thresholding technique followed was provided in Section 3.1.

- ➢ The warping of the mesh triangles from the original set of vertices to the final set of vertices has been done using the built in *warp* function in MATLAB. Texture mapping was not preferred as compared to uniform scaling since that has not produced any noticeably improved results for other researchers (discussed in Section 3.3) , and also, that MATLAB supports texture mapping for surface based objects rather than patch based objects.

The time for running the entire framework needs special mention here. This is because of the major reason that we use the code for region based segmentation (given in C++) and then from there take the output to further process our MATLAB algorithm. Since, the region based segmentation is a onetime process, as is the computation of the entire saliency map, the major time taken by the algorithm is running the optimization procedures. Typically for 600x400 sized images, the optimization framework takes around 200 – 250 seconds to give the results. We don't formally give a table or a chart citing our timing results for different images, since we cannot compare our results with the other approaches (to which we are comparing our results). This is because the other approaches have generally given their timing results after running a C/C++ version of their program on an Intel based CPU. Such timing results cannot be compared to MATLAB program implementation. However, given the nature of the algorithm and the general complexity that some of the other algorithms have, we can surmise that our algorithm is well implementable in real time, may be in a more feasible way with some factorization methods for running optimization.

The major contributions of our work are highlighted in *Fig. 3-15*. We regard the combination of points (4) and (5) for **efficient feature** preservation, the points of (6), (7), (9), (11), (12), (13) mainly deal with the efficient **semantics preservation** along with **feature maintenance** in the resized content.

1) Compute the Graph Based Visual Saliency (GBVS) Map of the input image $I$, to get $G$.
2) Compute the *L1* Gradient Energy of the input image $I$, to get $E$.
3) Compute the Hough transform map from the input image $I$, to get $H$.

4) Compute the *L2* gradient energy of the input image, to get $Y$.
$$E(i,j) = 0.1; \forall Y(i,j) < \alpha * mean(Y)$$

Here, $i$ and $j$ represent the indices for representing the image along rows and columns respectively. After thresholding combine as follows:
$$W = E + \beta Y$$

$\alpha$ and $\beta$ are adjustable parameters. For most cases, we have assigned the following values to these parameters.
$$\alpha = 1.2 \ ; \ \beta = 1.5$$

5) Compute the final importance map for the given input image as $M$, by the following combination rule.
$$M = \aleph\big(\aleph(G) * \aleph(E) + \aleph(W) + \gamma \aleph(H)\big)$$

$\gamma$ is an adjustable parameter whose value is typically chosen as $\gamma = 2$. Here, $\aleph$ is the normalization operator which normalizes all the values of the image between 0 and 1.

6) Compute Graph Based Color Image Segmentation to obtain the pixels labelled in various regions so segmented. Let the set of segmented regions be represented as follows:
$$\mathbb{R} = \bigcup_{r=1}^{N_R} R_r$$

Here, $N_R$ is the total number of regions segmented with $R_r$ being the individual regions. The algorithm requires a parameter for deciding on the number of segmented regions, which we set typically set as $k = 1000 \ ; \ \sigma = 0.5$.

7) Form a region based importance map $M_R$ using the following expressions.

$$M_R = \left( \bigcup_{r=1}^{N_R} (R_r * w_{R_r}) \right) \forall R_r \in \aleph(I)$$

Here, $w_{R_r}$ is the average of the region $R_r$, taken from the importance map $M$.

8) Form a uniform initial mesh on the input image $I$, using Delaunay constrained triangular mesh grid. The grid is only constrained within the image boundaries and not constrained anywhere within the image.

9) Divide the entire set of triangles $\mathbb{Q}$ in the initial mesh grid into two categories, thereby forming the following sets.

$$\wp_{R_r} = \bigcup_{\mathbb{Q}} \{q, q_{R_{r_n}}\}; \; M(i,j) \geq \mu \max(M_{R_r}) \forall \{i,j\} \in q$$

$$\mathcal{L}_{R_r} = \bigcup_{\mathbb{Q}} q \in R_r \; ; 1 \leq r \leq N_R$$

Here, $q_{R_{r_n}}$ is the set of triangles that have at least two labels of pixels obtained from region segmentation, with one of the them being $R_r$. The parameter $\mu$ is an adjustable parameter chosen as $\mu = 0.9$.

10) Let the set $c$ represent the set of control points of the initial mesh, or all vertices in $\mathbb{Q}$ and $c'$ the set of control points of the mesh found after optimization (iteration wise). The following mesh deformation functions are defined for solving the optimization model.

$$E_1 = \sum_{q \in \mathbb{Q}} w_q \sum_{\{i,j\} \in \mathcal{E}(q)} \left\| (c'_i - c'_j) - \vartheta_q (c_i - c_j) \right\|^2$$

$$\vartheta_q = \frac{\sum_{\{i,j\} \in \mathcal{E}(q)} (c_i - c_j)^T (c'_i - c'_j)}{\sum_{\{i,j\} \in \mathcal{E}(q)} \left\| (c_i - c_j) \right\|^2}$$

Here, the set $\mathcal{E}(q)$ is the set of edges of the triangles $q \in \mathbb{Q}$, while $\mathcal{E}$ is the set of all edges contained in the set $\mathbb{Q}$. $w_q$ is the average weight of all pixels within the triangle $q$ according to $M$.

$$E_2 = \sum_{\{i,j\} \in \mathcal{E}} \left\| (c'_i - c'_j) - \left( \frac{\|c'_i - c'_j\|}{\|c_i - c_j\|} \right) (c_i - c_j) \right\|^2$$

11) Define the region based semantic preservation energy function and the feature protection driven energy functions as follows. These are also for inclusion in the optimization model.

$$E_3 = \sum_{r=1}^{N_R} w_{R_r} \left( \sum_{\{i,j\} \in \mathcal{L}_{R_r}} \left\| (c'_i - c'_j) - \left( \frac{p'_{ij}}{p_{ij}} \right) (c_i - c_j) \right\|^2 + \tau \left\| (c'_i - c'_j) - (c_i - c_j) \right\|^2 \right)$$

Here, $p_{ij}$ is the perpendicular on to the edge $(i,j)$ from the opposite vertex in the original mesh, and $p'_{ij}$ the perpendicular from the opposite vertex in the new generated mesh at each iteration. $\tau$ is an adjustable parameter whose value is set for our simulations as 0.4.

12) After the $\vartheta_q$ is found as in (10) above, update these by solving the following optimization criterion. This is for smoothing of only the feature specific regions and thus to avoid distortion. Let $\vartheta_q^u$ be the updated values.

$$A_I = \sum_{r=1}^{N_R} \sum_{q \in (\wp_{R_r})} \sum_{q_n \in (\mathbb{N}(q) \cap \wp_{R_r})} \left( (\vartheta_q^u - \vartheta_q)^2 + \frac{1}{2} (w_q + w_{q_n}) (\vartheta_q^u - \vartheta_{q_n})^2 \right)$$

Here, $\mathbb{N}(q)$ is the set of the triangles which share their edges with at least one of the edges of $q$.

13) The final energy function for optimization is now given as
$$E_o = E_1 + E_2 + E_3$$

14) The following constraints are defined for the optimization model apart from the image boundary constraints. These constraints are standard constraints to avoid flip over and fold over in a triangular mesh grid optimization.

$$(c'_{i_x} - c'_{j_x})(c_{i_x} - c_{j_x}) \geq \varepsilon_t$$
$$(c'_{i_y} - c'_{j_y})(c_{i_y} - c_{j_y}) \geq \varepsilon_t$$
$$(c'_{z_x} - p'_{ij_x})(c_{z_x} - p_{ij_x}) \geq \varepsilon_p$$
$$(c'_{z_y} - p'_{ij_y})(c_{z_y} - p_{ij_y}) \geq \varepsilon_p$$

Subscripts *x* and *y* indicate the x and y co-ordinates of the corresponding vertex. $c_z$ is the other vertex apart from $c_i$ and $c_j$ in a triangle. $p_{ij}$ is the perpendicular from on to the edge defined by $c_i$ and $c_j$. $\boldsymbol{\varepsilon_t}$ and $\boldsymbol{\varepsilon_p}$ are adjustable parameters whose values are typically set as $\boldsymbol{\varepsilon_t = 0.02}$ ; $\boldsymbol{\varepsilon_p = 0.05.}$

15) Solve the optimization framework using standard Newton method for multi grid solver to obtain the final control points of the mesh as *c'*.

**Fig. 3-15** A Complete outline of our entire algorithm for content aware image retargeting It includes all major equations, nomenclature used, and an idea of how the algorithm works step wise step.

# 4 Results and Discussions

This chapter presents the results or our proposed algorithm for content aware image resizing. The results are presented for various kinds of images and are also compared with the other already implemented techniques. The comparison is limited in the sense that for the implementations such as [66], [65], we do not have any executable or full running code, using which we could generate their implementation results for our input images. Thus, the images for which the results are shown in their papers are only used for a comparison with these techniques. For the techniques such as [12], [61] whose executable / running code is available, we generate their results using our set of input images and use those for comparison with our results for the same set of inputs. Also, the results include relevant pieces of discussion, wherever surmised that they might be needed.

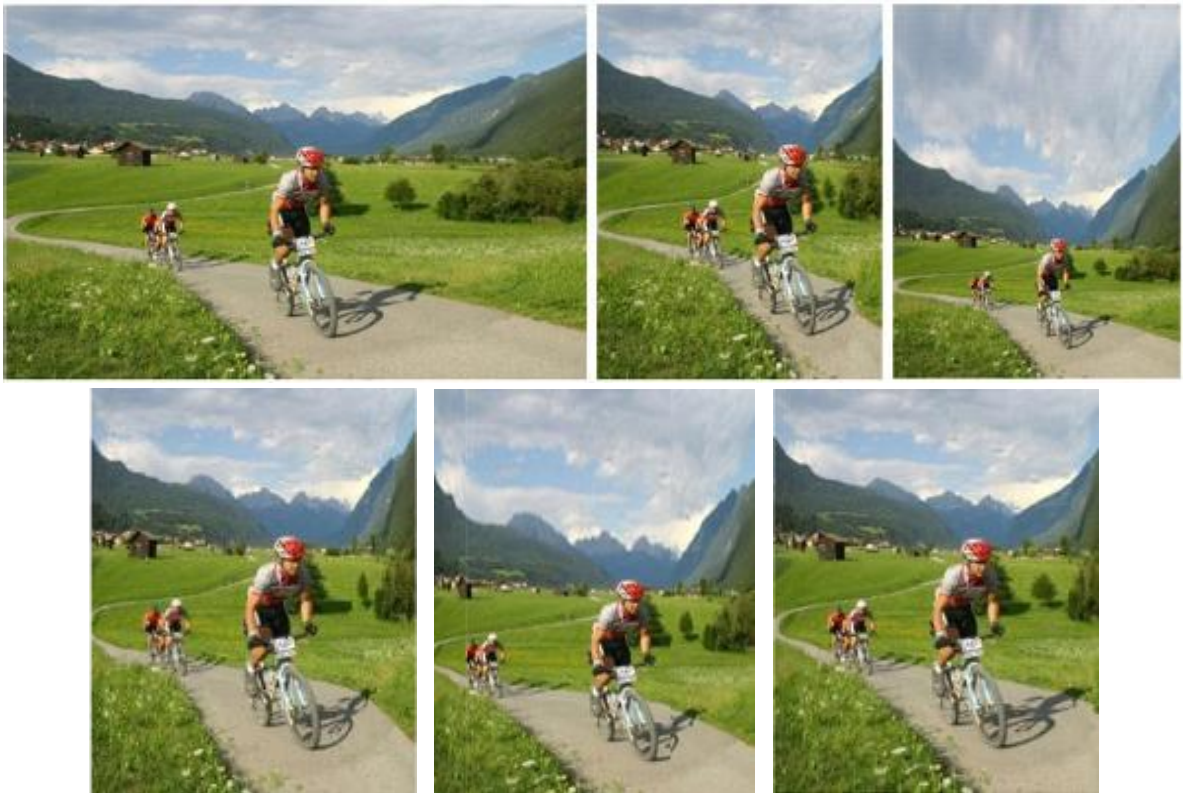

**Fig. 4-1** *(Top Row - Left to Right)* Input Image *[Image Adapted from [66] (© Springer)]*, Retargeted Image using seam carving of [12], Retargeted Image using warping of [61]. *(Bottom Row – Left to Right)* Retargeted Image using [66] *[Image Adapted from [66] (© Springer)]*, Retargeted Image **using our method** when we don't consider texture thresholding (points 4,5) and region based edge smoothing (point 12), Retargeted Image **using our full algorithm.** The retargeting is done for a 50% reduction in the width with no change in the height of the input image.

From *Fig. 4-1*, we can see that the results using our algorithm produce better results that the other state-of-art techniques. Noticing the two shadows, the grass regions around the road and the shape of the mountains, one can clearly see the difference. Also, we have presented our result when we did not use the thresholding technique (for valuing the fine texture of the grass less) and the region based edge smoothing (neglecting to warp the region boundaries by approximately the same factor). The middle figure in the bottom row of *Fig. 4-1,* we can clearly see the tilt in the road towards the end and also the distortion in the shadows.

This is basically due to the fact that when we quantify texture as more salient, the corresponding grass edges cannot be retained since the grass texture also cannot be scaled beyond a limit. Also, when we avoid region based edge smoothing, the shadows (which formed separate regions in the segmented image), are not scaled smoothly throughout their edges. One can also see slight enlargement in the sky region when we avoid these two contributions in our algorithm. This is due to the fact that with the salient texture, it still tries to preserve the aspect ratios of the most salient regions, which it is not able to do to the extent of retaining almost the original size. Note that if remove region based segmentation criteria, our algorithm more or less gives results similar to that of [61]. So, in a way, region based segmentation forms the core of our algorithm.

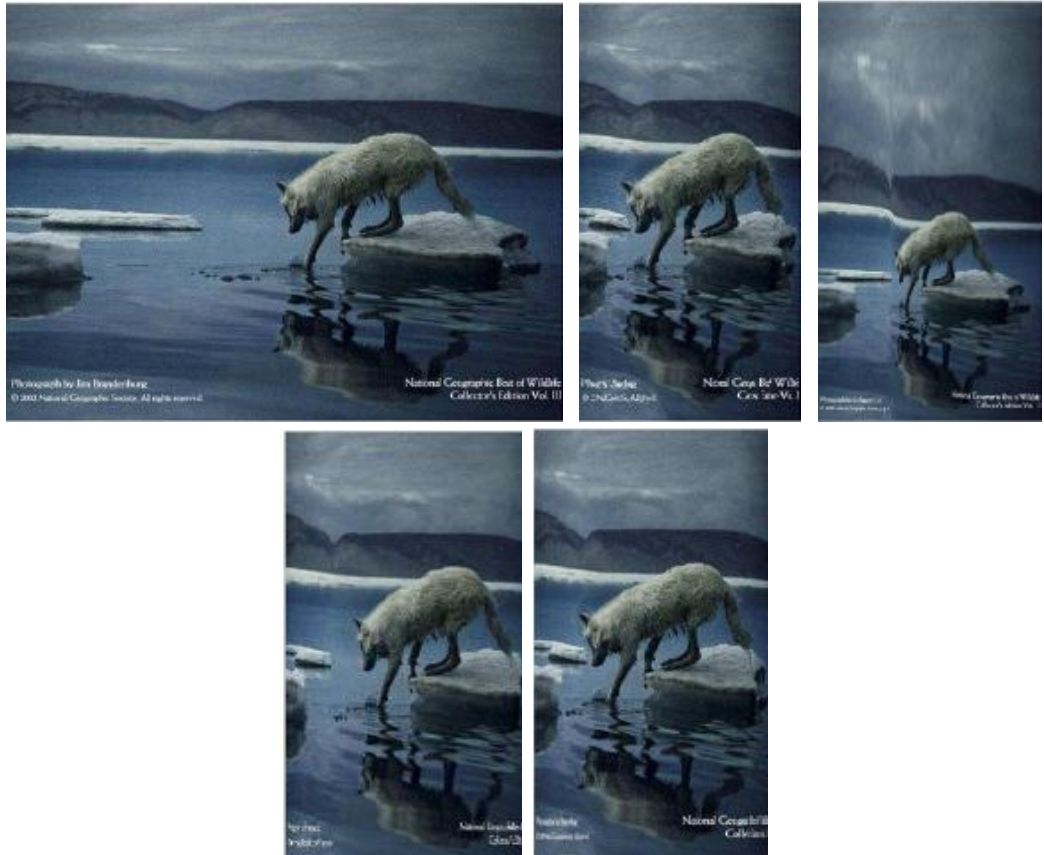

**Fig. 4-2** *(Top Row - Left to Right)* Input Image *[Image Adapted from [66] (© Springer)]*, Retargeted Image using seam carving of [12], Retargeted Image using warping of [61]. *(Bottom Row – Left to Right)* Retargeted Image using [66] *[Image Adapted from [66] (© Springer)]*, Retargeted Image **using our full algorithm.** The retargeting is done for a 30% reduction in the width with no change in the height of the input image.

Looking at *Fig. 4-2*, we can see that seam carving and our algorithm produces the best results from the point of view of retaining the size of the animal (Note the slight difference in the shape of the mountains in seam carving output and our output). We perform perhaps a bit inferior in terms of retaining the white edge below the mountains as compared to the method of [66].

...

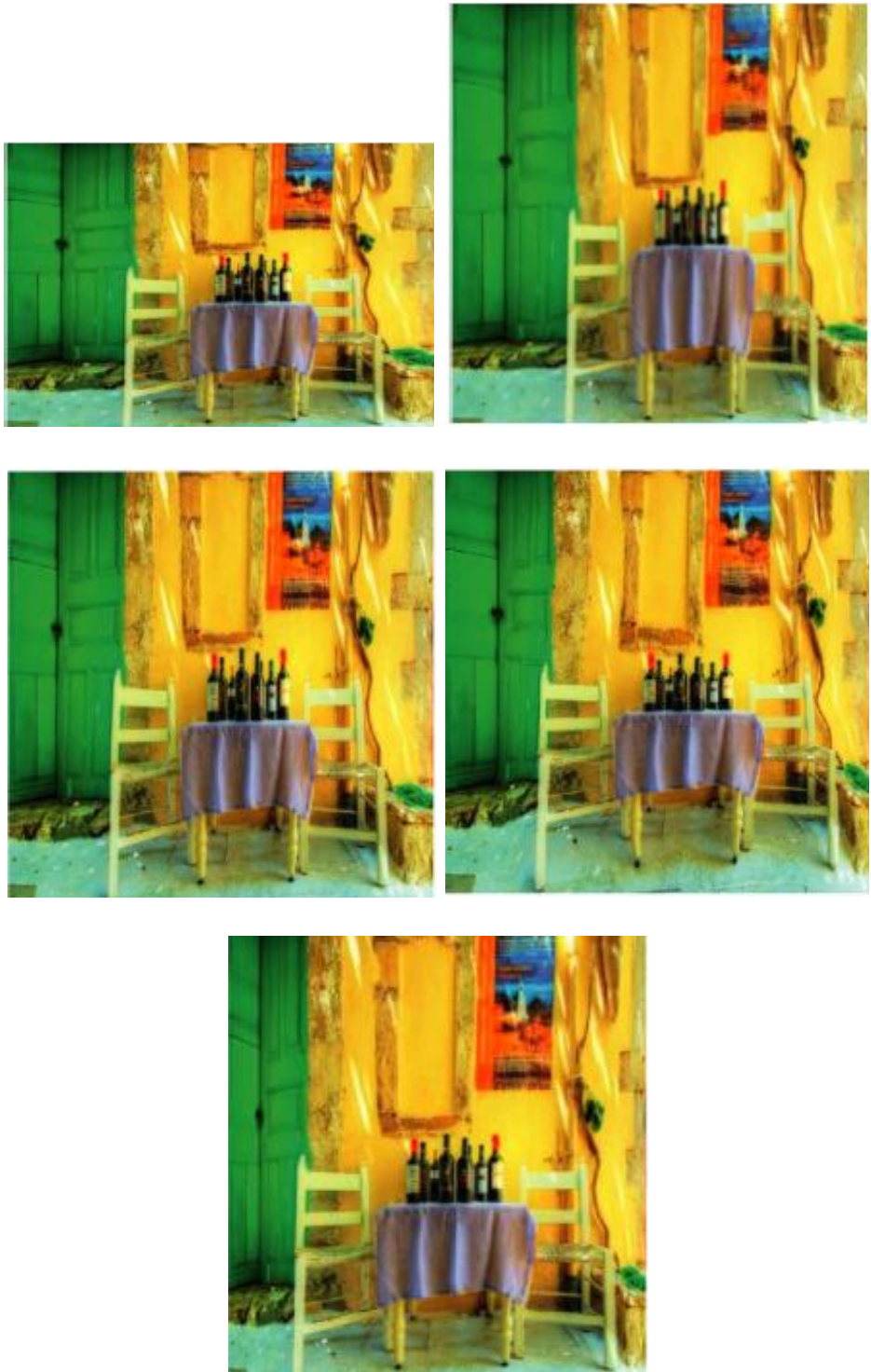

**Fig. 4-3** *(Top Row - Left to Right)* Input Image *[Image Adapted from [65] (©IEEE)],* Retargeted Image using seam carving of [12], (Middle Row – Left to Right) Retargeted Image using warping of [61], Retargeted image using [65] *[Image Adapted from [65] (©IEEE)], (Bottom Row)* Retargeted Image **using our full algorithm.** The retargeting is done for a 30% increase in the height with no change in the width of the input image.

It can be seen from *Fig. 4-3* that our output does better than [65]; but, is perhaps marginally inferior to that of [61] in terms of the entire edge preservation of the main window towards the top of the image. However, we preserve the aspect ratio of the table in a better way, as the result of [65] also suggests.

From here on, we compare our results to the method of seam carving of [12] and the method of warping of [61] since the respective authors have made utilities/code available for us to generate their results with our set of input images. For other images given in [65], [66], our results are more or less the same, and researchers in these papers have generally given their results for only those set of input images, where their algorithm works well or almost . The failure cases are generally not given in the publications. Thus, without the availability of their utilities/ code for testing their results for our set of input images, it becomes difficult to compare out approach to theirs with true essence.

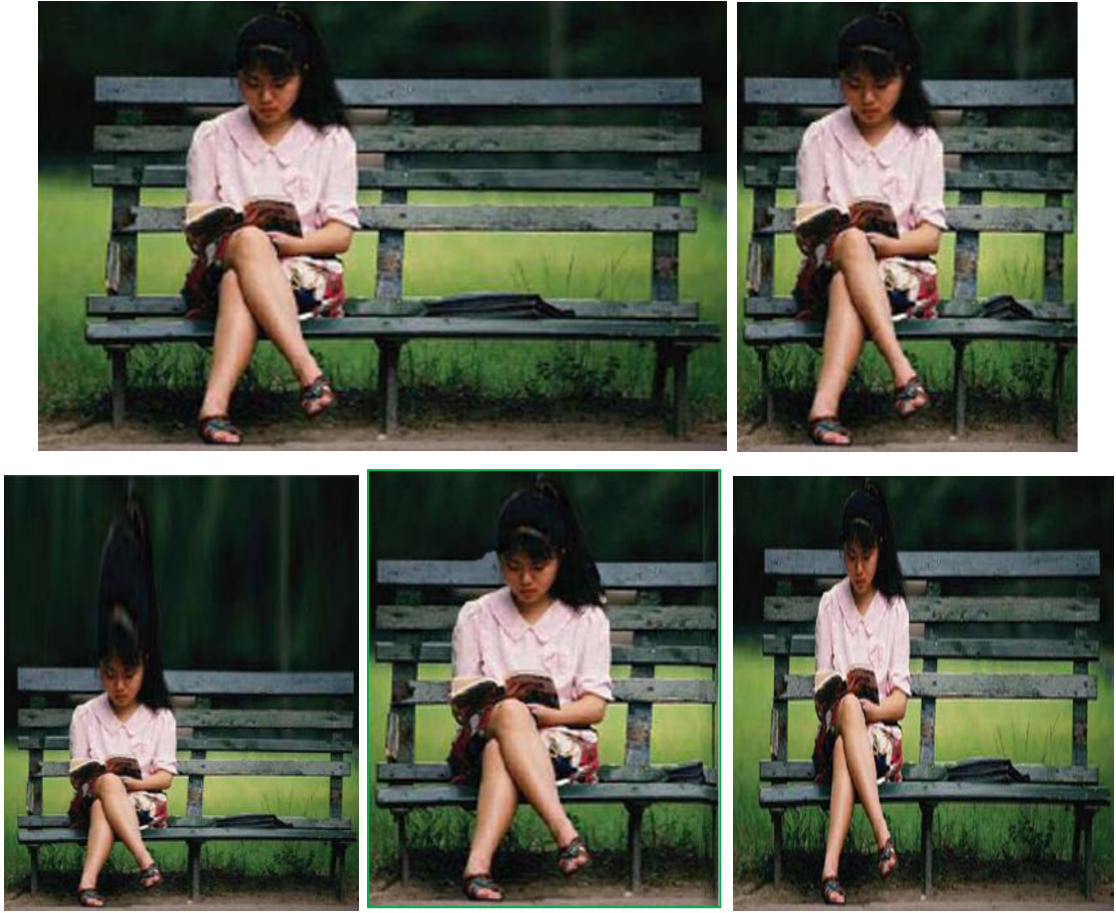

**Fig 4-4** *(Top Row – Left to Right)* Input Image *[Image adapted from http://www.asse.com/united_states/thailand.htm]*, Retargeted Image using seam carving of [12] *(Bottom row – Left to Right)* Retargeted image using warping of [61], Retargeted Image **using our algorithm,** Uniform Scaling Result. The retargeting is done for a 50% reduction in the width of the input image without a change in the original height.

It can be seen from *Fig. 4-4* that our result improves upon the warping of [61] in a huge way. The result of seam carving is the best, with our result nearly there. There occurs a slight distortion in our image near the right end in the edges. In the *Fig. 4-5* below, our algorithm depicts better results than both of seam carving and warping based algorithm of [61]. Our algorithm is very nicely able to distribute the distortion in both directions by enlarging the building portion of the image. However, the shoe of the girl gets stretched while the floor gets elongated. This can be attributed to a failure case where the region based segmentation essentially fails with the shoe of the girl being of the same color as the floor.

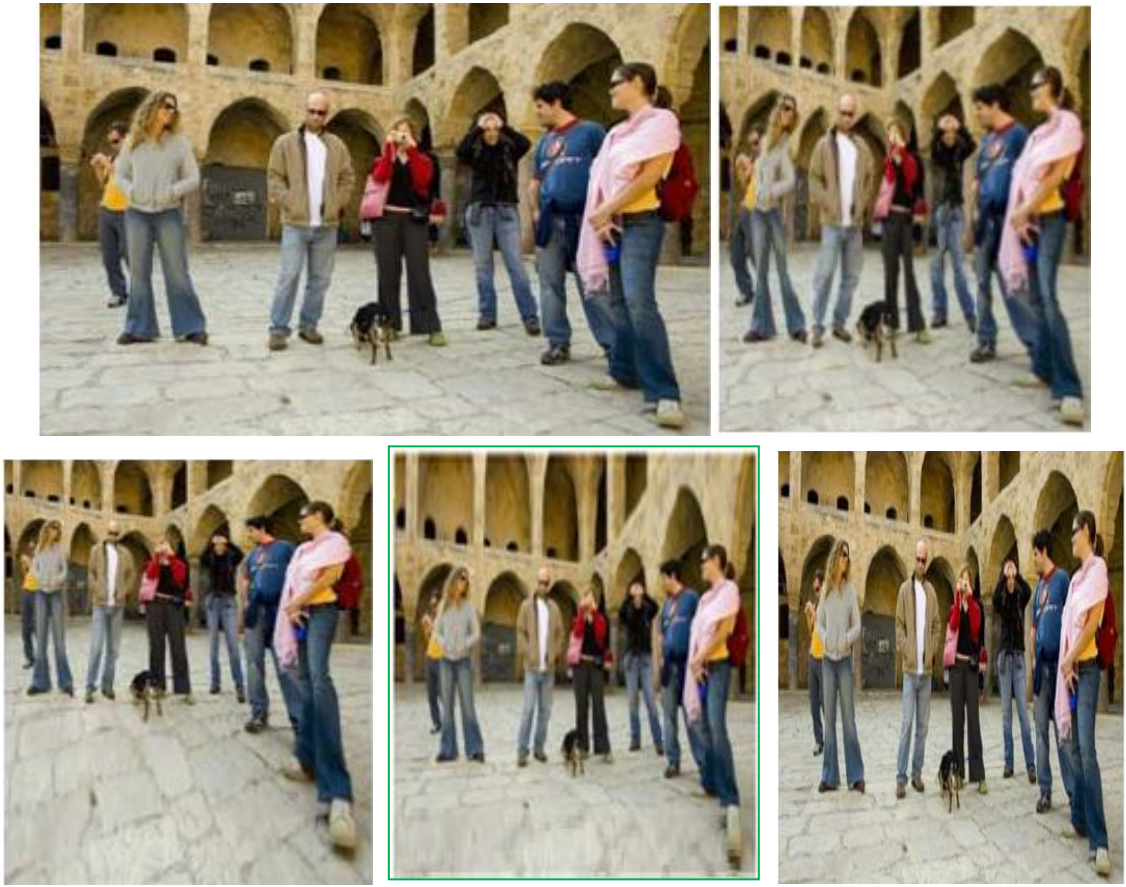

**Fig 4-5** *(Top Row – Left to Right)* Input Image *[Image adapted from [74] (© IEEE)]* Retargeted Image using seam carving of [12] *(Bottom row – Left to Right)* Retargeted image using warping of [61], Retargeted Image **using our algorithm,** Uniform Scaling Result. The retargeting is done for a 50% reduction in the width of the input image without a change in the original height.

*Fig. 4-6* depicts one of the other examples where our algorithm clearly performs better than seam carving of [12] and the optimized scale and stretch warping method of [61]. *Fig. 4-7* presents one of the cases where one can say that our algorithm performs better; however given the nature of the image, the decision making more or less depends on the analysing subject. In *Fig. 4-10*, it can be seen that our algorithm produces almost similar results as of uniform scaling; however, given the nature of the image, scaling looks best.

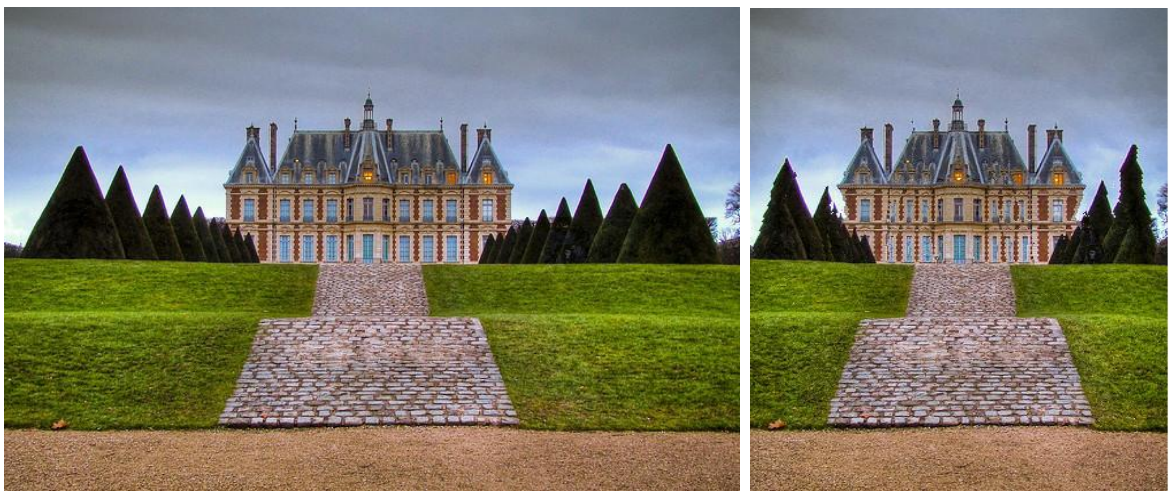

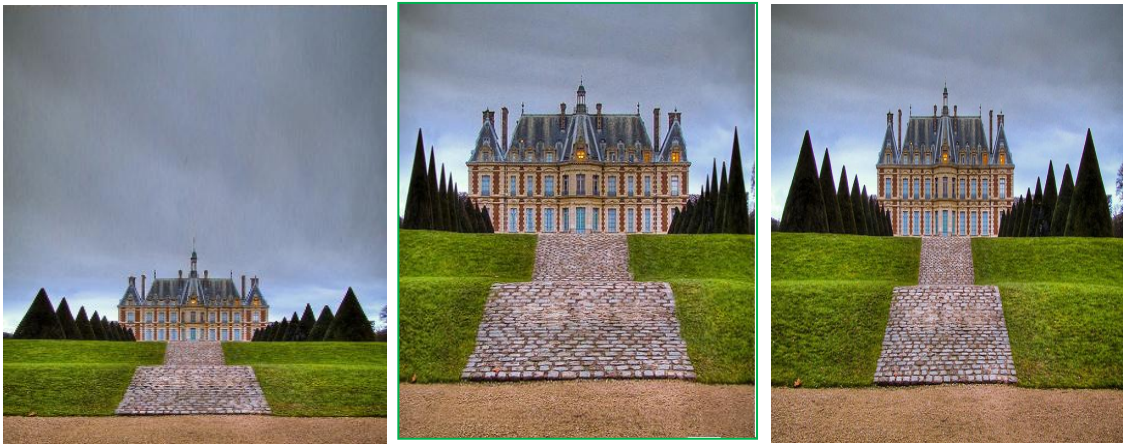

**Fig 4-6** *(Top Row – Left to Right)* Input Image *[Image courtesy – http://www.flickr.com/photos/ayushbhandari/2172625468 (Château in the Parc de Sceaux)]* Retargeted Image using seam carving of [12] *(Bottom row – Left to Right)* Retargeted image using warping of [61], Retargeted Image **using our algorithm,** Uniform Scaling Result. The retargeting is done for a 50% reduction in the width of the input image without a change in the original height.

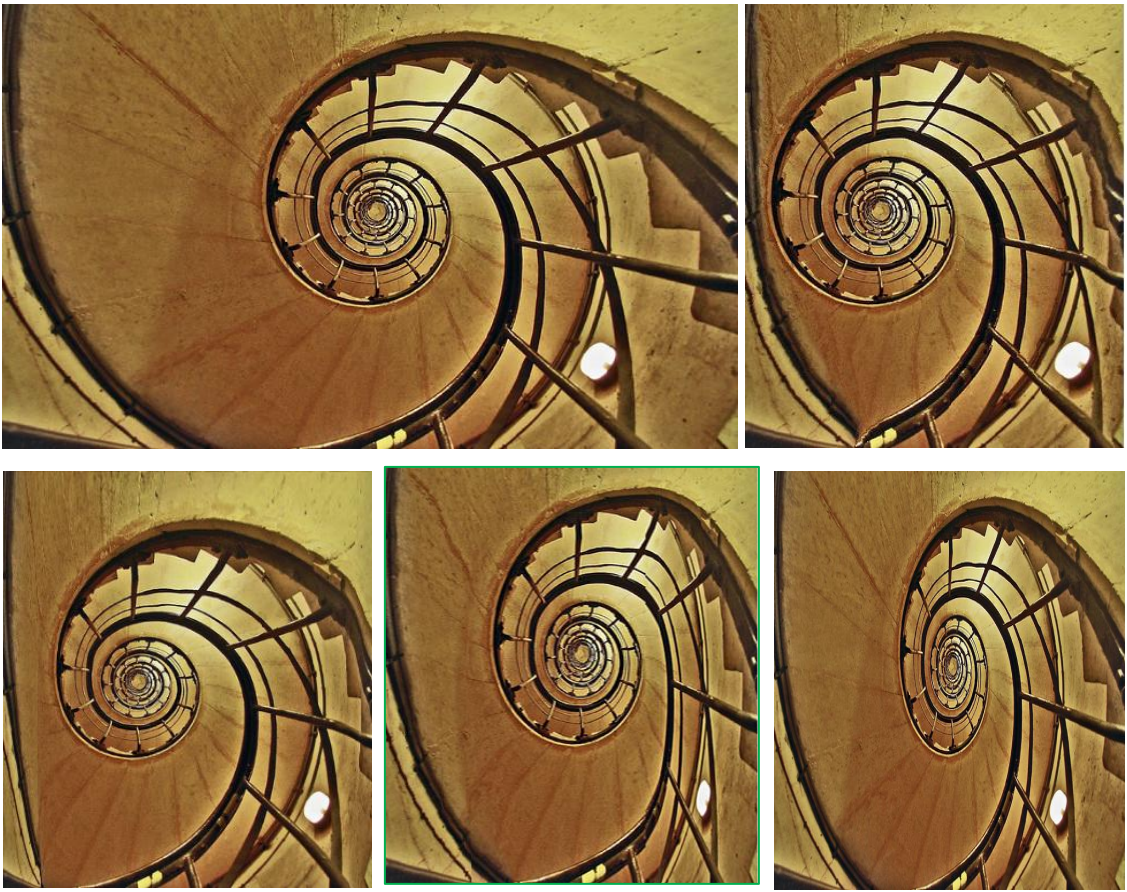

**Fig 4-7** *(Top Row – Left to Right)* Input Image *[Image Courtesy – http://www.flickr.com/photos/ayushbhandari/2164951646/ (Spiral Stairway to the top of Arc de Triomphe)]*, Retargeted Image using seam carving of [12] *(Bottom row – Left to Right)* Retargeted image using warping of [61], Retargeted Image **using our algorithm**, Uniform Scaling result The retargeting is done for a 50% reduction in the width of the input image without a change in the original height.

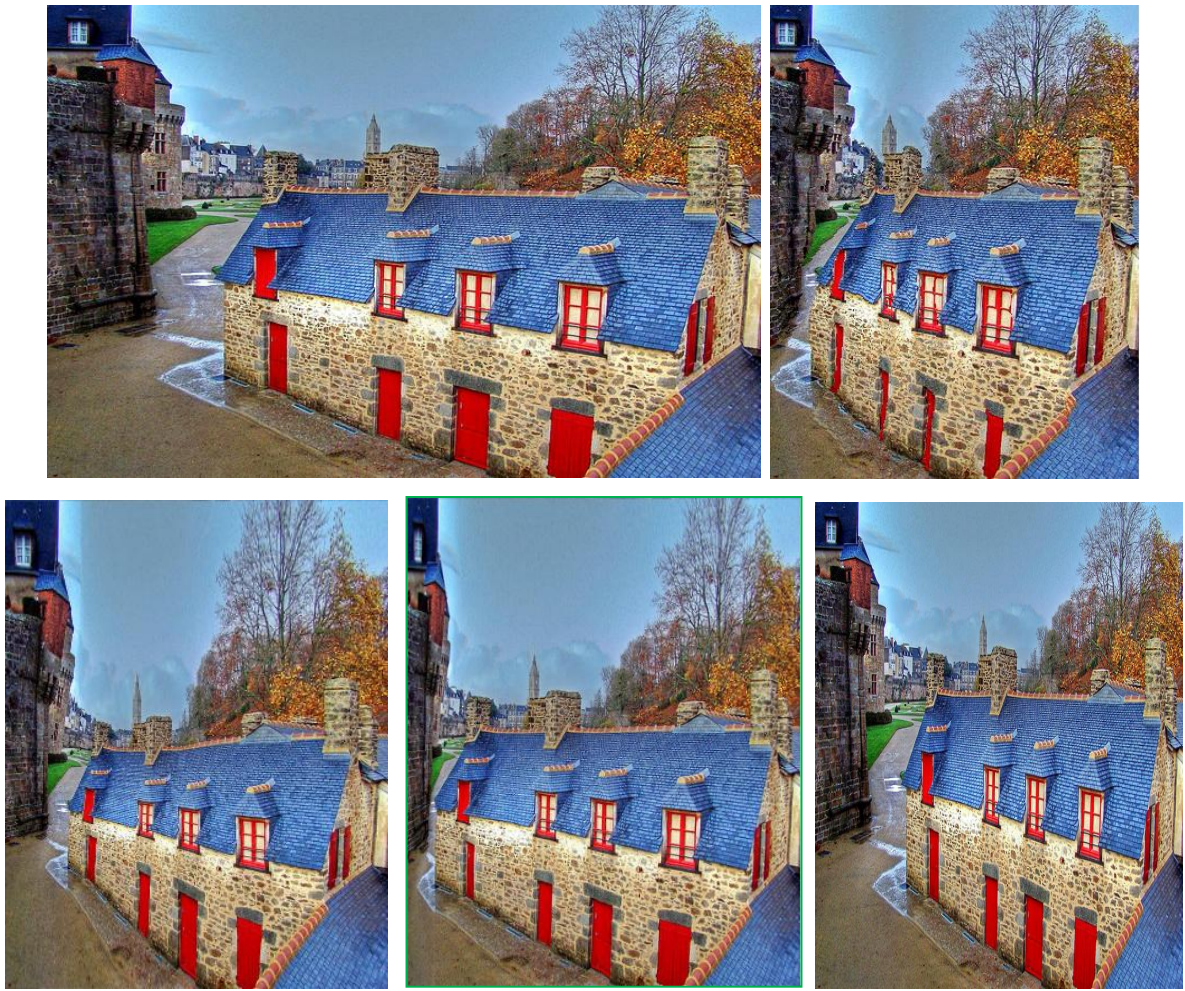

**Fig 4-8** *(Top Row – Left to Right)* Input Image *[Image Courtesy – http://www.flickr.com/photos/ayushbhandari/2397969443/ (House with Red Doors, Vannes)],* Retargeted Image using seam carving of [12] *(Bottom row – Left to Right)* Retargeted image using warping of [61], Retargeted Image **using our algorithm,** Uniform Scaling results. The retargeting is done for a 50% reduction in the width of the input image without a change in the original height.

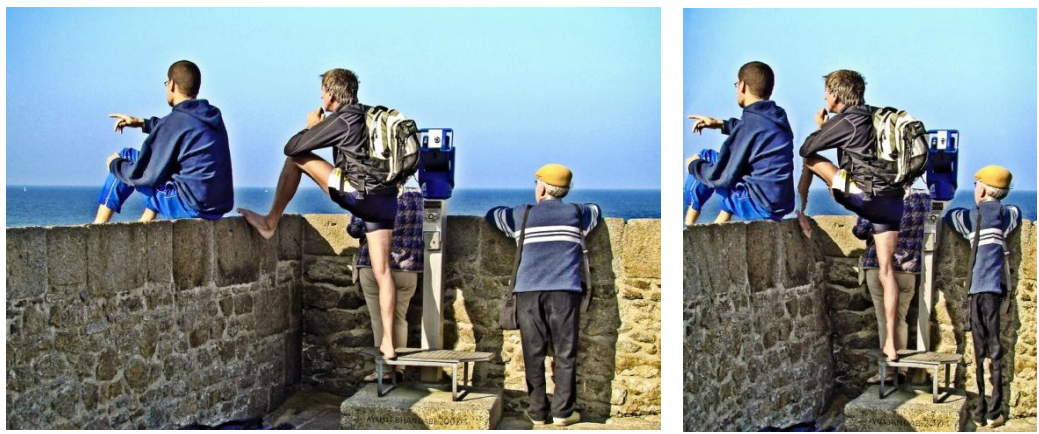

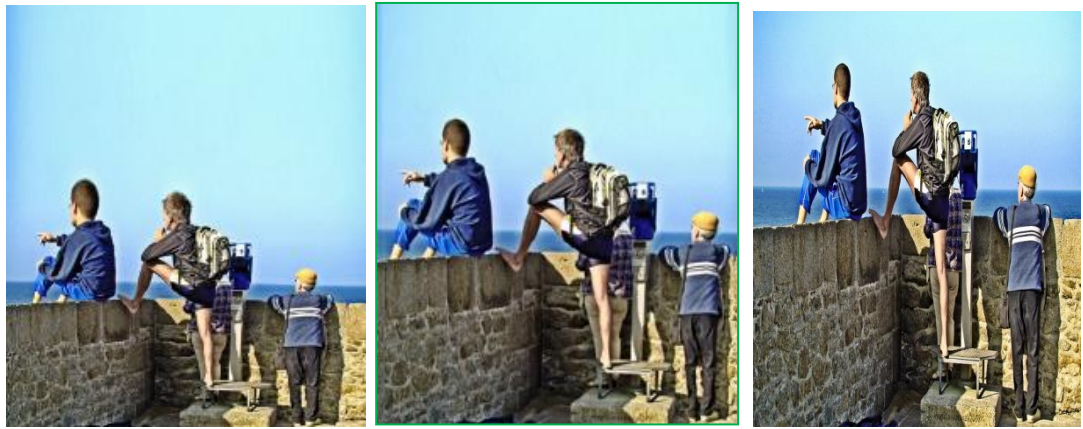

**Fig 4-9** (*Top Row – Left to Right*) Input Image *[Image courtesy - http://www.flickr.com/photos/ayushbhandari/2054189454/ (The World Tomorrow)],* Retargeted Image using seam carving of [12] *(Bottom row – Left to Right)* Retargeted image using warping of [61], Retargeted Image **using our algorithm**, Uniform scaling result. The retargeting is done for a 50% reduction in the width of the input image without a change in the original height.

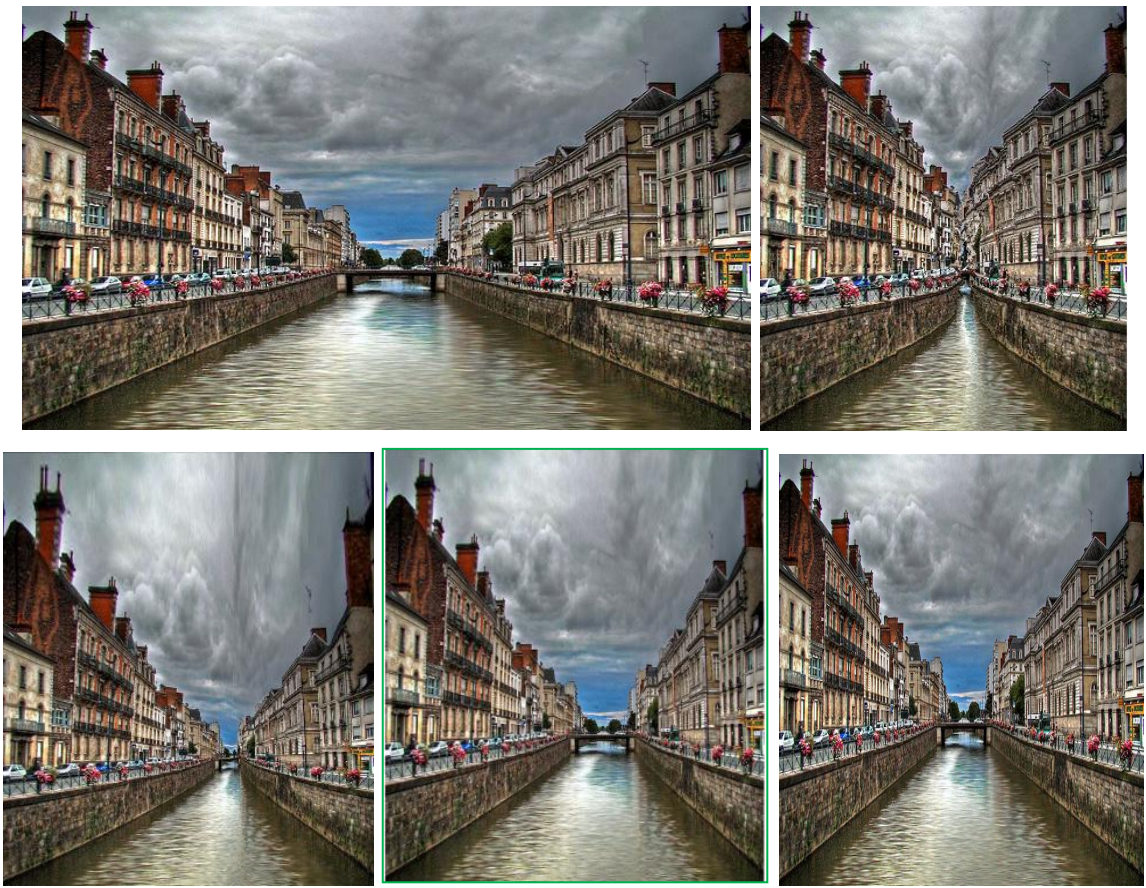

**Fig 4-10** *(Top Row – Left to Right)* Input Image *[Image courtesy – http://www.flickr.com/photos/ayushbhandari/2330980893/ (Stormy Sky over Rennes!)].* Retargeted Image using seam carving of [12] *(Bottom row – Left to Right)* Retargeted image using warping of [61], Retargeted Image **using our algorithm**, Uniform scaling result. The retargeting is done for a 50% reduction in the width of the input image without a change in the original height.

# 5 Conclusions and Future Work

We have discussed the various techniques used for content aware image retargeting including the state-of-the-art algorithms, and proposed our novel perspective of content aware image retargeting using a region based segmentation driven approach. Besides, we have also delineated the commonly occurring problems in the state-of-the-art algorithms and presented the nuances of our algorithm to minimize those problems as far as possible.

In conclusion, it would not be wrong to say that given an image and a desired target aspect ratio, a single algorithm cannot be deemed to work satisfactorily. This is because of the major constraint that content aware image retargeting problem has not been approached by analysing the real semantics of the images. Analysing semantics in tough and unreliable besides being computationally complex; however, for content aware image retargeting, the knowledge of the semantics need to be only limited most of the times for acceptable results. In our approach, we have presented a segmentation based method for exploiting the semantics of the image before resizing. As shown in the results, our algorithm normally performs better than other widely used algorithms for the same purpose.

Our perspective presents one of the initial methods for a semantics driven content aware image retargeting, which works for a better variety of images. We have also used novel smoothing approaches for protecting the features as far as possible. However, the proposed algorithm can be improved by integrating efficient line and object boundary detection techniques for better feature preservation. Also, the method may be combined with a better segmentation approach which can segment the regions more efficiently and in a content aware manner. The perspective can be modified to include various symmetry considerations (including those which comprise of occlusions as well) along with an efficient method of segregating the symmetry and non-symmetry based regions. With all of these, the constraints and the functions thus designed can be extended for arbitrary manifolds instead of rectangular shapes of input images only.